\documentclass[a4paper,11pt]{article}
\usepackage[utf8]{inputenc}
\usepackage{amsmath, amssymb}
\usepackage{geometry}
\usepackage{graphicx}
\usepackage{booktabs}
\usepackage{caption}
\usepackage{hyperref}
\usepackage{array}
\usepackage{braket}
\usepackage[numbers]{natbib}
\usepackage{placeins}
\numberwithin{figure}{section}

\title{The Mathematics of Questions}
\author{Ruadhan O'Flanagan\thanks{rof@ruadhan.net}}
\date{March, 2025}

\begin{document}

\maketitle

\begin{abstract}
{\footnotesize I report the existence of exactly one non-trivial solution to the equation \( i(A,B) + i(A,\neg B) + i(\neg A,B) + i(\neg A,\neg B) = 0 \), where \( i(A,B) = \log \frac{P(A \text{ and } B)}{P(A)P(B)} \), and \( P(A) \) is the probability of the proposition \( A \). The equation specifies an information balance condition between two logical propositions, which is satisfied only by full independence and by this new solution. The solution is a new elementary informational relationship between logical propositions, which we denote as \( A \sim B \).

The \( \sim \) relation cannot be expressed as a relationship between probabilities without the use of complex numbers. It can, however, be greatly simplified by expressing each proposition as a combination of a question and an answer, for example, writing, ``All men are mortal'', as (Are all men mortal?, Yes).

We will study the mathematics of questions and find out what role the \( \sim \) relationship plays inside the algebra. We will find that, like propositions, questions can act on probability distributions. A proposition, \( X \), can be given, setting \( P(X) \) to 1. The question of \( X \) can be raised, setting \( P(X) \) to \( 1/2 \). Giving the proposition adds information to the probability distribution, but raising the question takes information away. Introducing questions into probability theory makes it possible to represent subtraction of information as well as addition.

We will then study how questions can be oriented in space and related to each other geometrically. Remarkably, the simplest way of orienting questions in space has the same structure as the simplest quantum system---the two-state system. We will find that the essential mathematical structure of the two-state quantum system can be derived from the mathematics of questions, including non-commutativity, complementarity, wavefunction collapse, the Hilbert space representation and the Born rule, as well as quantum entanglement and non-locality.}
\end{abstract}

\section*{Introduction}
Imagine a committee of bureaucrats who have no interest in the truth, and only care about which questions have been settled. They would say things like, ``If question 1 has been settled, and question 2 has been settled, then question 3 has been settled.'' If they were asked about the answer to question 3, they would say, ``We don't keep track of that. We can only tell you that the answer to question 3 is knowable. There is enough information to determine it.''

We can acknowledge that the bureaucrats are using a valid form of reasoning, and we can acknowledge that it's useful to know what is knowable, but we've never had a reason to believe that this form of reasoning would be worth studying. We've never had an incentive to study the mathematics of questions.

Today, we do have an incentive. We have found an unexpected solution to the equation \( i(A,B) + i(A,\neg B) + i(\neg A,B) + i(\neg A,\neg B) = 0 \), where \( i(A,B) = \log \frac{P(AB)}{P(A)P(B)} \) and \( AB \) means \( A \) and \( B \).

As we will see, this solution is an elementary relationship between questions, which we denote as \( \sim \). The solution is, however, very complex, and to understand exactly what \( \sim \) is, we will need to undertake a systematic examination of the mathematics of questions: Their algebras, the quantities associated with them, and how they act on probability distributions.

After we have studied questions, we will understand what the constraint means, and how \( \sim \) is used in symbolic reasoning. We will also be able to simplify the mathematical expression for \( \sim \) so that it becomes intelligible as a sequence of transformations on shapes in two-dimensional space and in the complex plane. We will see that it is using cubic functions to perform origami in the complex plane to achieve the symmetry required by the constraint, \( i(A,B) + i(A,\neg B) + i(\neg A,B) + i(\neg A,\neg B) = 0 \).

Meaningful real and complex quantities related by these transformations are difficult to identify. The \( \sim \) relation has more to tell us in the future. So far, it has told us that we should study questions. When we do this, we discover the distinction between unsigned (pure) questions and signed (askable) questions.

The pure question of \( A \) is the question of whether or not \( A \) is true. It's a topic, not a query. It can't be given a Yes/No answer because it doesn't specify what is being asked.

The askable question of \( A \) is the question, ``Is \( A \) true?''. It differs from the topic (whether or not \( A \) is true) by specifying that what is being asked is whether \( A \) is true, and not whether \( A \) is false. When we write ``All men are mortal'' as (Are all men mortal?, Yes), the question, ``Are all men mortal?'', is an askable question.

Inside a proposition, then, we find the following parts:

\begin{itemize}
    \item A pure question,
    \item A specification of what to ask, which, together with the pure question, makes an askable question, and
    \item An answer, which, together with the askable question, makes a proposition.
\end{itemize}

When we study the algebra of pure questions, we will also learn that a pure question can be further broken down into a subject and a predicate, which are specific abstract objects within the algebra of pure questions. Subjects have an algebra of their own, and we will see that the \( \sim \) relation is most clearly understood as a way to combine questions together into natural subjects.

The algebra of pure questions will also indicate an optimal way to write a question in terms of other questions. This optimal method of representation, when extended to propositions (by specifying what to ask and then adding an answer), is already known, and is called Algebraic Normal Form.

We will then turn our attention to askable questions, examining their algebra and how it relates to the algebra of pure questions. We will find, using the mathematics of askable questions, that the complex expression for the \( \sim \) relation can be simplified and partially understood.

The structure of the \( \sim \) relation will reveal that a property of the pure question and a property of the askable question can be combined into a single complex number. These complex-valued properties of questions impose a strong constraint on the complex functions that can relate them, namely \( f(-z^*) = -f(z)^* \). The simplest non-linear function that satisfies this constraint is \( f(z) = z^3 \). We will understand what this implies about the way that complex-valued properties of questions can be related to each other.

Performing a measurement corresponds to asking an askable question. The result of the measurement is an answer to the question. Together, the question and answer form a proposition, which updates the probability distribution by being given. The algebra of possible physical measurements is consequently related to the algebra of askable questions, which will we study here. We will see that the quantities which are natural properties of the measurement are those that relate to the askable question, rather than to the proposition that asserts the answer.

By considering askable questions in the context of geometry, we will make the most striking discovery: The simplest way of orienting askable questions in space has the same structure as the simplest quantum system. This is the two-state quantum system, which describes how a spin-1/2 particle behaves when the direction of its spin is measured.

We will also learn that incorporating questions into probability theory makes it possible to mathematically represent the collapse of the wavefunction that occurs after a quantum measurement. What distinguishes classical measurements from quantum measurements is that, in classical measurements, information is gained and no information is lost. This can be represented mathematically as \( P(W) \to P(W|X) \). A proposition is given.

In quantum measurements, information is gained and other information is lost. The loss of information from a probability distribution cannot be represented by giving a proposition. However, a pure question can act on a probability distribution by setting the corresponding probabilities to \( 1/2 \). We refer to this action as raising the question, \( x \), using lower-case letters to represent the pure questions of the corresponding propositions:

\[ P(W) \to \frac{P(W|X) + P(W|\neg X)}{2} = P(W|x). \]

The collapse of the wavefunction was mysterious, in part because it involved subtracting information from a probability distribution, and we had no way to represent that. We did not know which algebraic object was involved, or what its numerical properties were. Now we know, at least in the case of the two-state system, that the subtraction can be represented by raising questions, and we understand the algebra of questions and know their numerical properties.

We will see that the Hilbert space representation of the two-state system can be derived from the simplest way to orient askable questions in space. We'll also see that quantum amplitudes can be understood as complex-valued properties of questions.

Examining how questions relate to wavefunctions will reveal the first major negative result that comes from introducing questions into quantum mechanics: The value, \( \psi(x) \), is not a valid complex-valued property of the question, ``Is the particle at position \( x \)?''. It's a property of a different question.

By using the requirements of complex-valued properties, we'll find a strong indication that \( \psi(x) \) is a property of the question of whether the particle's presence at \( x \) should be treated as a preparation or a detection.

The Born rule, which states that the probability of getting a result is the square of the magnitude of the quantum amplitude, can be understood as a consequence of the fact that askable questions, not propositions, are the objects that are oriented in space during quantum measurements. The rule expresses a trigonometric relationship between askable questions and propositions.

In the case of quantum entanglement in two-state systems, entangled states can be expressed using the simplest natural group operation on questions. The violations of Bell's inequalities can be understood as consequences of raising non-local questions. These cannot be used to signal faster than light, because they only take information away.

Our results, such as the derivation of the Hilbert space and the Born rule, and the algebraic representation of the collapse of the wavefunction, apply only to the two-state quantum system. We will not investigate wave mechanics, quantum interference, time-evolution, quantum field theory, higher-spin systems or quantum gravity.

Our task here is to understand the elementary mathematics of questions and how it immediately and obviously relates to quantum mechanics. A solid foundation will make it possible to approach more complex systems in the future.

We start off without any foundation, without knowing anything about questions. All we have is an unexpected and very complicated solution to a very simple equation, which seems to describe something at the center of the intersection of logic and information. We begin this investigation, not as physicists, but as mathematicians.

Section \ref{sec:information-logical-propositions} provides an expository introduction to the quantity, \( i(A,B) \), distinguishing it from mutual information. Section \ref{sec:unexpected-discovery} presents the novel non-trivial solution to \( i(A,B) + i(A,\neg B) + i(\neg A,B) + i(\neg A,\neg B) = 0 \). Sections \ref{sec:introducing-questions} to \ref{sec:structure-knowledge} develop the algebraic and numerical properties of questions, and their relationship to the \( \sim \) relation. 

The remaining sections relate the geometry of questions to the two-state quantum system, covering measurement, the Hilbert space formulation, the Born rule, and quantum entanglement. A quick glimpse at the connection between questions and quantum physics can be achieved by reading sections \ref{subsec:signed-probability-gaps} and \ref{sec:geometry-questions}.

On a first reading, the reader should skip any mathematics that appears tiresome. The code used to generate the figures in the paper is available at \url{https://github.com/roflanagan/questions}, along with code that verifies important results.

\subsection*{Main Results}

Before we turn our attention to physics, we complete a systematic study of the mathematics involved. From the mathematical point of view, the main results can be summarized as:

\begin{itemize}
    \item A novel, highly complex, solution to the equation, $i(A,B) + i(A,\neg B) + i(\neg A,B) + i(\neg A,\neg B) = 0$.
    \item A detailed specification of the group structures formed by questions, and their associated numerical properties.
    \item The identification of the exact role that the solution plays in the algebra of questions.
    \item A simplification of the complex solution when reformulated in terms of questions.
    \item The simplification reveals how complex numbers can be used to combine properties of real and askable questions to yield complex-valued properties of questions, and allows us to identify which complex functions coherently relate these properties.
    \item Introducing questions into probability theory generates a non-commutative algebra of actions on probability distributions, which can represent subtraction as well as addition of information. We identify the role of the new solution in this algebra.
\end{itemize}

When that mathematics is applied to physics, the main results are:

\begin{itemize}
    \item Askable questions which are related to each other geometrically, through cosine projections, produce the same probabilities as the two-state quantum system, in a much simpler way.
    \item Adding questions into probability theory makes it possible, at least in two state systems, to represent the effects of quantum measurements on the probability distribution algebraically, in terms of raising questions.
    \item The Hilbert space structure can be derived from a sphere of askable questions. The probabilities encoded in the sphere of questions match up with the probabilities predicted by the Hilbert space formulation, making it truly a connection between the Hilbert space and a sphere of questions, rather than just a sphere.
    \item The Born rule can be derived from the geometric relationships between the questions, without having to assume it as a postulate or rely on symmetry arguments.
    \item Entangled states can be represented using the natural group operation on questions, and quantum non-locality can be understood as the consequence of raising non-local questions.
    \item How quantum non-locality can coexist with special relativity is partially clarified by the observation that the non-local action on the probability distribution is a pure subtraction of information, making it impossible to use that effect to send a signal faster than light.
\end{itemize}

The algebraic structures and equations derived are presented in a systematic format in the two appendices, with Appendix A covering pure mathematics, and Appendix B covering physics.

\pagebreak
\tableofcontents

\pagebreak

\section{Information and Logical Propositions}
\label{sec:information-logical-propositions}

Everything that follows comes as a consequence of the decision to study the equation: 

\begin{align}
i(A,B) + i(A,\neg B) + i(\neg A,B) + i(\neg A,\neg B) = 0. 
\end{align}

The fact that this equation has a non-trivial solution is unexpected. Before we look at the solution, let's understand what the equation says.

The foundational concept is the relationship between probability and information: The logarithm of the reciprocal of a probability is a measure of information \cite{Shannon1948}: \( i = \log \frac{1}{P} \).

If the probability of a proposition, \( A \), is \( 1/2 \), then the amount of information, \( i(A) \), gained by learning that \( A \) is true would be \( \log_2 \frac{1}{P(A)} = \log_2 2 = 1 \) bit of information. If \( P(A) = 1/4 \), then learning that \( A \) is true would provide two bits of information, and so on.

Information can be measured in different units, depending on the base of the logarithm used. It's convenient for us to use base 2, because then we can measure information in bits.

We can use either \( i(A) \) or \( P(A) \) to describe what we know about \( A \), because they are related through the formula \( i(A) = -\log P(A) \). Which one is more convenient to use depends on how the propositions involved are related:

\begin{itemize}
    \item Probability adds up for mutually exclusive propositions joined together by OR: \\ If \( P(A \text{ and } B) = 0 \), then \( P(A \text{ or } B) = P(A) + P(B) \).
    \item Information adds up for independent propositions joined together by AND: \\ If \( P(A \text{ and } B) = P(A)P(B) \), then \( i(A \text{ and } B) = i(A) + i(B) \).
\end{itemize}

If we're dealing with a collection of mutually exclusive propositions, we do our reasoning using probability, because then we can add probabilities. If we're dealing with independent propositions, we use information, because that's the quantity that adds up.

We can conclude that \( A \) is true if \( P(A) \) reaches 1 or \( i(A) \) reaches zero. So we can reason, and reach rational conclusions, using either probability or information \cite{Jaynes2003}.

We don't need to adopt a particular interpretation of whether probability, when applied in the real world, inherently describes frequencies, randomness, or subjective uncertainty \cite{Cox1961,Jaynes2003}. The fact that probabilities add up for mutually exclusive propositions entitles us to use probabilities in reasoning.

Likewise, information adds up for independent propositions, and that makes it useful for reasoning. Information specifies things, and the size of a piece of information measures how much it can specify.

In our case, we only need to clarify that we are dealing with information in the context of logical propositions and reasoning. When information theory is applied to real-world problems, such as communication, it usually deals with aggregate quantities of information, such as the entropy of a message, or the mutual information between two probability distributions.

We will be dealing with individual quantities of information, not averages or aggregates. Each quantity of information will be related to specific propositions, and will relate those propositions in a specific way.

Let's look at the foundational example of a quantity of information that relates two propositions:

\begin{align}
i(A,B) = i(A) + i(B) - i(AB)
\end{align}

In words, this quantity is equal to the amount of information that would be gained if \( A \) was discovered to be true, plus the corresponding amount for \( B \), minus the amount of information that would be gained if it was discovered that both \( A \) and \( B \) are true.

Although it has the form, \( \log \frac{P(AB)}{P(A)P(B)} \), of a term that occurs in the formula for mutual information, it can be positive, negative or zero, and it is not an average. It's a quantity that relates propositions, not sets, messages or probability distributions.

If this quantity is equal to zero, then \( i(AB) = i(A) + i(B) \), which implies that \( P(AB) = P(A)P(B) \), meaning that \( A \) and \( B \) are independent.

If it is positive, then \( i(AB) \) is smaller than \( i(A) + i(B) \). This means that there is an overlap between \( i(A) \) and \( i(B) \). There is some information asserted by both \( A \) and \( B \).

For example, if \( C \), \( D \) and \( E \) are independent propositions, and \( A = CD \) and \( B = DE \), then the proposition \( D \) expresses the information that \( A \) and \( B \) have in common. \( AB = CDE \), which implies:

\begin{align}
i(A,B) &= i(A) + i(B) - i(AB) \\
& = i(C) + i(D) + i(D) + i(E) - (i(C) + i(D) + i(E)) \\
& = i(D). 
\end{align}

So, if \( i(A,B) \) is positive, it's the amount of information that \( A \) and \( B \) have in common. More precisely, it's the amount of information that they have in common which is not already known to be true: If \( D \) is known to be true, then \( i(D) = 0 \), so \( i(A,B) = 0 \).

The maximum value \( i(A,B) \) can have is the minimum of \( i(A) \) and \( i(B) \). If \( A \) implies \( B \), then \( i(AB) = i(A) \), and \( i(A) \) is greater than or equal to \( i(B) \). So:

\begin{align}
i(A,B) = i(A) + i(B) - i(AB) = i(A) + i(B) - i(A) = i(B).
\end{align}

This expresses the fact that \( A \) and \( B \) have the maximum amount of information in common when the information that one expresses is contained entirely within the information the other expresses.

Now let's consider what happens when \( i(A,B) \) is negative. In this case, \( i(AB) \) is larger than \( i(A) + i(B) \). If we learn that both \( A \) and \( B \) are true, we gain more than the sum of the information that \( A \) and \( B \) would provide separately.

For example, if \( n \) is a whole number between 1 and 100, then \( A \) could be the proposition that \( n \) is even, and \( B \) could be the proposition that \( n \) is prime. \( A \) provides 1 bit of information, and \( B \) provides 2 bits, since there are 25 prime numbers less than 100. \( A \) and \( B \) in combination, however, imply that \( n = 2 \), which provides \( \log_2 100 = 6.6 \) bits of information.

As a result, the value of \( i(A,B) \) is \( 1 + 2 - 6.6 = -3.6 \) bits, indicating that the conjunction of \( A \) and \( B \) provides 3.6 more bits of information than it would if \( A \) and \( B \) were independent. \( A \) alone would provide 1 bit, narrowing down the pool of possible candidates for \( n \) by 50\%. \( B \) alone would narrow it down to 25\%. In combination, they would reduce it to 12.5\% if they were independent, providing 3 bits of information in total.

In fact, their combination provides 6.6 bits of information, with the additional 3.6 bits of information coming from the fact that two is the only even prime number, so \( AB \Rightarrow n = 2 \).

More generally, when \( i(A,B) \) is negative, it measures how much extra information is implied by the combination of \( A \) and \( B \), beyond what is implied by \( A \) and what is implied by \( B \). Its value becomes \( -\infty \) when \( A \) and \( B \) are exclusive: \( P(AB) = 0 \Rightarrow i(AB) = \infty \Rightarrow i(A,B) = -\infty \).

In summary:

\begin{itemize}
    \item If \( i(A,B) \) is positive, then \( A \) and \( B \) have some information in common.
    \item If it's negative, then \( AB \) implies additional information not implied by \( A \) or by \( B \).
    \item The maximum value it can have is the minimum of \( i(A) \) and \( i(B) \), which occurs when one proposition implies the other.
    \item The minimum value it can have is \( -\infty \), which occurs when \( A \) and \( B \) are exclusive.
    \item If \( i(A,B) = 0 \), then \( A \) and \( B \) are independent.
\end{itemize}

The informational quantity \( i(A,B) \) is able to express logical relations between \( A \) and \( B \), such as independence, exclusivity, and implication \cite{Boole1854}. It shows how logical relations between propositions can be expressed as arithmetic relations between quantities of information.

This means that we can explore the possible logical relations between propositions by exploring equations that relate quantities of information.

\section{An Unexpected Discovery}
\label{sec:unexpected-discovery}

Now that we understand the meaning of \( i(A,B) \), let's look at the original equation:
\begin{align*}
i(A,B) + i(A,\neg B) + i(\neg A,B) + i(\neg A,\neg B) = 0
\end{align*}

The first term, \( i(A,B) \), is how much information \( A \) and \( B \) have in common, if it's positive, and it's how much extra information they imply in conjunction with each other if it's negative.

If it's zero, then \( A \) and \( B \) are independent, which implies that all the other terms are zero too. So independence is a solution to the equation.

Let's suppose that \( i(A,B) \) is positive. Then \( A \) and \( B \) have information in common, \( i(AB) < i(A) + i(B) \), and \( P(AB) > P(A)P(B) \). This implies that \( \neg A \) and \( \neg B \) are also correlated: \( P(\neg A \neg B) > P(\neg A)P(\neg B) \), so \( i(\neg A, \neg B) \) is also positive.

It also implies that \( i(A, \neg B) \) and \( i(\neg A, B) \) are negative, because, for \( A \) to be positively correlated with \( B \), it must be negatively correlated with \( \neg B \).

For the equation above to be satisfied, the positive terms, \( i(A,B) \) and \( i(\neg A, \neg B) \), would need to exactly cancel the negative terms:

\begin{align}
i(A,B) + i(\neg A, \neg B) = -i(A, \neg B) - i(\neg A, B) = |i(A, \neg B)| + |i(\neg A, B)|
\end{align}

This says that the amount of information shared by \( A \) and \( B \), plus the amount shared by \( \neg A \) and \( \neg B \), is equal to the sum of the amounts of excess information implied by the conjunctions of \( A \) with \( \neg B \) and \( \neg A \) with \( B \).

So in order for \( A \) and \( B \) to satisfy the equation, \( i(A,B) + i(A, \neg B) + i(\neg A, B) + i(\neg A, \neg B) = 0 \), and not be independent, the amount of excess information implied by the combinations which are less likely than they would be if they were independent (\( A \neg B \) and \( \neg A B \) in this case), must be equal to the amount of redundant information in the combinations that are more likely, \( AB \) and \( \neg A \neg B \).

The consequences of those quantities of information cancelling out are far from obvious, and it's not obvious that it's even possible. The trivial solution to the equation is independence. If independence is the only solution, then the equation is simply a statement that \( A \) and \( B \) are independent. If there's another solution, then it means that \( A \) and \( B \) can be related in such a way that the information excesses and redundancies in their possible joint outcomes cancel out.

To try to solve the equation, we rewrite it in terms of \( x = P(AB) \), \( a = P(A) \) and \( b = P(B) \):

\begin{align}
&\ \  \frac{P(A B)}{P(A) P(B)} \cdot \frac{P(A \neg B)}{P(A) P(\neg B)} \cdot \frac{P(\neg A B)}{P(\neg A) P(B)} \cdot \frac{P(\neg A \neg B)}{P(\neg A) P(\neg B)} = 1 \\
\Rightarrow & \ \ \frac{x}{a b} \cdot \frac{a-x}{a(1-b)} \cdot \frac{b-x}{b(1-a)} \cdot \frac{1-a-b+x}{(1-a)(1-b)} = 1 \\
\Rightarrow &\ \ x(a-x)(b-x)(1-a-b+x) = a^2 b^2(1-a)^2(1-b)^2
\end{align}

This final equation is a quartic equation, which we can solve using Wolfram Alpha \cite{wolframalpha}. There are four solutions. One is independence, \( x = ab \). The other three solutions are highly complex. The general form of the solutions, involving cubes and cube roots, can be understood as a result of the fact that the quartic equation reduces to a cubic equation when divided by the known solution, \( x - ab = 0 \).

The first non-trivial solution is not a possible probability distribution because \( x \) is greater than \( \min(a, b) \), which means that \( x \) can't be interpreted as \( P(AB) \). The second non-trivial solution can't be a probability distribution either, because \( x \) is negative.

\begin{figure}[t]
    \centering
    \includegraphics[width=0.6\textwidth]{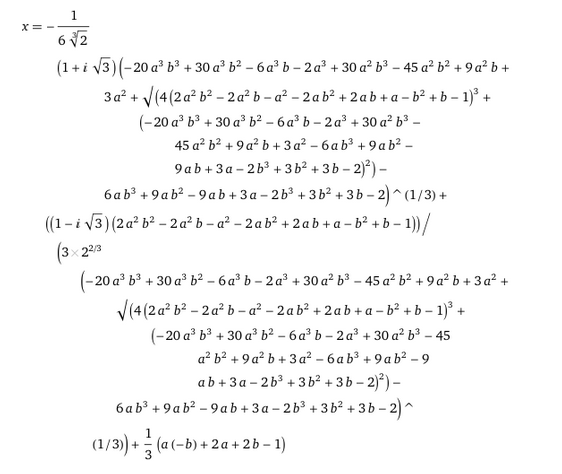}
    \caption{\footnotesize The unique valid non-trivial solution to the constraint, \( x (a - x) (b - x) (1 - a - b + x) = a^2 b^2 (1 - a)^2 (1 - b)^2 \), as specified by Wolfram Alpha. The quartic equation has four solutions, including trivial independence, \( x = ab \), the above solution, and two other solutions of similar complexity, neither of which satisfy both \( x \leq \min(a, b) \) and \( x \geq 0 \), which are requirements for \( x \) to be equal to \( P(AB) \), where \( a = P(A) \) and \( b = P(B) \). The solution above is the only non-trivial valid probability distribution that satisfies the constraint.}
    \label{fig:figure2.1}
\end{figure}

\begin{figure}[t]
    \centering
    \includegraphics[width=0.8\textwidth]{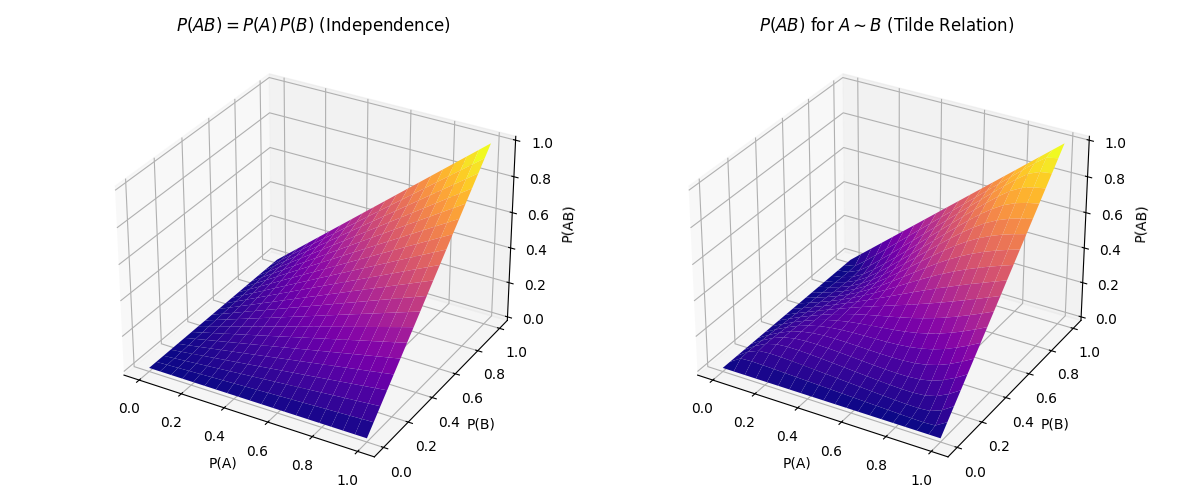}
    \caption{\footnotesize Surface plots showing how \( P(AB) \) depends on \( P(A) \) and \( P(B) \) in two cases. Independence, \( P(AB) = P(A)P(B) \), is shown on the left, and the tilde relation is shown on the right. The surfaces are very similar, but not exactly identical. These two surfaces are the only solutions to the equation, \( i(A,B) + i(A, \neg B) + i(\neg A, B) + i(\neg A, \neg B) = 0 \).}
    \label{fig:figure2.2}
\end{figure}

The third solution is a valid probability distribution. We refer to it as \( \sim \), or the tilde relation. It is shown in Figure~\ref{fig:figure2.1}, and plotted in Figure~\ref{fig:figure2.2}, along with the trivial solution \( x = ab \) for comparison.

\begin{figure}[t]
    \centering
    \includegraphics[width=0.95\textwidth]{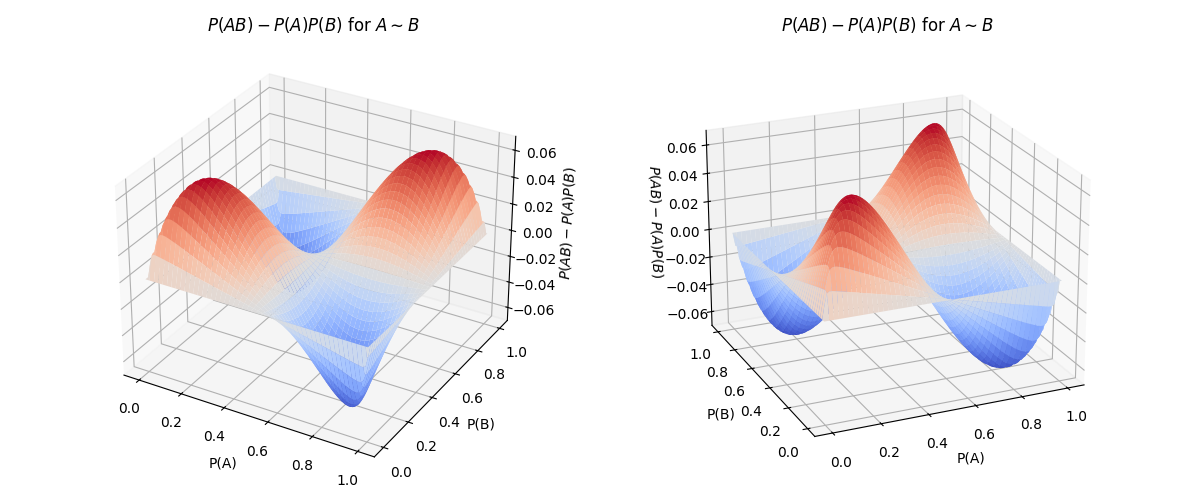}
    \caption{\footnotesize The discrepancy, \( P(AB) - P(A)P(B) \), between the \( \sim \) relation and independence, as a function of \( P(A) \) and \( P(B) \). The discrepancy is exactly 0 when either \( P(A) \) or \( P(B) \) is equal to 0, \( 1/2 \), or 1. When both \( P(A) \) and \( P(B) \) are greater than \( 1/2 \), or both are less than \( 1/2 \), \( P(AB) \) exceeds \( P(A)P(B) \) by an amount whose maximum value is about 0.0674. When \( P(A) \) and \( P(B) \) are on opposite sides of \( 1/2 \), \( P(AB) \) is smaller than \( P(A)P(B) \) by a corresponding amount. The entire surface forms a smooth, symmetric wave over the unit square of probabilities of \( A \) and \( B \).}
    \label{fig:figure2.3}
\end{figure}

The surface formed by \( \sim \) in the 3D space with \( P(A) \), \( P(B) \) and \( P(AB) \) axes is very similar to the surface corresponding to independence. The difference between the tilde relation and independence is shown in Figure~\ref{fig:figure2.3}.

As Figure~\ref{fig:figure2.3} shows, the tilde relation differs from independence by a small discrepancy of \( P(AB) \) from \( P(A)P(B) \), which forms a symmetric wave-like surface over the square of possible values of \( P(A) \) and \( P(B) \). The discrepancy is zero when \( P(A) \) or \( P(B) \) are equal to 0, 1 or \( 1/2 \). The maximum size of the discrepancy corresponds to a change in probability of about 0.0674.

The formula for \( \sim \) provided by Wolfram Alpha can be simplified to:

\begin{align}
\Phi_1 =& -20 a^3 b^3 + 30 a^3 b^2 - 6 a^3 b - 2 a^3 + 30 a^2 b^3 - 45 a^2 b^2 + 9 a^2 b + 3 a^2 \notag \\
& -6 a b^3 + 9 a b^2 - 9 a b + 3 a - 2 b^3 + 3 b^2 + 3 b - 2, \\
\Phi_2 =&\ 2 a^2 b^2 - 2 a^2 b - a^2 - 2 a b^2 + 2 a b + a - b^2 + b - 1.
\end{align}
\begin{align}
\Psi_1 &= \sqrt{4 (\Phi_2)^3 + (\Phi_1)^2}, \\
\Psi_2 &= (\Phi_1 + \Psi_1)^{1/3},
\end{align}
\begin{align}
x(a,b) &= -\frac{(1 + \mathrm{i} \sqrt{3}) \cdot \Psi_2}{6 \cdot 2^{1/3}} + \frac{(1 - \mathrm{i} \sqrt{3}) \cdot \Phi_2}{3 \cdot 2^{2/3} \cdot \Psi_2} + \frac{1}{3} \bigl(-ab + 2a + 2b - 1\bigr).
\end{align}

The code provided verifies that this simplification is correct. It's not easy to see from the form of the solution that it is the only way for \( x \) to satisfy the equation, \( x (a - x) (b - x) (1 - a - b + x) = a^2 b^2 (1 - a)^2 (1 - b)^2 \), apart from \( x = ab \). However, the fact that there is exactly one valid non-trivial solution to this equation can be confirmed with a calculator, by checking that, for \( a = b = 0.25 \), the only values of \( x \) between 0 and 0.25 that satisfy the equation are \( x = ab = 0.0625 \), and \( x = 0.12299828119582 \), which is the result of using the formula above.

The existence of this solution is surprising, and its properties are remarkable:

\begin{itemize}
    \item It is the only way that \( A \) and \( B \) can satisfy \( i(A,B) + i(A,\neg B) + i(\neg A,B) + i(\neg A,\neg B) = 0 \) and not be independent.
    \item It's a weak interaction, very close to independence.
    \item It looks like a type of logical entanglement between \( A \) and \( B \).
    \item It can't be expressed without using complex numbers.
    \item It's a very complicated solution to a very simple equation.
\end{itemize}

It's complicated and uses complex numbers because it's a root of a cubic equation, and the formula for the roots of a cubic equation is complicated and necessarily involves complex numbers.

When we say that it necessarily uses complex numbers, we don't mean that it can't be expressed in a much more complicated form that uses only real numbers. Every application of complex numbers can be expressed in a more complicated form using real numbers, and using trigonometry instead of arithmetic.

What we mean is that, even if it is expressed using a representation of a single complex number as two real numbers, the real numbers naturally combine into a complex number, which is used in complex arithmetic in the formula.

The expression for \( \sim \) provided by Wolfram Alpha is difficult to simplify further. There are large polynomials in \( a \) and \( b \) which occur multiple times in the formula, but there is no obvious way to simplify those polynomials or express them in a way that would allow them to be interpreted as meaningful quantities.

The irremovable presence of complex numbers indicates that the mathematical formula for \( \sim \) is likely to be difficult to understand as a relationship between intelligible quantities. The complexity of the expression makes the challenge even harder. There is little hope of identifying a simple relationship between simple quantities which is stated in symbols by the formula for the tilde relation.

\begin{figure}[t]
    \centering
    \includegraphics[width=0.95\textwidth]{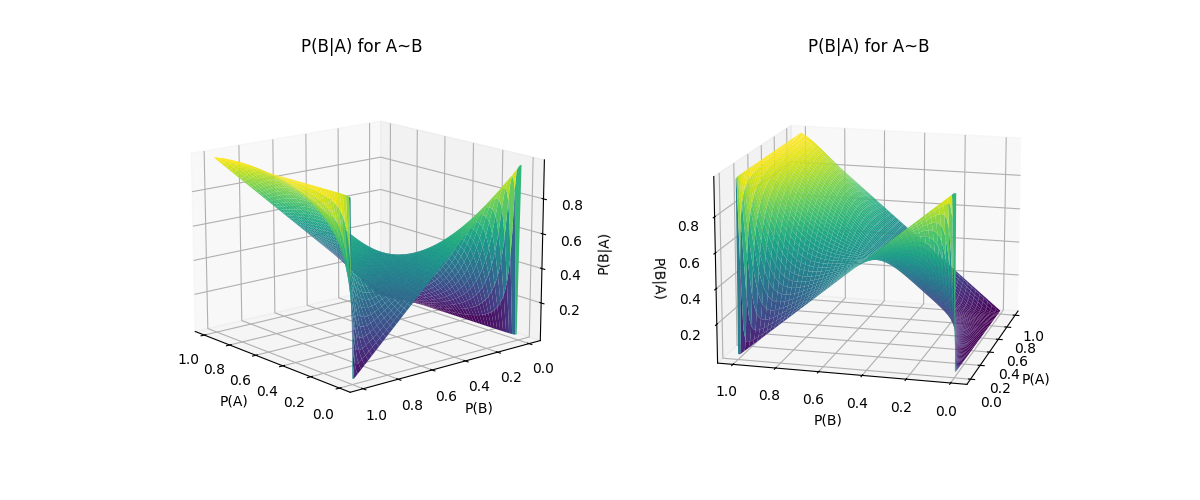}
    \caption{\footnotesize Surface plots of \( P(B|A) \) as a function of \( P(A) \) and \( P(B) \) for the tilde relation. They show that \( P(B|A) = P(B) \) at \( P(A) = 1 \), and \( P(B|A) \) approaches \( P(\neg B) \) as \( P(A) \) approaches 0, except at \( P(B) = 0 \) or 1. The value of \( P(B|A) \) at \( P(A) = 1/2 \) is exactly equal to \( P(B) \). The surface is a cubic interpolation between the three discrete rules that specify the values of \( P(B|A) \) at \( P(A) = 0 \), 1 and \( 1/2 \).}
    \label{fig:figure2.4}
\end{figure}

We can, however, gain further insight into how \( \sim \) relates \( A \) and \( B \) by plotting \( P(B|A) \) as a function of \( P(A) \) and \( P(B) \). This is shown in Figure~\ref{fig:figure2.4}.

The result is the most enlightening insight into \( \sim \) that we have achieved so far. The surface plot reveals that, when \( 0 < P(B) < 1 \):

\begin{itemize}
    \item If \( P(A) = 1 \) then \( P(B|A) = P(B) \).
    \item If \( P(A) = 1/2 \) then \( P(B|A) = P(B) \).
    \item If \( P(A) = 0 \) then \( P(B|A) = P(\neg B) \).
\end{itemize}

The final statement above is a continuation of the plotted surface shown in Figure~\ref{fig:figure2.4}, to the case when \( P(A) = 0 \). The figure shows that, as \( P(A) \) approaches 0, \( P(B|A) \) approaches \( P(\neg B) \). That is, even though the formula \( P(B|A) = P(AB)/P(A) \) can't be used when \( P(A) = 0 \), a unique value of \( P(B|A) \) at \( P(A) = 0 \) is specified by the value that \( P(B|A) \) approaches as \( P(A) \) goes to zero. The complex formula for \( \sim \) can be used to calculate \( P(B|A) \) in the form \( x/a \) only when \( a = P(A) \) is greater than zero, but the relation \( \sim \) extends uniquely to the case when \( P(A) \) is zero, specifying a value for \( P(B|A) \) through another route.

The three statements above are rules about how \( P(B) \) should be updated if \( A \) is given. If we know that \( A \sim B \), then we can use these simple rules in our calculations. We don't need to use the complex formula for \( \sim \) if \( P(A) \) is 1, \( 1/2 \) or 0. The formula for \( \sim \) is a cubic interpolation between these three rules, which extends \( \sim \) to all possible values of \( P(A) \) and \( P(B) \).

If we suppose that, initially, \( P(A) \) is 1, \( 1/2 \), or 0, then we can say that \( A \sim B \) means that, if \( A \) is given, \( P(B) \) doesn't change unless \( P(A) \) was 0, in which case it changes to \( P(\neg B) \). If we interpret the probability as a measure of subjective uncertainty, then what \( \sim \) says is that, if we are certain that \( A \) is true, but then find out that we were wrong (\( P(A) = 0 \) and then \( A \) is given), then we were also wrong about \( B \), and we need to correct this by replacing \( P(B) \) with \( P(\neg B) \). Otherwise (i.e.\ \( P(A) = 1 \) or \( 1/2 \)), finding out that \( A \) is true has no effect on \( P(B) \).

Replacing \( P(B) \) with \( P(\neg B) \) is equivalent to replacing \( i(B) \) with \( i(\neg B) \). That is, the quantity of information, \( i(B) \), becomes associated with \( \neg B \) instead of \( B \), and \( i(\neg B) \) becomes associated with \( B \) instead of \( \neg B \). So we can say that, if we were certain about \( A \) and learn that we were wrong, then we conclude that we were also wrong about whether the quantity \( i(B) \) should be attributed to \( B \) or to \( \neg B \).

If \( A \sim B \), then \( B \sim A \), and \( A \sim \neg B \) and \( \neg A \sim B \) and \( \neg A \sim \neg B \). This is because of the symmetries of \( \sim \) and the equation it satisfies. This means that the three rules above also apply if \( A \) is replaced by \( \neg A \), or \( B \) is replaced by \( \neg B \) or \( A \) and \( B \) are switched. For example, if \( P(B) = 1 \) and \( 0 < P(A) < 1 \), then \( P(A|\neg B) = P(\neg A) \).

So based on the three rules above and the symmetries involved, we can say that \( \sim \) is a relationship between \( A \) and \( B \) which behaves like independence, except when we are certain about one proposition and learn that we were wrong, in which case we need to switch the probability of the other proposition with the probability of its negation.

If we interpret the probabilities involved as frequencies of events, then what the rules above say is that the events \( A \) and \( B \) occur independently when they occur close to half of the time. If, however, \( A \) is very rare (\( P(A) \) is close to zero), then an occurrence of \( A \) will change the probability of an occurrence of \( B \) from \( P(B) \) to \( P(\neg B) \). So if \( A \) is very rare, and \( B \) is rare, then \( B \) will frequently occur when \( A \) occurs.

That is, if \( \sim \) is understood as a relationship between frequencies, it describes a correlation which is strong for events which are both rare, but which disappears when either event occurs close to 50\% of the time.

As Figure~\ref{fig:figure2.4} shows, the rule that if \( P(A) = 0 \) then \( P(B|A) = P(\neg B) \) only applies when \( P(B) \) is strictly between 0 and 1. If \( P(B) \) is 1 or 0 and \( P(A) = 0 \), then \( P(B|A) \) can take any value between 0 and 1.

When \( P(B) \) and \( P(A) \) are both close to zero, the value of \( P(B|A) \) depends on which one is closer. If \( P(A) \ll P(B) \), then \( P(B|A) \) is approximately \( P(\neg B) \), which is close to 1. But if \( P(B) \ll P(A) \), then \( P(B|A) \) is approximately \( P(B) \).

In summary, we have achieved some insights into what \( \sim \) is:

\begin{itemize}
    \item It is the only non-trivial way in which there can be an information balance between two propositions, in which, for the four possible conjunctions, \( AB \), \( A \neg B \), \( \neg A B \) and \( \neg A \neg B \), the information redundancies for the two more-likely-than-independence conjunctions cancel out the excess implied information from the two less-likely-than-independence conjunctions.
    \item It is a cubic interpolation of a set of simple rules for updating the probability of one proposition when the other is given.
    \item It describes a relationship in which \( A \) and \( B \) are mostly independent, except that if one of them has probability 0 and is given, or has probability 1 and its negation is given, the probability of the other proposition switches place with the probability of its negation.
    \item It has a complicated mathematical formula, involving complex numbers and cubic operations, but it can be used for symbolic reasoning without the formula in the special cases when the probability of one of the propositions is equal to 1, 0 or \( 1/2 \).
\end{itemize}

It remains mysterious, however. We do not understand what the formula for \( \sim \) is saying about relationships between meaningful quantities. We know that \( \sim \) can be used inside a system of symbolic reasoning, but we don't know what the rest of this system consists of. It seems that \( \sim \) belongs to a branch of mathematics that is unfamiliar to us.

The formula for the tilde relation, and its properties, are also reminiscent of quantum mechanics. There are irremovable complex numbers in relationships between probabilities. \( \sim \) seems to describe a type of logical entanglement between propositions, reminiscent of entanglement between quantum states. It seems plausible that, if we can fully understand the branch of mathematics to which \( \sim \) belongs, we might gain some insights into quantum mechanics.

This motivates us to identify and thoroughly study that branch of mathematics.

\section{Introducing Questions}
\label{sec:introducing-questions}

Since the formula for $\sim$ is too complicated, we look instead at the original equation: $i(A,B)+i(A,\neg B)+i(\neg A,B)+i(\neg A,\neg B)=0$. $\sim$ is the only solution to this equation apart from the trivial solution in which all the terms are zero. 

We observe that the quantity $i(A,B)+i(A,\neg B)+i(\neg A,B)+i(\neg A,\neg B)$ is symmetric under interchange of $A$ with $\neg A$ and $B$ with $\neg B$. If we call this quantity $c(A, B)$, then $c(A, B)=c(\neg A,B)=c(A,\neg B)=c(\neg A,\neg B)$.

The function $c$ takes two propositions as inputs, but it doesn't depend on whether $A$ or $\neg A$ is provided as one of the inputs. $A$ and $\neg A$ are equivalent, from $c$'s point of view.

When we have a collection of distinct objects which are treated as equivalent, we can break up the information needed to specify one of the objects into two parts: The information needed to specify the equivalence class to which the object belongs, and the information needed to specify which object within that class it is.

In our case, we can specify the proposition $A$ by first specifying the equivalence class $\{A, \neg A\}$, and then by specifying which of the two elements of the class is the one we're referring to.

The function $c$ ignores which element is specified. It only depends on the equivalence class. It is a function of the set, $\{A, \neg A\}$, rather than a function of $A$.

So the quantity, $i(A,B)+i(A,\neg B)+i(\neg A,B)+i(\neg A,\neg B)$, measures a relationship between the equivalence classes, $\{A, \neg A\}$ and $\{B, \neg B\}$, rather than between the propositions $A$ and $B$.

Sufficient reflection reveals that the word for the part of a proposition that it has in common with its negation is the ``question'' of that proposition. It contains all of the information in the proposition except whether what it describes is asserted to be true or false. 

We can write a proposition as a combination of a question and an answer:
\begin{itemize}
    \item $A$: All men are mortal $\rightarrow$ (Are all men mortal?, Yes)
    \item $\neg A$: Not all men are mortal $\rightarrow$ (Are all men mortal?, No)
\end{itemize}

Negating the proposition only changes the answer. The question is the part of the proposition that is left unchanged by negation.

Now we know that $i(A,B)+i(A,\neg B)+i(\neg A,B)+i(\neg A,\neg B)$ is a quantity that describes a relationship between questions, and $\sim$ is a special relationship between questions, for which this quantity vanishes.

\subsection{The Group of Pure Questions}
\label{subsec:group-pure-questions}

What we immediately want to know is:
\begin{itemize}
    \item Is there a binary operation that combines questions in the way that AND and OR combine propositions?
    \item What algebra do questions form?
    \item What role does $\sim$ play in the algebra?
    \item What's the relationship between the algebra of questions and the algebra of propositions (logic)?
    \item What are the numerical properties of questions?
\end{itemize}

Let's look for a binary operation. We have two questions, represented by the sets, $\{A, \neg A\}$ and $\{B,\neg B\}$. We want a binary operation that results in a set containing two opposite propositions.

We could try $\{AB, \neg(AB)\}$. This is the question of whether or not both $A$ and $B$ are true. Unfortunately, we cannot construct this set from $\{A, \neg A\}$ and $\{B,\neg B\}$, because they don't change when $A$ is replaced by $\neg A$ or $B$ is replaced by $\neg B$, but $\{AB, \neg(AB)\}$ does.

A set which can be constructed from $\{A, \neg A\}$ and $\{B,\neg B\}$, and which does have the right symmetries, is $\{AB \text{ or } \neg A\neg B, A\neg B \text{ or } \neg AB\}$. This corresponds to the question of whether the two questions have the same answer, $AB$ or $\neg A\neg B$, or different answers, $A\neg B$ or $\neg AB$. It can be written as $\{\neg(A \text{ xor } B), A \text{ xor } B\}$. 

If we use the notation $q(A)$ to refer to question of $A$, then we can call this binary operation $*$, and define it as:
\begin{align}
q(A) * q(B) = q(A \text{ xor } B).
\end{align}

Next, we ask what $q(A) * q(A)$ is. We can use the definition to find:
\begin{align}
q(A) * q(A) = q(A \text{ xor } A) = q(\text{False}) = \{\text{True, False}\}.
\end{align}

That is, combining a question with itself using $*$ yields the trivial question, $I=\{\text{True, False}\}$. This means that, under the $*$ operation, each question is its own inverse.

We can verify that $I$ acts as an identity element:
\begin{align}
q(A) * I = q(A \text{ xor True}) = q(A).
\end{align}

We can say what kind of algebra questions form by combining what we know about $*$ and the collection of questions that it operates on:
\begin{enumerate}
    \item $*$ is associative, because xor is associative: $(q(A)*q(B))*q(C) = q(A \text{ xor } B) * q(C) = q(A \text{ xor } B \text{ xor } C)$.
    \item Every question has an inverse, which is the question itself.
    \item The trivial question, $\{\text{True, False}\}$, acts as an identity element.
\end{enumerate}

These three facts together imply that questions form a group under the $*$ operation. It's an Abelian group, because $*$ is commutative. It's a group of involutions, meaning that every element is its own inverse.

Next, let's look at the global structure of a group of questions. If we start with $N$ independent questions, $q_1, q_2, \ldots, q_N$, then by using $*$, we can generate a full group, and examine its structure.

Every question is its own inverse, so each question can appear at most once in an expression like $q_1*q_3*q_7$. When we express a general question in the group as a string of generating questions joined together by $*$, each generating question is either there or not there. 

Because the generating questions are independent, none of them can cancel with any combination of the others. Any element of the group can be uniquely expressed as a collection of generating questions connected by $*$. 

This means that every question in the group consists of a subset of the $N$ generating questions, joined together by $*$. For every subset of the set of generating questions, there's a corresponding group element, with $I$ corresponding to the empty subset, and $q_1*q_2*\ldots*q_N$ corresponding to the full set.

So the group of questions generated by $N$ independent questions has $2^N$ elements, corresponding to the number of subsets of a set of $N$ elements. The $*$ operation on questions within the group corresponds to the symmetric difference operation on the corresponding subsets of generating questions: $(q_1*q_2*q_3)*(q_2*q_4) = q_1*q_3*q_4$ corresponds to $\{q_1,q_2,q_3\} \oplus \{q_2,q_4\} = \{q_1,q_3,q_4\}$.

This shows us that the $*$ operation, which combines two questions to generate a third, by asking, ``Do these two questions have the same answer?'', organizes any given collection of logically independent questions into a group. 

We understand the structure of the group, how many questions it contains, and how they are related to each other algebraically. Next, we'll look at the quantitative, rather than algebraic, properties of questions, and understand how these quantities relate to the algebra.

\subsection{Numerical Properties of Questions}
\label{subsec:numerical-properties-questions}

The pure question, $q(A)$, of a proposition, $A$, is the part of $A$ that remains unchanged when $A$ is replaced by $\neg A$: $q(A)=q(\neg A)$.

If there is a function $s$, which associates a number, $s(A)$, with a proposition, $A$, then $s$ is a property of the pure question, $q(A)$, if and only if $s(A)=s(\neg A)$.

That is, numbers which change when $A$ is replaced by $\neg A$ are not properties of the pure question of $A$, because the pure question of $A$, and all of its numerical properties, are unchanged when $A$ is replaced by $\neg A$.

We call the object whose properties don't change when $A$ is negated the pure question of $A$, because we will see later that there are also signed questions, whose numerical properties change sign when the proposition is negated. For now, we focus on pure questions.

We can identify numerical properties of a pure question by checking what happens when $A$ is replaced by $\neg A$, and we can find them by finding ways to symmetrize the numerical properties of the proposition.

A proposition has a probability, $P(A)$. This is not a numerical property of the pure question of $A$, because $P(A)$ is not equal to $P(\neg A)$ in general. However, we can take $P(A)$ and use it to form symmetric quantities. These quantities are genuinely properties of the question of $A$.

Two major examples of quantities which are properties of the pure question of $A$ are:
\begin{itemize}
    \item $|P(A)-P(\neg A)|$
\end{itemize}

This is the unsigned probability gap associated with the pure question of $A$. It's the magnitude of the difference between the probability of $A$ and the probability of $\neg A$. Its minimum value is 0 at $P(A)=1/2$, and its maximum value is 1 when $P(A)=1$ or 0. We will soon see that the probability gap is related to the group structure of questions under the $*$ operation.
\begin{itemize}
    \item $\sqrt{P(A)P(\neg A)}$
\end{itemize}

This is the geometric mean probability of the pure question of $A$. Its minimum value is 0 when $P(A)=1$ or 0, and its maximum value is $1/2$, which occurs when $P(A)=1/2$. The geometric mean probability will play an important role later on, in the algebra of subjects. We will see that the geometric mean probability is the key to understanding the significance of the tilde relation.

For any numerical property, $s(A)$, of a proposition, $A$, we can find corresponding properties of the question, $q(A)$, by taking symmetric combinations of $s(A)$ and $s(\neg A)$, such as $s(A)*s(\neg A)$ or $s(A)+s(\neg A)$. 

For convenience, we will use lower-case letters such as $a$, $b$, and so on, to refer to the pure questions of the corresponding propositions, $a=q(A)$, $b=q(B)$, etc.

Then we can use the definition: The probability gap, $\text{gap}(a)$ of an unsigned question, $a$, is:
\begin{align}
\text{gap}(a) = |P(A)-P(\neg A)|
\end{align}
where $a=\{A, \neg A\}$.

Let's consider the value of $\text{gap}(a*b)$, in the case when $a$ and $b$ are independent questions:
\begin{align}
\text{gap}(a*b) &= \text{gap}(q(A \text{ xor } B)) \notag \\
& = |P(A \text{ xor } B) - P(\neg(A \text{ xor } B))| \notag \\
&= |P(AB \text{ or } \neg A\neg B) - P(A\neg B \text{ or } \neg AB)| \notag \\
&= |P(AB) + P(\neg A\neg B) - P(A\neg B) - P(\neg AB)| \notag \\
&= |P(A)P(B) + P(\neg A)P(\neg B) - P(A)P(\neg B) - P(\neg A)P(B)| \notag \\
&= |P(A)P(B) + (1-P(A))(1-P(B)) \notag \\
&\quad - P(A)(1-P(B)) - (1-P(A))P(B)| \notag \\
&= |P(A)P(B) + 1 - P(A) - P(B) + P(A)P(B) \notag \\ 
&\quad - P(A) + P(A)P(B)- P(B) + P(A)P(B)| \notag \\
&= |1 - 2P(A) - 2P(B) + 4P(A)P(B)|.
\end{align}

Now let's compare this to $\text{gap}(a)\text{gap}(b)$:
\begin{align}
|P(A)-P(\neg A)||P(B)-P(\neg B)| & = |2P(A)-1||2P(B)-1| \notag \\
& = |(2P(A)-1)(2P(B)-1)| \notag \\
& = |1 - 2P(A) - 2P(B) + 4P(A)P(B)|.
\end{align}

This proves that $\text{gap}(a*b)=\text{gap}(a)\text{gap}(b)$ when $a$ and $b$ are the pure questions of independent propositions, $A$ and $B$.

In fact, $\text{gap}(a*b)=\text{gap}(a)\text{gap}(b)$ implies and is implied by $P(AB)=P(A)P(B)$. Propositions are independent if and only if their pure questions are independent.

For propositions, probability is the primary quantity, and, if $A$ and $B$ are independent, then $P(AB)=P(A)P(B)$. Probabilities multiply when independent propositions are joined together by AND.

For pure questions, the probability gap, $\text{gap}(a)=|P(A)-P(\neg A)|$, is the primary quantity, and gaps multiply when independent questions are joined together using the $*$ operation: $\text{gap}(a*b)=\text{gap}(a)\text{gap}(b)$.

In the group of questions generated by $N$ independent questions, we can calculate the probability gap of any question, such $q_1*q_2*q_3$, by multiplying the probability gaps of the questions that appear in the expression: $\text{gap}(q_1*q_2*q_3) = \text{gap}(q_1)\text{gap}(q_2)\text{gap}(q_3)$.

We now understand the structure of the group generated by $N$ independent pure questions, and we can calculate the probability gap of any question in the group.

We cannot use this form of quantitative reasoning to find out whether a proposition is probably true or probably false, but we can use it to find out whether the probability gap of the question is close to zero (completely uncertain) or close to 1 (settled).

If every question is settled, $\text{gap}(q_1)=\text{gap}(q_2)=\text{gap}(q_3)=1$, then their combination is also settled, $\text{gap}(q_1*q_2*q_3)=1$. On the other hand, if a question is completely uncertain, e.g.\ $\text{gap}(q_3)=0$, then any combinations that include it are completely uncertain.

We can also see that there is a quantity similar to information that builds up as more and more questions are added to a string of questions connected by $*$. The probability gap of the resulting question is the product of the gaps of the individual questions in the expression. This means that the log of the resulting question's gap is equal to the sum of the logs of the individual gaps.

If we use $d(x)$ to represent $-\log(\text{gap}(x))$, then:
\begin{align}
d(q_1*q_2*q_3) = d(q_1)+d(q_2)+d(q_3).
\end{align}

With independent propositions, the log of the probability adds up as more propositions are combined using AND, and this quantity is called information. For questions, it's the log of the probability gap that adds up. This is a log of a gap, not a log of a probability, so it's not information, even though it has similar properties. More propositions means more information. More questions means something else. We will see what it means and find an appropriate word to call it later on, when signed questions provide an insight into what $d$ measures.

\subsection{The Group of Pure Questions Generated by $N$ Independent Propositions}
\label{subsec:group-pure-questions-n-propositions}

Propositions can be combined using AND or OR. They can also be negated. The algebra of propositions, namely propositional logic, is a rich structure, more complicated than the group generated by $N$ independent questions.

The question $q(AB)$, of whether or not both $A$ and $B$ are true, cannot be expressed as a combination of $q(A)$ and $q(B)$ using the $*$ operation, because $q(A)*q(B)=q(A \text{ xor } B)$, not $q(AB)$. The question $q(AB)$ is not in the group generated by the questions $q(A)$ and $q(B)$. That group has only four elements: $I$, $q(A)$, $q(B)$ and $q(A)*q(B)$.

Let's consider the full group of pure questions generated by $N$ independent propositions, $A$, $B$, $C$, \ldots, $Z$. Every possible logical combination, using AND or OR, of the generating propositions, and their negations, yields a proposition with a corresponding pure question.

We can observe that the generating propositions, $A$, $B$, and so on, have corresponding questions, $a=q(A)$, $b=q(B)$, and so on, which together generate the group described in the previous subsection, with $2^N$ elements.

This group consists of all combinations of the questions $a$, $b$, $c$, \ldots, $z$, using the $*$ operation. It is a subgroup of the full group generated by $N$ propositions, because the full group contains $q(AB)$, which can't be generated by any combination of questions in the subgroup.

We can refer to this subgroup as the group of rank-1 questions, or the group of one-proposition questions. This reflects the fact that $q(AB)$, which is a two-proposition question, has a higher rank in the full group than the one-proposition questions. Let's use $Q_1(N)$ to denote the group of one-proposition questions generated by $N$ independent propositions.

We can examine what happens when a one-proposition question is combined with a two-proposition question using $*$:
\begin{align}
q(A)*q(AB) = q(A \text{ xor } AB) = q(A\neg B).
\end{align}

This means that $q(AB)$ and $q(A\neg B)$ differ only by a rank-1 question, $q(A)$. We also find that:
\begin{align}
q(A)*q(B)*q(AB) & = q(A \text{ xor } B \text{ xor } AB) \notag \\
& = q(A \text{ xor } \neg AB) = q(\neg A \text{ xor } \neg AB) = q(\neg A\neg B).
\end{align}

So the four rank-2 questions $q(AB)$, $q(A\neg B)$, $q(\neg AB)$, and $q(\neg A\neg B)$, differ from each other by rank-1 questions. They are, therefore, not all generators in the group of rank-2 questions. Introducing one of them is enough to generate all of them, in combination with the rank-1 questions.

Let's consider the question, $q(AB)$, that doesn't have any negated propositions in it, to be the rank-2 question that generates the other three questions, $q(A\neg B)$, $q(\neg AB)$, and $q(\neg A\neg B)$.

Then we can count the number of generators of rank 2: Every combination of 2 of the $N$ generating propositions, $A$, $B$, $C$, \ldots, $Z$, yields a two-proposition question that acts as a generator. The number of these is $N(N-1)/2$.

More generally, every subset of the set of $N$ generating propositions corresponds to a generator in the group, which is the question of their conjunction. The subset $\{A, B\}$ of the $N$ propositions corresponds to $q(AB)$. There is one generator in the group for each non-empty subset of the set of generating propositions.

This means that we can count the number of generators in the group: It's equal to the number of subsets of a set with $N$ elements, minus 1, because the empty set is a subset but corresponds to the identity element of the group instead of a generator. There are $2^N$ subsets in total, so the number of generators in the group is $2^N-1$.

This allows us to calculate the total number of elements in the group. Each element in the group is expressible as a string of generating elements connected by the $*$ operation. For example $q(A\neg B)=q(A)*q(AB)$, where $q(A)$ and $q(AB)$ are generators. Each generator can appear at most once, because two occurrences would cancel out since each question is its own inverse.

Each element in the group therefore corresponds to a subset of the set of generators. There are $2^N-1$ generators, so there are $2^{2^N-1}$ elements in the group in total.

We can check that this is the correct result by observing that there are two propositions for every question, and so the number of propositions is two multiplied by $2^{2^N-1}$, which is $2^{2^N}$. This is already known to be the number of propositions that can be generated from $N$ independent propositions, using AND, OR and NOT.

In summary:
\begin{itemize}
    \item The group, $Q(N)$, of pure questions generated by $N$ independent propositions has $2^{2^N-1}$ elements, corresponding to the number of combinations of the $2^N-1$ generators.
    \item Each element is expressible as a combination, using $*$, of questions about conjunctions of one or more of the $N$ propositions, such as $q(A)*q(AB)*q(CDE)$.
    \item The $2^N-1$ generators of $Q(N)$ are the questions of the possible conjunctions, $A$, $AB$, etc.\ of the $2^N-1$ non-empty subsets of the generating propositions.
    \item $Q(N)$ consists of $N$ nested groups: $Q_1(N) \subset Q_2(N) \subset Q_3(N) \ldots \subset Q_N(N)=Q(N)$.
    \item The number of generators in $Q_m(N)$ which aren't in $Q_{m-1}(N)$, is $\left({}^N_m\right)$. 
    \item The generators are questions about the conjunctions that don't have any negated propositions in them.
    \item A question about a conjunction with a negation, such as $q(A\neg B)$, can be expressed as a combination using $*$ of the same conjunction without negations, and a question with a lower rank. For example, $q(A\neg B) = q(A)*q(AB)$.
\end{itemize}

\section{Subjects}
\label{sec:subjects}

\subsection{Quotient Groups}
\label{subsec:quotient-groups}

If there's an Abelian (commutative) group, $G$, which contains a subgroup, $H$, then it is possible to form a new group called the quotient group, $G$ modulo $H$, written as $G/H$, in which every element of $H$ is represented by the identity element. There's a homomorphism, $\pi$, from $G$ to $G/H$, which assigns a representative, $\pi(x)$, in $G/H$ to every element, $x$, of $G$.

As a homomorphism, $\pi$ preserves the group operation, $*$:
\[
\pi(x*y) = \pi(x) * \pi(y).
\]

If $x$ and $y$ are elements of $G$, and $x = y * j$, where $j$ is an element of $H$, then $x$ and $y$ are represented by the same element in $G/H$:
\[
\pi(x) = \pi(y * j) = \pi(y) * \pi(j) = \pi(y).
\]

Each element of the quotient group corresponds to an equivalence class of elements of the full group. For example, the identity element of the quotient group, $G/H$, corresponds to the set of all the elements in $H$. 

Every other element of the quotient group corresponds to a coset of $H$. For a given element, $x$, which is in $G$ but not in $H$, the coset containing $x$ is the set of all the elements in $G$ that are expressible as $x * h$, where $h$ is in $H$. 

The coset is not a group, because it doesn't contain an identity element. $x$ is not in $H$ so its inverse is not in $H$, so $x * h$ is never equal to the identity element. The coset is denoted as $x * H$.

The entire group, $G$, can be thought of as the product of the quotient group and the subgroup: $G \cong G/H \times H$. Every element of $G$ can be specified by first specifying an element of $G/H$, which corresponds to a coset of $H$, and then specifying an element within that coset.

Theoretically, we can write an element, $x$, of $G$ in the form $(\pi(x), h)$, where $\pi(x)$ is an element of the quotient group, $G/H$, and $h$ is an element of the subgroup, $H$.

However, in order to do this, we need to choose a canonical representative element of the equivalence class $\pi(x)$, which we can then combine with $h$ using the group operation, $*$. It would be convenient if we could choose $x$ as the representative of $\pi(x)$, but we can't do that because $\pi(y) = \pi(x)$ for every element $y$ which is of the form $y = x * h$ for any element $h$ of $H$. If the representative of $\pi(x)$ is $x$, then the representative of $\pi(y)$ will also be $x$.

There might be numerical properties associated with the group elements which naturally select a representative element from the coset corresponding to an element of $G/H$, such as the largest or smallest element within that coset. Otherwise, we can just pick one. 

If we use $r(x)$ to denote the representative of the coset corresponding to $\pi(x)$, then $r(x)$ is in the same equivalence class as $x$, so it differs from $x$ by an element, $h$, of $H$: $x = r(x) * h$.

This means that we can write $x$ as $(r(x), h)$. So an element of the full group, $G$, can be written as a pair of group elements, one from the quotient group and one from the subgroup. Under the $*$ operation, such pairs combine component-wise: $(r, h) * (s, k) = (r * s, h * k)$.

\subsection{Quotient Groups Inside the Group of Questions Generated by $N$ Independent Propositions}
\label{subsec:quotient-groups-questions}

In the case of questions, we have the group, $Q(N)$, of questions generated by $N$ independent propositions, which has a nested sequence of subgroups, $Q_1(N)$, $Q_2(N)$, and so on up to $Q_N(N) = Q(N)$.

$Q_2(N)$, which is the group of pure questions up to and including rank 2, contains $Q_1(N)$ as a subgroup. We can therefore form the quotient group, $Q_2(N)/Q_1(N)$, which is the group of questions about conjunctions of up to 2 propositions modulo one-proposition questions.

Two questions, $x$ and $y$, from $Q_2(N)$, become equivalent if they differ by a one-proposition question: $c$ in $Q_1(N)$ and $y = x * c$ implies $\pi(x) = \pi(y)$.

For example, $q(A) * q(AB) = q(A\neg B)$, so $q(AB)$ and $q(A\neg B)$ are equivalent in $Q_2(N)/Q_1(N)$: $\pi(q(AB)) = \pi(q(A\neg B))$. In fact, $q(AB)$, $q(A\neg B)$, $q(\neg AB)$, and $q(\neg A\neg B)$ are all equivalent within $Q_2(N)/Q_1(N)$. 

This means that, in the quotient group, $Q_2(N)/Q_1(N)$, adding negations into a conjunction doesn't change the group element. $\pi(q(AB))$ is a property of the questions $a$ and $b$, not the propositions $A$ and $B$, because it doesn't change when $A$ or $B$ are negated.

Let's use $s(ab)$ to denote this element, and indicate that it is a property of the combination of the questions $a$ and $b$.

This way of combining questions is not the same as the $*$ operation. $a * b$ is the rank-1 question of whether or not $a$ and $b$ have the same answer. There are two alternatives: they do or they don't. The subject, $s(ab)$, has four alternatives, $AB$, $A\neg B$, $\neg AB$, and $\neg A\neg B$.

\subsection{Subjects}
\label{subsec:subjects-definition}

$s(ab)$ is an element of the quotient group $Q_2(N)/Q_1(N)$, so it is not a pure question, but an equivalence class of pure questions. It is a topic, in the sense that it is something about which questions can be asked.

We can call $s(ab)$ the subject whose subject matter consists of the questions $a$ and $b$. 

$s(ab)$ contains no references to any specific outcomes or any logical relations between $a$ and $b$. By contrast, the question, $q(A\neg B)$, involves logical relations: $q(A\neg B) = \{A\neg B, \neg A \text{ or } B\} = \{A\neg B, A \Rightarrow B\}$. It is not a pure specification of the subject matter; it's a representation of a logical relation.

We use the word subject because $s(ab)$ is a neutral specification of $a$ and $b$ as the subject matter, about which questions can be asked by choosing an element of the equivalence class, $\{q(AB), q(A\neg B), q(\neg AB), q(\neg A\neg B)\}$, which yields a pure question, such as $q(AB)$.

We use $S_2(N)$ to denote the group $Q_2(N)/Q_1(N)$ and indicate that it is the group whose generators are two-question subjects, such as $s(ab)$.

The number of generators in $S_2(N)$ is $\left({}^N_{\:2}\right)$ = $N(N-1)/2$, corresponding to the number of choices of two questions from the $N$ questions, $q(A)$, $q(B)$, etc.\ of the $N$ independent propositions, $A$, $B$, etc.

Each element in $S_2(N)$ is expressible as a string of two-question subjects combined using $*$, such as $s(ab) * s(cd)$. The number of these expressions is $2^{N(N-1)/2}$.

This extends to higher ranks: $q(AB) * q(ABC) = q(AB\neg C)$, so $q(ABC)$ and $q(AB\neg C)$ differ by a rank-2 question. The group $Q_3(N)/Q_2(N)$ is the group $S_3(N)$ whose generators are 3-question subjects, such as $s(abc)$.

Generally, $Q_m(N)/Q_{m-1}(N)$ is the group $S_m(N)$ of $m$-question subjects and their combinations. It has $\left({}^N_m\right)$ generators, and $2^{\left({}^N_m\right)}$ elements.

\subsection{The Group of Subjects}
\label{subsec:group-subjects}

A rank-1 pure question, $a$, about one of the independent propositions, $A$, can be thought of as a one-question subject. This can be combined with another such question, $b$, using the $*$ operation to yield another rank-1 question, or it can be combined with $b$ by forming the subject, $s(ab)$, which is a two-question subject.

We can call this new way of combining questions the subject-wise combination, and denote it by $\circ$. Given two subjects of possibly different ranks with non-overlapping subject matter, we can combine them into a bigger subject:
\begin{align}
s(ab) \circ s(cd) = s(abcd).
\end{align}

We can make this binary operation on subjects into a group operation by defining the result of $\circ$ when subject matter overlaps as the subject whose subject matter is the symmetric difference between the subject matters of the two overlapping subjects:
\begin{align}
s(ab) \circ s(bcd) = s(acd).
\end{align}

This means that each subject is its own inverse, $s(ab) \circ s(ab) = I$, where $I$ is the identity element in the group of subjects, corresponding to an empty subject with zero questions. 

This makes the subjects generated by $N$ independent propositions into a group. We can call this group $S(N)$. It uses a group operation, $\circ$, which is not the same as the group operation, $*$, used in the group $S_m(N)$ generated by $m$-question subjects.

We can in fact drop the $s$ from the expression $s(ab)$, and just use $ab = a \circ b$ to identify the subject, since it contains only a list of pure questions, which is all that is necessary to specify the subject.

Within $S_m(N)$, the subjects act as generators, and not every element in the group is a subject. $ab * cd$ is in the group $S_2(N)$, but it is not a subject. The groups $S_m(N)$ are not groups of subjects but groups generated by subjects.

In the group of subjects, $S(N)$, though, every element, apart from the identity element, $I$, is a non-empty subject: $ab \circ cd = abcd$.

The number of elements in $S(N)$ is equal to $2^N$, because there is one subject for every subset of the $N$ questions, $a$, $b$, etc.\ from the $N$ independent propositions.

In fact, $S(N)$ has exactly the same structure as $Q_1(N)$, the group of rank-1 questions. This is because, in both groups, every element corresponds to a subset of the $N$ independent questions, and the group operation, $*$ or $\circ$, corresponds to the symmetric difference operation on the corresponding sets:
\begin{itemize}
    \item The group, $S(N)$, of subjects generated by $N$ independent propositions is isomorphic to the group generated by the $N$ questions of those $N$ propositions.
\end{itemize}

\subsection{Geometric Mean Probability}
\label{subsec:geometric-mean-probability}

Propositions have probabilities, and, when independent propositions are combined using AND, their probabilities multiply. Questions have probability gaps, and when independent questions are combined using $*$, their probability gaps multiply.

Subjects form a group, and can be combined using the $\circ$ operation. It makes sense to ask: What is the quantity associated with a subject, which multiplies when independent subjects are combined using $\circ$?

When we introduced pure questions, we identified two quantities associated with a question: The probability gap, $|P(A) - P(\neg A)|$, and the geometric mean probability, $\sqrt{P(A)P(\neg A)}$.

Subjects don't have probability gaps, but they have geometric mean probabilities:
\begin{align}
\text{gmp}(ab) = (P(AB)P(A\neg B)P(\neg AB)P(\neg A\neg B))^{1/4}.
\end{align}

If $a$ and $b$ are independent, then:
\begin{align}
\text{gmp}(ab) &= (P(A)P(B)P(A)P(\neg B)P(\neg A)P(B)P(\neg A)P(\neg B))^{1/4} \notag \\
&= (P(A)P(\neg A)P(B)P(\neg B))^{1/2} \notag \\
&= (P(A)P(\neg A))^{1/2} (P(B)P(\neg B))^{1/2} \notag \\
&= \text{gmp}(a) \, \text{gmp}(b).
\end{align}

This shows the important result:
\begin{itemize}
    \item If $a$ and $b$ are independent questions, then the geometric mean probability of the subject $ab$ is equal to the product of the geometric mean probabilities of $a$ and $b$.
\end{itemize}

If two subjects have overlapping questions in their subject matter, then combining them with $\circ$ does not result in simple multiplication:
\begin{align}
\text{gmp}(ab \circ b) = \text{gmp}(a) = \text{gmp}(a)\text{gmp}(b) / \text{gmp}(b) = \text{gmp}(ab) \text{gmp}(b) / \text{gmp}(b)^2.
\end{align}

That is, the product of the geometric mean probabilities needs to be divided by the square of the geometric mean probability of the subject corresponding to the overlap of the two subject matters.

Overall, for subjects, the geometric mean probability plays the role that probability plays for propositions and that probability gaps play for questions.

\subsection{Two Ways of Reasoning About Questions Using Numbers}
\label{subsec:two-ways-reasoning}

The group of subjects and the group of rank-1 questions are both isomorphic to the group of subsets of $N$ questions, with the symmetric difference being the group operation, and are therefore the same structure. 

This single group structure manifests itself twice in the algebra of questions, once horizontally, relating questions with the same rank, and once vertically, relating subjects with different ranks.

These two manifestations each have an associated numerical measure, which assigns a number to each group element, with multiplication of these numbers corresponding to the group operation. 

Such a numerical measure is called a character. It is a homomorphism from the group of combinations of questions to the real numbers, with multiplication acting as the group operation for numbers.

If we consider the log of the geometric mean probability, we see that it is an additive quantity:
\begin{align}
\log(\text{gmp}(ab)) = \log(\text{gmp}(a)) + \log(\text{gmp}(b)).
\end{align}

This is a quantity of information, because it is a log of a probability. If we refer to $-\log \text{gmp}(s)$ as the information value, $i(s)$, of the subject, $s$, then we have:
\begin{align}
i(ab) = i(a) + i(b)
\end{align}
when $a$ and $b$ are independent.

Earlier, we introduced $d(a) = -\log \text{gap}(a)$, and found that independence implies $d(a * b) = d(a) + d(b)$.

Both $\text{gap}$ with $*$ and $\text{gmp}$ with $\circ$ are one-dimensional group representations of the group of combinations of $N$ questions. 

Both quantities, $i$ and $d$, make it possible to reason about independent questions using simple addition and subtraction of numbers, in the same way as probability and information can be used to reason about propositions.

If $d(a) = 0$, then $a$ is settled: $\text{gap}(a) = 1$, and if $d(a) = \infty$, then $a$ is completely uncertain, $\text{gap}(a) = 0$.

If $i(a) = 1$ bit, then $\text{gmp}(a) = 1/2$, so $\text{gap}(a) = 0$. If $i(a) = \infty$, then $\text{gmp}(a) = 0$, so $\text{gap}(a) = 1$.

This shows that $i(a)$ can be used in calculations when $d(a)$ can't, and vice versa: $i(a)$ is infinite if $\text{gap}(a) = 1$, and $d(a)$ is infinite when $\text{gap}(a) = 0$.

Suppose that we want to calculate $\text{gap}(a)$. We can use either of the following equations to deduce $\text{gap}(a)$ from other quantities:
\begin{enumerate}
    \item $d(a) = d(a * b) - d(b)$
    \item $i(a) = i(ab) - i(b)$
\end{enumerate}

In the second case, we actually calculate $\text{gmp}(a)$, which determines $\text{gap}(a)$ through a non-linear relationship:
\begin{align}
\sqrt{p(1-p)} = \frac{1}{2} \sqrt{1 - |2p - 1|^2} \quad \Rightarrow \quad \text{gmp}(a) = \frac{1}{2} \sqrt{1 - \text{gap}(a)^2}.
\end{align}

In the group, $Q(N)$, there are:
\begin{itemize}
    \item Questions whose gap is zero,
    \item Questions whose gap is one, and
    \item Questions whose gap is strictly between zero and one.
\end{itemize}

We can use $i$ but not $d$ to reason about the questions whose gap is zero, and we can use $d$ but not $i$ for the questions with a gap of 1, and we can use both for the remaining questions. So it's possible to reason about every question in the group using at least one of the two systems.

If we're reasoning about independent questions using additive quantities, then, for the questions which are settled, we use $d$ and we combine questions using $*$. For the questions which are completely uncertain, we use $i$ and we use subject-wise combination, $\circ$.

\subsection{The Role of $\sim$ in the Algebra of Subjects}
\label{subsec:role-tilde}

We can write the equation $i(A,B) + i(A,\neg B) + i(\neg A,B) + i(\neg A,\neg B) = 0$ in the form:
\begin{align}
P(AB)P(A\neg B)P(\neg AB)P(\neg A\neg B) &= P(A)P(B)P(A)P(\neg B)P(\neg A)P(B)P(\neg A)P(\neg B)\notag \\
&= P(A)^2 P(\neg A)^2 P(B)^2 P(\neg B)^2.
\end{align}

Taking the fourth root of both sides yields:
\begin{align}
(P(AB)P(A\neg B)P(\neg AB)P(\neg A\neg B))^{1/4} = (P(A)P(\neg A))^{1/2} (P(B)P(\neg B))^{1/2}.
\end{align}

This equation says that $\text{gmp}(ab) = \text{gmp}(a) \text{gmp}(b)$. So the condition for geometric mean probabilities to multiply for questions when they are combined into a subject is $i(A,B) + i(A,\neg B) + i(\neg A,B) + i(\neg A,\neg B) = 0$.

This has two solutions: 
\begin{itemize}
    \item $P(AB) = P(A)P(B)$, and
    \item $a \sim b$.
\end{itemize}

What this means is that the equation $i(ab) = i(a) + i(b)$ is satisfied if $a \sim b$, as well as if $a$ and $b$ are independent. If we only know that $i(ab) = i(a) + i(b)$, then we don't know whether $a$ and $b$ are independent. They could be independent or they could be related by $\sim$.

That is, when $a \sim b$, $a$ and $b$ behave as though they were independent, from the point of view of geometric mean probabilities. Their information values add together when they're combined into a two-question subject. If we are reasoning about subjects by adding and subtracting information values, then $a$ and $b$ can be treated as though they're independent, even though they're related by $\sim$.

Tilde is a relationship between questions that allows them to be non-independent, and yet be reasoned about as if they were independent, using geometric mean probabilities. It is, in fact, the only such relationship.

\begin{itemize}
    \item If $a$ and $b$ are independent questions, then we can use either logs of probability gaps with $*$, or information values with $\circ$, to reason about them using addition and subtraction.
    \item If $a \sim b$, then we can only use information values with $\circ$ to reason about them using addition and subtraction. We can't use gaps, because $\text{gap}(ab)$ is not equal to $\text{gap}(a)\text{gap}(b)$, which would imply independence.
\end{itemize}

The tilde relation can then be thought of as a dependency between $a$ and $b$ that combines them into a ``natural subject". It makes it possible to reason about them using information values and subject-wise combination. We can use $i(a) = i(ab) - i(b)$ to calculate $\text{gmp}(a)$, which determines $\text{gap}(a)$. But we can't use gaps and $*$ because they're not independent.

So $\sim$ forces us to reason using geometric mean probabilities and information values along with subject-wise combination of questions. It makes $ab$ into a natural subject, in the sense that, if $a$ and $b$ were independent, we could reason about them using the subject $ab$ and its geometric mean probability, but we would have another option (probability gaps). When $a \sim b$, however, we must use the subject $ab$ if we want to reason using addition and subtraction.

When $a \sim b$, $a$ and $b$ can be considered to be independent as subjects, but not as questions.

\section{Askable Questions}
\label{sec:askable-questions}

If there's a quantity, $s(A)$, which is a property of a proposition, $A$, then it can be written as the sum of a symmetric part and an antisymmetric part:
\begin{align}
s(A) = \frac{s(A) + s(\neg A)}{2} + \frac{s(A) - s(\neg A)}{2}.
\end{align}

The symmetric part, $\frac{s(A) + s(\neg A)}{2}$, is a property of the pure question, $q(A)$, because it remains unchanged when $A$ is replaced by $\neg A$. This reflects the fact that $q(A)$ is completely independent of the answer. The answer means nothing to the pure question.

An answer combined with a pure question is not enough to constitute a full proposition. The pure question of $A$ is the question of whether $A$ is true or $A$ is false. Providing the answer ``Yes" to this question does not tell us whether $A$ is true or false. 

We can construct the set, $\{\{A, \neg A\}, \text{Yes}\}$, to represent the combination of the pure question and the answer. It clearly doesn't single out either $A$ or $\neg A$. What's missing is the specification of which of the alternatives the ``Yes" applies to.

The antisymmetric part of $s(A)$ is $\frac{s(A) - s(\neg A)}{2}$, which changes sign when $A$ is replaced by $\neg A$. This is a property of a part of the proposition for which the answer means everything. We can see that the object whose properties are the antisymmetric functions of $A$ contains the pure question inside it, because:
\begin{itemize}
    \item An antisymmetric quantity, $s(A) = -s(\neg A)$, can be converted into a symmetric quantity by removing its sign: $s_+(A) = |s(A)|$.
    \item To convert a symmetric quantity, $s(A) = s(\neg A)$ into an antisymmetric quantity, it is necessary to specify additional information, namely whether $s(A)$ or $s(\neg A)$ should be given the negative sign.
\end{itemize}

This means that the pure question can be obtained by removing information from this object. We call this part of $A$ the askable question, to indicate that it can be combined with an answer to yield a proposition. It asks, ``Is $A$ true rather than false?". 

The askable question of $A$ specifies the pure question, and also specifies which alternative is asked about. The set that accomplishes this is $\{\{A\}, \{A, \neg A\}\}$. This is the representation as a set of the ordered pair, $(A, \neg A)$. 

So, while the pure question of $A$ is represented by the set $\{A, \neg A\}$, the askable question is represented by the ordered pair, $(A, \neg A)$.

We can use the notation $A?$ to denote the askable question of $A$. This notation clearly distinguishes askable questions from pure questions, which we write in lower case, and propositions, which we write in upper case.

\subsection{Signed Probability Gaps}
\label{subsec:signed-probability-gaps}

A proposition, $A$, has a probability, $P(A)$. The corresponding property for an askable question is the signed probability gap:
\begin{align}
\text{gap}(A?) = P(A) - P(\neg A) = 2P(A) - 1.
\end{align}

We use the same function name, $\text{gap}$, to denote the probability gaps of pure and askable questions. In the case when the argument to $\text{gap}$ is a pure question, $\text{gap}(a)$, it yields an unsigned probability gap. If the argument is an askable question, $\text{gap}(A?)$, then it yields a signed gap.

$\text{gap}(A?)$ can take any value in between $-1$ and $1$. When $\text{gap}(A?) = 1$, $P(A) = 1$, and when $\text{gap}(A?) = -1$, $P(A) = 0$, while $\text{gap}(A?) = 0$ implies $P(A) = 1/2$.

In the case of pure questions, the unsigned probability gaps multiply when independent questions are combined using the $*$ operation: $\text{gap}(a * b) = \text{gap}(a) * \text{gap}(b)$. The pure question, $a * b$, is represented by the set $\{AB \text{ or } \neg A\neg B, A \text{ xor } B\}$. 

There are two askable questions corresponding to $a * b$, namely $(AB \text{ or } \neg A\neg B)?$ and $(A \text{ xor } B)?$. To be able to define a binary operation on askable questions that matches the $*$ operation on pure questions, we need to decide whether $A? * B?$ is $(AB \text{ or } \neg A\neg B)?$ or $(A \text{ xor } B)?$. We can make this decision by using the requirement that $\text{gap}(A? * B?)$ should be equal to $\text{gap}(A?) * \text{gap}(B?)$ when $A?$ and $B?$ are independent. 

Suppose $\text{gap}(A?) = \text{gap}(B?) = 1$. Then $\text{gap}((A \text{ xor } B)?) = -1$, because if $A$ and $B$ are true, then $A \text{ xor } B$ is false. 

On the other hand, $\text{gap}((AB \text{ or } \neg A\neg B)?) = \text{gap}(A?)\text{gap}(B?)$ when $A?$ and $B?$ are independent:
\begin{align}
\text{gap}((AB \text{ or } \neg A\neg B)?) &= 2P(AB \text{ or } \neg A\neg B) - 1 \notag \\
&= 2(P(AB) + P(\neg A\neg B)) - 1 \notag \\
&= 2(P(A)P(B) + P(\neg A)P(\neg B)) - 1 \notag \\
&= 2(P(A)P(B) + (1 - P(A))(1 - P(B))) - 1 \notag \\
&= 2(2P(A)P(B) - P(A) - P(B) + 1) - 1 \notag \\
&= 4P(A)P(B) - 2P(A) - 2P(B) + 1 \notag \\
&= 2(P(A) - 1) \cdot 2(P(B) - 1) \notag \\
&= \text{gap}(A?)\text{gap}(B?).
\end{align}

This shows that the binary operation that naturally combines askable questions is, ``Do these questions have the same answer?", which is $(AB \text{ or } \neg A\neg B)?$. This is the operation for which signed probability gaps multiply when independent askable questions are combined. 

We therefore define $A? * B?$ to be the question, $(AB \text{ or } \neg A\neg B)?$.

\subsection{The Weight of Evidence}
\label{subsec:weight-evidence}

A proposition, $A$, has an associated quantity of information, $i(A)$, which is the amount of information that $A$ asserts beyond what is known. It's the amount of information that would be added to the probability distribution if $A$ was given.

When we antisymmetrize this, we get:
\begin{align}
e(A?) &= i(A) - i(\neg A) \notag \\
&= \log P(\neg A) - \log P(A) \notag \\
&= \log \frac{P(\neg A)}{P(A)} \notag \\
&= -\log \frac{P(A)}{P(\neg A)}.
\end{align}

$P(A)/P(\neg A)$ is the odds in favor of $A$, and $P(\neg A)/P(A)$ is the odds against $A$. The log of the odds in favor of $A$ is called the weight of evidence in favor of $A$, and the log of the odds against $A$ is called the weight of evidence against $A$ \cite{Good1950}.

Looking at the equation above, using this terminology, we can say that, when we antisymmetrize the quantity of information that $A$ asserts beyond what is known, we get the weight of evidence against $A$. The weight of evidence against $A$ is therefore a property of the signed question of $A$.

When $e(A?)$ is positive, $P(A) < 1/2$, because the weight of evidence against $A$ is positive. If $e(A?)$ is negative, then $P(A) > 1/2$, because the weight of evidence is in favor of $A$.

Like probability, information, logs of probability gaps, and logs of geometric mean probabilities, evidence is a quantity that permits a form of reasoning using addition and subtraction.

If $B$ is given, then the odds of $A$ change by a multiplicative factor:
\begin{align}
\frac{P(A|B)}{P(\neg A|B)} = \frac{P(A)}{P(B)} \times \frac{P(B|A)}{P(B|\neg A)}.
\end{align}

This means that the evidence against $A$ decreases by an additive quantity:
\begin{align}
e(A?|B) = e(A?) - \log \frac{P(B|A)}{P(B|\neg A)}.
\end{align}

Another way of saying the same thing is that the evidence in favor of $A$ increases by $\log \frac{P(B|A)}{P(B|\neg A)}$. The quantity $\log \frac{P(B|A)}{P(B|\neg A)}$ is called the weight of evidence in favor of $A$ provided by $B$. We can denote it as $e(B \to A)$, and then we have:
\begin{align}
e(A?|B) = e(A?) - e(B \to A).
\end{align}

If two propositions, $B$ and $C$, are both given, then they are said to provide independent evidence for $A$ if $B$ and $C$ are independent given $A$ and independent given $\neg A$, which implies:
\begin{align}
\frac{P(BC|A)}{P(BC|\neg A)} = \frac{P(B|A)P(C|A)}{P(B|\neg A)P(C|\neg A)}.
\end{align}

\noindent which then implies:
\begin{align}
e(A?|BC) &= e(A?) - \log \frac{P(B|A)}{P(B|\neg A)} - \log \frac{P(C|A)}{P(C|\neg A)} \notag \\
&= e(A?) - e(B \to A) - e(C \to A).
\end{align}

If $e(A?)$ reaches zero, then the weight of evidence is neutral, and $P(A) = 1/2$. If $e(A?)$ reaches infinity, then the weight of evidence against $A$ is infinite, and $P(A) = 0$. If it reaches negative infinity, then the weight of evidence in favor of $A$ is infinite, and $P(A) = 1$.

Evidence, then, is a property of an askable question, which permits reasoning using addition and subtraction when an independence condition is satisfied, namely independence of the contributing propositions given either answer to the askable question.

\subsection{The Group of Askable Questions}
\label{subsec:group-askable-questions}

We've defined the binary operation, $*$, on askable questions as:
\[
A? * B? = (AB \text{ or } \neg A\neg B)? = (\neg(A \text{ xor } B))?.
\]

\noindent and we've shown that unsigned probability gaps multiply when combined using this operation: $\text{gap}(A? * B?) = \text{gap}(A?)\text{gap}(B?)$.

To prove that askable questions form a group under this operation, we need to show that:
\begin{itemize}
    \item There is an identity element,
    \item Each askable question has an inverse, and
    \item The $*$ operation is associative
\end{itemize}

When we calculate $A? * A?$, we find:
\begin{align}
A? * A? = (A \text{ or } \neg A)? = \text{True}?.
\end{align}

The trivial question, $\text{True}?$, does act as an identity element:
\begin{align}
A? * \text{True}? = (A \text{ and True or } \neg A \text{ and False})? = A?.
\end{align}

So every askable question is its own inverse, yielding the trivial question, $\text{True}?$, when combined with itself using $*$. Finally, we need to check if $(A? * B?) * C? = A? * (B? * C?)$. We can do this by breaking it down into two cases:

\begin{enumerate}
    \item The answer to $B?$ is Yes.

    In this case:
    \begin{itemize}
        \item $(A? * B?) * C? = A? * C?$
        \item $A? * (B? * C?) = A? * C?$
    \end{itemize}

    So if the answer to $B$ is Yes, then both $(A? * B?) * C?$ and $A? * (B? * C?)$ ask if $A?$ and $C?$ have the same answer.

    \item The answer to $B?$ is No.

    In this case:
    \begin{itemize}
        \item $(A? * B?) * C? = (\neg A)? * C? = A? * C? * \text{False}?$
        \item $A? * (B? * C?) = A? * (\neg C)? = A? * C? * \text{False}?$
    \end{itemize}

    If the answer to $B$ is No, then they both ask if $A?$ and $C?$ have different answers.
\end{enumerate}

This proves that the $*$ operation is associative, and that askable questions therefore form a group.

\subsection{Doubt}
\label{subsec:doubt}

The $*$ operation generates a new question, $A? * B?$, from two askable questions, $A?$ and $B?$, by asking whether they have the same answer: $A? * B? = (AB \text{ or } \neg A\neg B)?$.

If we are careful, we can write this as $A? * B? = (A = B)?$, where we use $A = B$ to denote $AB \text{ or } \neg A\neg B$. The Boolean operation that maps the values of $A$ and $B$, each of which is either True or False, to the value of $AB \text{ or } \neg A\neg B$, is called XNOR: 
\[
A \text{ xnor } B = AB \text{ or } \neg A\neg B.
\]

It can, however, be thought of as $=$ in the sense that, if $A \text{ xnor } B$ is True, then $A$ and $B$ have the same truth value.

The expression $A? * B? * C?$ can then be written as:
\[
A? * B? * C? = ((A = B) = C)?.
\]

This question doesn't ask if $A$, $B$, and $C$ all have the same truth value. It asks whether the truth value of $C$ is the same as the truth value of $A = B$. That is, it asks whether $C$ tells us whether $A = B$ or not.

Similarly, $A? * B? * C? * D? = (((A = B) = C) = D)?$, which asks whether $D$ tells us whether $C$ tells us if $A = B$ or not.

There is a danger, however, associated with using this notation. If we write $A = B = C$, just as we can write $A? * B? * C?$, then we are in danger of thinking that $A = B = C$ is the proposition that $A$, $B$, and $C$ all have the same truth value. In fact, $(A = B) = C$ is true if $C$ is false and $A$ and $B$ have different truth values.

We are technically allowed to remove the brackets from $(A = B) = C$ because $=$ is associative, but we shouldn't, so we should use XNOR instead of $=$ whenever we remove the brackets:
\[
((A = B) = C)? = (A \text{ xnor } B \text{ xnor } C)? = A? * B? * C?.
\]

Let's suppose that $A$, $B$, and $C$ all have probabilities greater than $1/2$, and are independent questions, so that $\text{gap}(A? * B? * C?) = \text{gap}(A?) \text{gap}(B?) \text{gap}(C?)$, where all of the numbers involved are positive.

Then the log of the gap adds up linearly as we add new questions to the string connected by $*$:
\begin{align}
\log \text{gap}(A? * B? * C?) = \log \text{gap}(A?) + \log \text{gap}(B?) + \log \text{gap}(C?).
\end{align}

The log of a positive gap is a non-positive number, because a gap is either 1, in which case the log is zero, or less than 1, in which case the log is negative.

The quantity:
\begin{align}
-\log \text{gap}(A? * B? * C?) &= -\log \text{gap}(A?) - \log \text{gap}(B?) - \log \text{gap}(C?) \notag \\
&= d(A? * B? * C?) = d(A?) + d(B?) + d(C?).
\end{align}

\noindent is then a non-negative quantity that adds up as more independent questions are added to the string.

What we can observe is that, when all the propositions are likely to be true, $P(A) > 1/2$ etc., $\text{gap}(A? * B? * C?)$ is a positive number. That is, $P((A = B) = C) > 1/2$.

It remains more likely than not that $((A = B) = C)$ is true, although the certainty that it is true decreases with each additional proposition added. The quantity that accumulates, as we make the string of questions connected by $*$ longer, can be called doubt.

Doubt is a more specific thing than uncertainty: A specific proposition is doubted, $((A = B) = C)$ in this case. Uncertainty, on the other hand, refers to a symmetric lack of certainty, which applies equally to a proposition and its negation.

For propositions with probability 1, $P(A) = 1$, there is no doubt that they are true: 
\[
\text{gap}(A?) = 1 \Rightarrow d(A?) = -\log \text{gap}(A?) = -\log 1 = 0.
\]

\noindent while, for propositions with probability $1/2$, there is infinite doubt:
\[
\text{gap}(A?) = 0 \Rightarrow d(A?) = -\log 0 = \infty.
\]

As we add more askable questions to a string connected by $*$, in which the answer to each question is probably Yes, we become more and more doubtful that the answer to the resulting question, $(((A = B) = C)\ldots)$, is Yes.

As explained in the section on pure questions, more independent propositions connected by AND results in more information. More independent questions, however, connected by their binary operation, $*$, results in more doubt.

In the case when one of the propositions has a probability less than $1/2$, e.g.\ $\text{gap}(A?) < 0$, while $\text{gap}(B?) > 0$ and $\text{gap}(C?) > 0$, the quantity of doubt associated with $A? * B? * C?$ is complex:
\begin{align}
d(A? * B? * C?) &= -\log \text{gap}(A?) - \log \text{gap}(B?) - \log \text{gap}(C?) \notag \\
&= -\log -|\text{gap}(A?)| - \log |\text{gap}(B?)| - \log |\text{gap}(C?)| \notag \\
&= -\log |\text{gap}(A?)| - \log |\text{gap}(B?)| - \log |\text{gap}(C?)| + i \pi.
\end{align}

The quantity $d(A?) = -\log \text{gap}(A?)$, therefore has a real part which measures how much doubt there is, and an imaginary part which specifies whether it is doubted that the answer is Yes, or that the answer is No.

So if $d(A?)$ has an imaginary part which is equal to $\pi$, or an odd multiple of $\pi$, then $A$ is probably false, and the real part of $d(A?)$ measures how doubtful it is that $A$ is false.

\subsection{The Group of Askable Questions Generated by $N$ Independent Propositions}
\label{subsec:structure-group-askable-questions}

The group, $Q(N)$, of pure questions generated by $N$ independent propositions was studied earlier. It has $2^{2^N-1}$ elements, which is half of the $2^{2^N}$ propositions that can be generated by combining the $N$ independent propositions using AND, OR and NOT.

In the group, $K(N)$, of askable questions, there are twice as many elements, $2^{2^N}$, matching the number of propositions. For every proposition, $X$, there is a corresponding askable question, $X?$.

An askable question, $X?$, in $K(N)$ can be specified by specifying the corresponding pure question, $x$, and then specifying one of the two alternatives, $X?$ or $(\neg X)?$. Also, negations cancel out when askable questions are combined using $*$: $(\neg A)? * (\neg B)? = A? * B?$.

This means that we can represent an askable question, $X?$, as a pair $(x, +)$, consisting of a pure question, $x$, and a sign, $+$ or $-$, indicating which of the two askable questions in $x$ is referred to. When we do this, the $*$ operation acts independently on the two components:
\[
(x, -) * (y, -) = (x * y, +).
\]

In other words, $K(N)$ has the structure of a direct product: $K(N) = Q(N) \times \mathbb{Z}_2$, where $\mathbb{Z}_2$ is the group with two elements.

\subsection{Answers}
\label{subsec:answers}

The mathematical representation of the askable question, $A?$, is the ordered pair, $(A, \neg A)$. When combined with an answer, the question yields a proposition, either $A$ or $\neg A$.

If the answer is Yes, then the result of combining the question with the answer is $A$. So Yes acts as a function that selects the first element of an ordered pair, while No selects the second element. If we define:
\begin{align}
\text{Yes}((x, y)) &= x \\
\text{No}((x, y)) &= y
\end{align}

\noindent then the functions Yes and No satisfy:
\begin{align}
\text{Yes}(A?) &= A \\
\text{No}(A?) &= \neg A
\end{align}

If we use the full set representation of an askable question, $A? = \{\{A\}, \{A, \neg A\}\}$, and consider how Yes and No select their results, then we can probe a bit deeper into the structures of the Yes and No functions.

The set representing an askable question, $A?$, contains a set with one proposition, $A$, and a set containing that proposition and its alternative, $\{A, \neg A\}$. The Yes function selects the proposition and ignores the set of alternatives. The No function selects the alternative to the proposition inside the set.

So the structure of Yes is:
\begin{itemize}
    \item Given an option and an alternative, select the option provided.
\end{itemize}

\noindent and the structure of No is:
\begin{itemize}
    \item Given an option and an alternative, select the alternative.
\end{itemize}

These are the unique functions that, together with askable questions, represented by ordered pairs, yield full propositions.

\subsection{Algebraic Normal Form}
\label{subsec:algebraic-normal-form}

Any question, $g$, in the group, $Q(N)$, of pure questions generated by $N$ independent propositions, can be expressed as a string of generators connected by $*$, such as $g = q(a) * q(ab)$.

$g$ is the question of whether or not $A \text{ xor } AB$ is true:
\begin{align}
q(a) * q(ab) = q(A \text{ xor } AB).
\end{align}

The proposition $A \text{ xor } AB$ is the same as $A\neg B$, but in the group of questions, it's naturally represented by a string of conjunctions connected by XOR, without any negated propositions appearing in the expression.

This way of writing a proposition is called Algebraic Normal Form \cite{Boole1854,HammingLogic}. In general, a proposition which is a combination of other propositions, using AND, OR and NOT, can be written either as a string of negation-free conjunctions connected by XOR:
\[
X = A \text{ xor } ABC \text{ xor } DCEF.
\]

\noindent or it can be written as one of these conjunctions XORed with True:
\[
\neg X = \text{True} \text{ xor } A \text{ xor } ABC \text{ xor } DCEF.
\]

False is the identity element for the XOR operation, so:
\[
X = \text{False} \text{ xor } A \text{ xor } ABC \text{ xor } DCEF.
\]

We can see that Algebraic Normal Form expresses a proposition as a combination of a question and an answer, and it expresses the question using its natural representation within the group of questions.

If $A$ and $B$ are independent, then $P(A \text{ xor } B) = P(A) + P(B) - 2P(A)P(B)$. So Algebraic Normal Form doesn't combine the terms in the expression in a way that's optimal for calculating probabilities. It does, however, express them in a way that's optimal for calculating probability gaps: If $X = A \text{ xor } B$, where $A$ and $B$ are independent, then $\text{gap}(x) = \text{gap}(a)\text{gap}(b)$.

From what we have learnt from the algebra of askable questions, we can say that XNOR, corresponding to the $*$ operation, can be used instead of XOR, and will produce an algebra capable of expressing everything that can be expressed using XOR. 

The askable question, $X?$, of an arbitrary proposition, $X$, which is a combination, using AND, OR and NOT, of independent propositions, can be expressed in the form:
\[
X? = (A \text{ xnor } ABC \text{ xnor } DCEF)?
\]
\noindent or
\[
X? = (\text{False} \text{ xnor } A \text{ xnor } ABC \text{ xnor } DCEF)?.
\]

Because the XNOR operation corresponds to the $*$ operation on askable questions, using XNOR has the advantage that:
\begin{align}
\text{gap}((A \text{ xnor } B)?) = \text{gap}(A? * B?) = \text{gap}(A?) \text{gap}(B?)
\end{align}

\noindent whereas:
\begin{align}
\text{gap}((A \text{ xor } B)?) = -\text{gap}(A?) \text{gap}(B?).
\end{align}

\noindent when $A$ and $B$ are independent. This makes the ``dual form" of Algebraic Normal Form, using XNOR instead of XOR, more aligned with a pure specification of arithmetic operations on probability gaps.

XNOR, like XOR, is a group operation on propositions. The identity element for the XNOR operation is True, while it is False for XOR. This means that we can express any proposition as a conjunction of propositions combined by XNOR:
\[
X = A \text{ xnor } ABC \text{ xnor } DCEF
\]

\noindent or as such a combination XNORed with False:
\[
X = \text{False} \text{ xnor } A \text{ xnor } ABC \text{ xnor } DCEF.
\]

So, while Algebraic Normal Form, using XOR as the connecting operation, is capable of expressing any logical proposition without negations and using only one connecting operation, its dual form, using XNOR, can do the same thing, but aligns more closely with the mathematics of askable questions, making it more convenient to use for calculating probability gaps.

\subsection{Predicates}
\label{subsec:predicates}

A subject isn't a pure question; it's an equivalence class of pure questions. To specify a pure question, it's necessary to specify, in addition to the subject, an element of the equivalence class. 

A rank-$m$ subject is an element of $S_m(N) = Q_m(N)/Q_{m-1}(N)$. This is an equivalence class which has the same number of elements as $Q_{m-1}(N)$. This means that we can specify a full question by specifying the pair, $(s, x)$, where $s \in S_m(N)$, and $x \in Q_{m-1}(N)$.

In order to write an expression such as $s * x$ for the question, however, we need to choose a canonical question to represent the subject $s$. If the subject is $abc$, then we can choose the pure question, $q(ABC)$, to represent it. The full question can then be specified as $q(ABC) * x$, where $q(ABC)$ is an element of $Q(N)$ representing the subject, and $x$ is an element of $Q(N)$ representing the predicate.

For example, in the expression:
\[
q(ABC) * q(AB) * q(C)
\]

\noindent $q(AB) * q(C)$ represents the predicate, and $q(ABC)$ represents the subject. 

In Algebraic Normal Form, when we write:
\[
Y = \text{False} \text{ xor } A \text{ xor } ABC \text{ xor } DCEF
\]

\noindent or when we use the alternative version with XNOR:
\[
Y = \text{False} \text{ xnor } A \text{ xnor } ABC \text{ xnor } DCEF
\]

\noindent the highest-ranking term, $DCEF$ in this case, corresponds to the subject, $dcef$. The combination of terms of lower rank, $A \text{ xor } ABC$ in one case and $A \text{ xnor } ABC$ in the other case, corresponds to the predicate. Together, the subject and the predicate constitute a question, while False corresponds to the answer.

This shows that Algebraic Normal Form already exhibits the internal structure that the algebra of questions induces on propositions. It expresses a proposition as a combination of a subject, a predicate, and an answer. A predicate has a natural expression as a combination of subjects.

\section{Actions On Probability Distributions}
\label{sec:actions-probability-distributions}

The conditional probability, $P(W|X)$, is the probability that $W$ would have if $X$ was given. If we learn that $X$ is true, then the value of $P(W)$ will change to $P(W|X)$.

So propositions can act on probability distributions by being given:
\begin{align}
P(W) \to P(W|X).
\end{align}

When a proposition is given, its probability is set to 1: $P(X) \to P(X|X) = 1$. 

Pure questions, askable questions, and subjects, are logical objects which, like propositions, can be used for reasoning. We can look at what actions they naturally have on probability distributions.

Pure questions have probability gaps, which can range between 0 and 1. Setting the probability gap to 0 and setting it to 1 are candidates for ways that pure questions act on probability distributions.

However, setting the probability gap to 1 requires a selection of one of the two possible outcomes. Setting the gap to 1 doesn't specify whether the probability should be 1 or 0, so setting the gap to 1 isn't an actual specification of a new probability distribution.

The only action that a pure question, $x$, can have that is symmetric between $X$ and $\neg X$ is to set $P(X)$ to $1/2$, setting $\text{gap}(x)$ to 0:
\begin{align}
P(W) \to P(W|x) = \frac{P(W|X) + P(W|\neg X)}{2}.
\end{align}

We call this action ``raising the question". We write it on the right-hand side of the conditional bar, using lower case to distinguish it from a proposition.

Raising a question can remove information from a probability distribution. If $P(X) = 1/2$ initially, then, when $X$ is given, $P(X)$ becomes 1, and the amount of information in the probability distribution increases by one bit. If the pure question of $X$ is then raised, $P(X)$ becomes $1/2$ again, removing that one bit of information.

Introducing the action of raising a question into probability theory results in a new algebra, which is the combined algebra of actions of propositions and pure questions on probability distributions.

We can write a general element of this algebra as a sequence of propositions and questions, such as $AbCDeF$, indicating the sequence of actions from left to right. 

In the expression, $P(AB|cD)$, what appears on the right-hand side of the conditional bar is an element, $cD$, of this algebra, rather than a list of given propositions. It indicates that this probability function is the one obtained from $P$ by first applying $P(W) \to (P(W|C) + P(W|\neg C))/2$, and then applying $P(W) \to P(W|D)$.

This algebra is not commutative: $P(X|Xx) = P(X|x) = 1/2$, but $P(X|xX) = P(X|X) = 1$. The order of the actions in the sequence matters. 

Allowing pure questions to be raised greatly increases our ability to navigate among probability distributions. If we can only act on probability distributions with propositions, then we can only add information, and, in the space of probability distributions, we can only move in the direction of increasing information.

Askable questions can't be raised, but they, together with an answer, result in a proposition being given. This corresponds to an answer-dependent action on the probability distribution:
\begin{align}
A? : P(W) \to 
\begin{cases} 
P(W|A) & \text{with probability } P(A) \\ 
P(W|\neg A) & \text{with probability } P(\neg A)
\end{cases}
\end{align}

Asking a question refers to requesting an answer, and consequently causing a proposition to be given.

We can't denote the action of asking a question as $P(W|A?)$, because, when the question is asked, there are two possible resulting probability distributions, depending on the answer. $P(W|A?)$ is not a well-defined single number.

This means that, on the right-hand side of the conditional bar, only propositions and pure questions will appear.

Subjects, like pure and askable questions, are logical objects which permit a form of quantitative reasoning. The only probability distribution that has the same symmetries as the subject itself is the distribution that is uniform over its possible alternatives. The subject $ab$ can be raised, setting $P(AB) = P(A\neg B) = P(\neg AB) = P(\neg A\neg B) = 1/4$:

\begin{align}
P(W) \to P(W|ab) = \frac{P(W|AB) + P(W|A\neg B) + P(W|\neg AB) + P(W|\neg A\neg B)}{4}.
\end{align}

Raising the subject $ab$ can be accomplished by first raising $a$ and then raising $b$. The notation $P(W|ab)$ can be regarded as indicating that the subject $ab$ is raised, or alternatively that $a$ is raised and then $b$ is raised. The result is the same.

\subsection{The Role of $\sim$ in the Algebra of Actions on Probability Distributions}
\label{subsec:role-tilde-actions}

As shown in Section \ref{sec:unexpected-discovery}, when $0 < P(B) < 1$, the $\sim$ relation satisfies:
\[
P(A) = 1 \Rightarrow P(B|A) = P(B), \text{ and}
\]
\[
P(A) = 1 \Rightarrow P(B|\neg A) = P(\neg B).
\]

This implies that, if $A \sim B$, and $P(A) = 1$, then:
\begin{align}
P(B|a) = \frac{P(B|A) + P(B|\neg A)}{2} = \frac{P(B) + P(\neg B)}{2} = \frac{1}{2}.
\end{align}

What this says is that, if $A \sim B$ and $P(A) = 1$, then raising the question of $A$ also raises the question of $B$. We can express this as: $A \sim B$ and $P(A) = 1 \Rightarrow P(W|a) = P(W|ab)$. Similarly, if $A \sim B$ and $P(A) = 0$, then $P(W|a) = P(W|ab)$.

This rule makes it possible to use $\sim$ in symbolic reasoning about actions on probability distributions. 

The $\sim$ relation is the unique relationship between pure questions for which:
\begin{itemize}
    \item The questions may be treated as independent subjects, when calculating geometric mean probabilities and information values of subjects.
    \item They can not be treated as independent questions, when calculating probability gaps.
    \item Raising a settled question raises the entire subject.
\end{itemize}

If $P(B) = 0$ or $P(B) = 1$, though, the relation $A \sim B$ doesn't put any constraint on $P(B|\neg A)$. If $a \sim b$ and both $a$ and $b$ are settled, then raising $a$ may or may not raise $b$. Raising a settled question is only guaranteed to raise the other question if it isn't settled, $\text{gap}(b) < 1$.

\subsection{Adding and Removing Information by Raising Pure Questions}
\label{subsec:adding-removing-info}

If we consider two independent propositions, $A$ and $B$, with $P(A) = P(B) = 1/2$ and $P(AB) = P(A)P(B)$, then the probability distribution is uniform, and the amount of information in $P$ is at a minimum.

Now suppose that $q(AB)$ is raised, setting $P(AB)$ to $1/2$. The distribution would no longer be uniform. $AB$ would have half of the probability, with the remaining three options, $A\neg B$, $\neg AB$, and $\neg A\neg B$, sharing the remaining half of the probability among them.

So raising $q(AB)$ adds information to $P$ if $P$ was initially uniform. This shows that raising questions doesn't always take information away from a probability distribution.

In a case such as $q(A)$, where $A$ is one of the generating propositions, it is clear that raising $q(A)$ only takes information away: $(P(W|A) + P(W|\neg A))/2$. It assigns half of the probability to $A$ being true, which corresponds to half of the possible alternatives, $AB$, $A\neg B$, $\neg AB$, $\neg A\neg B$. This has an averaging effect, bringing the distribution closer to uniform, and removing information.

Raising $q(AB)$, on the other hand, assigns half of the probability to $1/4$ of the possible alternatives. This would add information to a probability distribution that was initially uniform. It would take information away if $P(AB)$ was greater than $1/2$.

We can see that raising $q(A \text{ xor } B)$, where $A$ and $B$ are generating propositions, assigns half of the probability to half of the cases, namely when $A$ and $B$ have different answers. This means that, like $q(A)$, $q(A) * q(B)$ only takes information away when raised.

This means that:
\begin{itemize}
    \item Rank-1 questions, which are combinations using $*$ of questions from the $N$ generating propositions, always remove information from a probability distribution when raised.
\end{itemize}

Raising questions of other rank may add or remove information, depending on $P$.

However, for any proposition, $A$, raising the question, $a$, always removes information from the probability distribution that specifies $P(A)$ and $P(\neg A)$. It may add or remove information from other probability distributions that include more variables.

We can also see that raising a rank-1 question never increases the unsigned probability gap of another rank-1 question. Raising $q(A \text{ xor } B)$ pushes the probabilities of $A$ and $B$ closer to $1/2$.

It is possible to perform an operation on a probability distribution that removes information overall, but makes some questions more certain and other questions less certain. For example, if $A$ and $B$ are independent, and $P(A) = 1$ and $P(B) = 0.5$, then setting $P(A)$ to $1/2$ and $P(B)$ to $0.6$ will remove information overall, but it adds information about $B$ that wasn't there before.

Raising a rank-1 question doesn't do this. For all other rank-1 questions, it either brings their probabilities closer to $1/2$, or leaves them unchanged. It's a ``pure subtraction" of information from the probability distribution.

\section{Revisiting the Mathematical Expression for $\sim$}
\label{sec:revisiting-tilde}

Once again using $a = P(A)$, $b = P(B)$, and $x = P(AB)$ to denote the probabilities, and using:
\[
\text{gap}(A?) = 2a - 1
\]
\[
\text{gap}(B?) = 2b - 1
\]

\noindent the complex expression for $\sim$ given in Section \ref{sec:unexpected-discovery} breaks down into the following:
\begin{align}
\label{eq:T} T &= \frac{1}{8} \cdot (3 - \text{gap}(A?)^2) \cdot (3 - \text{gap}(B?)^2) - \frac{3}{2} \\
S &= -\frac{5}{32} \cdot \left( \left( \frac{9}{5} - \text{gap}(A?)^2 \right) \cdot \left( \frac{9}{5} - \text{gap}(B?)^2 \right) - \left( \frac{9}{5} \right)^2 + 9 \right) \\
Y &= \text{gap}(A?) \cdot \text{gap}(B?) \cdot S \\
U &= \sqrt{T^3 + Y^2} \\
\label{eq:V} V &= 2 \cdot w_2 \cdot \sqrt[3]{Y + U} \\
\label{eq:x} x - a \cdot b &= \frac{\text{Re}(V) - \text{gap}(A?) \cdot \text{gap}(B?)}{3}
\end{align}

\noindent where $w_2 = \frac{-1 - i \sqrt{3}}{2}$ is one of the cube roots of unity: $w_2^3 = 1$, and the cube root function takes the principal cube root, $\sqrt[3]{z} = \exp(\log(z)/3)$.

This way of representing $\sim$ is much simpler than the original expression. The code provided verifies that this method of calculating $x = P(AB)$ gives the correct solution.

The great simplification achieved by introducing probability gaps indicates that $\sim$ is indeed fundamentally a relationship between questions. However, the result is still quite intricate.

Equations \ref{eq:T}--\ref{eq:V} above describe a sequence of transformations that brings the pair of real numbers, $\text{gap}(A?)$ and $\text{gap}(B?)$, to a final single complex number, $V$. Equation \ref{eq:x} says that, when $A \sim B$, $P(AB)$ exceeds $P(A)P(B)$ by one-third of the amount by which the real part of $V$ exceeds $\text{gap}(A?)\text{gap}(B?)$.

It is worth noting that the following equation is always true:
\begin{align}
x - a \cdot b = \frac{\text{gap}(A? * B?) - \text{gap}(A?) \cdot \text{gap}(B?)}{4}.
\end{align}

When $A \sim B$, then, the real part of $V$ behaves like the probability gap of a question similar to $A? * B?$. The discrepancy of $\text{Re}(V)$ from $\text{gap}(A?) \text{gap}(B?)$ is $3/4$ of the discrepancy of the gap of $A? * B?$.

\begin{figure}[h!]
    \centering
    \includegraphics[height=4cm]{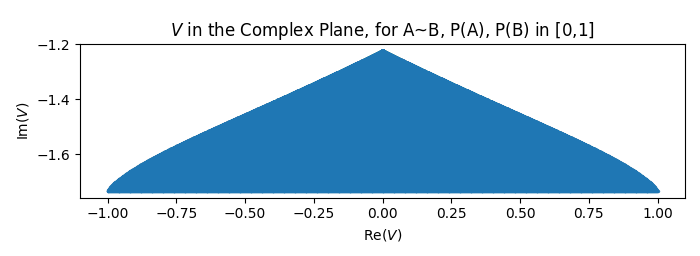}
    \caption{\footnotesize The shape formed in the complex plane by the values of $V$ for every possible value of $P(A)$ and $P(B)$. The unit square of possible probabilities is mapped by $V$ to a curved triangular shape in the complex plane. The real co-ordinate of a point inside this shape plays the role of a probability gap in the equation, $x - a \cdot b = (\text{Re}(V) - \text{gap}(A?) \cdot \text{gap}(B?))/3$.}
    \label{fig:figure-7.1}
\end{figure}

Figure \ref{fig:figure-7.1} shows the values of $V$ in the complex plane for all values of $P(A)$ and $P(B)$ between 0 and 1. It forms a curved triangular shape. The imaginary part of $V$ varies between $-\sqrt{3}$ and $-\sqrt{3/2}$, while the real part varies between $-1$ and 1, making it suitable to use as a probability gap. 

\begin{figure}[h!]
    \centering
    \includegraphics[width=0.5\textwidth]{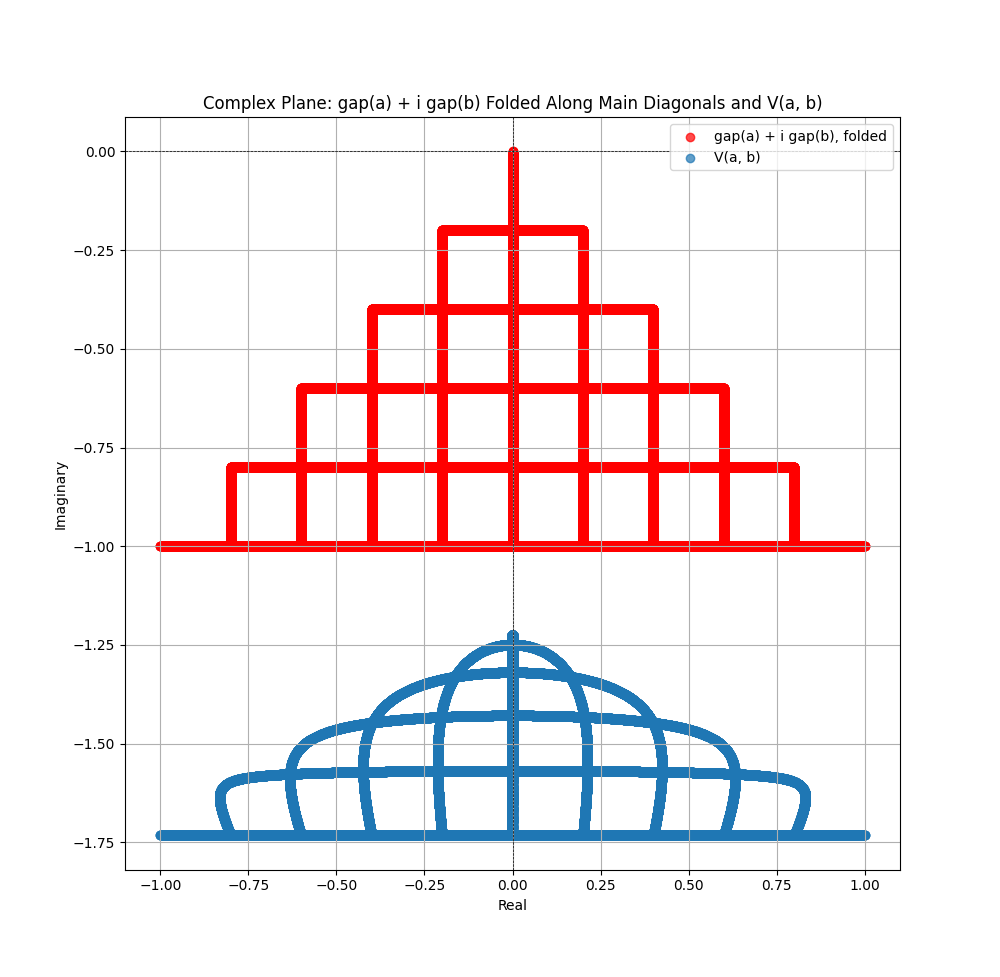}
    \caption{\footnotesize The shape at the top is the result of folding the lines $\text{gap}(A?) + i \text{gap}(B?)$, for $0 \leq P(A) \leq 1$ and $P(B) \in \{0, 0.1, 0.2, \ldots, 1\}$, along the main diagonals, $x = y$ and $x = -y$, where $z = x + i y$. The shape at the bottom shows the values of $V$ in the complex plane for those values of $P(A)$ and $P(B)$.}
    \label{fig:figure-7.2}
\end{figure}

To gain insight into its structure, we can show the values of $V$ when $P(A)$ takes any value and $P(B)$ is in $\{0, 0.1, 0.2, \ldots, 1\}$. This is shown in Figure \ref{fig:figure-7.2} at the bottom.

Also shown in Figure \ref{fig:figure-7.2}, at the top, is the result of folding the values of $\text{gap}(A?) + i \text{gap}(B?)$ along the two main diagonals. The figure makes it clear that the image of $V$ in the complex plane is a curved and displaced version of the result of folding the square of possible gaps, $(\text{gap}(A?), \text{gap}(B?))$.

\begin{figure}[h!]
    \centering
    \includegraphics[width=\textwidth]{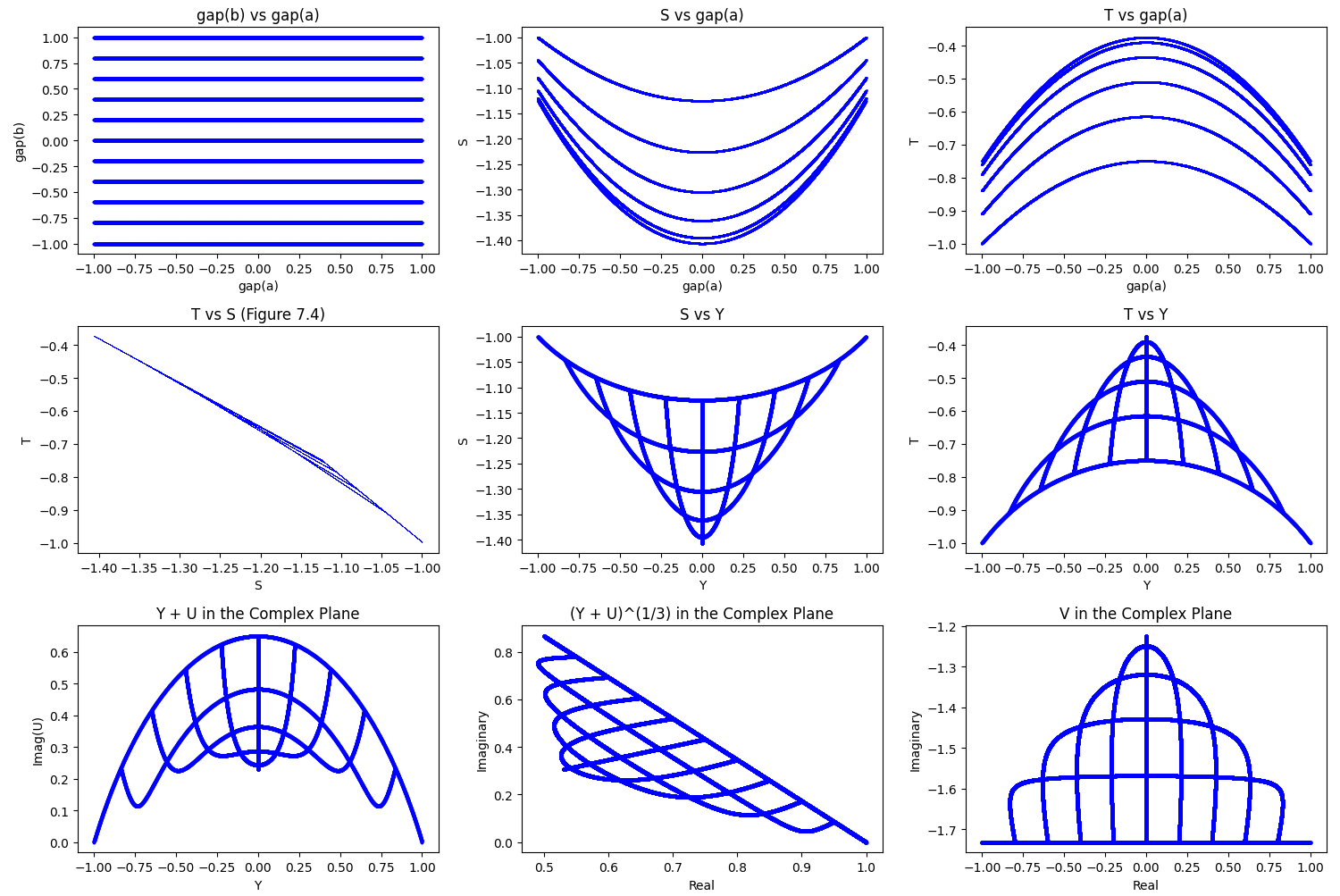}
    \caption{\footnotesize The transformations of the lines of values of $\text{gap}(A?)$ and $\text{gap}(B?)$ achieved by plotting the values of intermediate variables in the formula for $\sim$ against one another, and the resulting shapes in the complex plane. The final result can be thought of as being obtained through ``cubic origami" in the following way. Imagine that the lines in the top-left panel are drawn with marker on a transparent rubber sheet. The shape in the top-right panel is obtained by nailing the top-left corner of the sheet to the bottom-left corner, and also nailing the top-right corner to the bottom-right corner. The shape in the middle-right panel is then obtained by nailing the left-hand side of the top curve in the top-right panel to the right-hand side of the same curve, producing the vertical line in the middle of the middle-right panel. The final result is then obtained by nailing the curve at the bottom to a horizontal line.}
    \label{fig:figure-7.3}
\end{figure}

Figure \ref{fig:figure-7.3} shows the sequence of transformations represented by equations \ref{eq:T} to \ref{eq:V}, starting from the unit square of pairs of gaps in real 2-dimensional space, and ending with the corresponding values of $V$ in the complex plane.

\begin{figure}[h!]
    \centering
    \includegraphics[width=\textwidth]{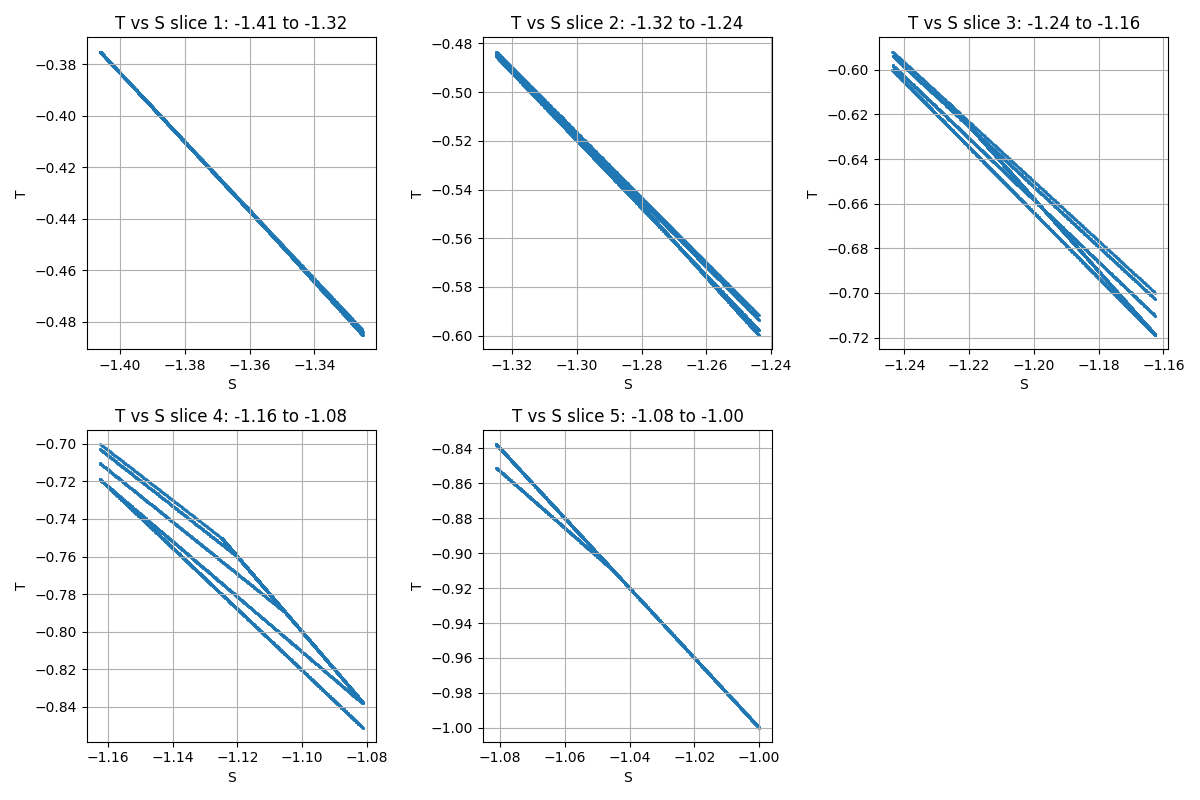}
    \caption{\footnotesize A detailed view of the shape formed by plotting $T$ versus $S$. $T$ uses $(3 - \text{gap}(A?)^2)(3 - \text{gap}(B?)^2)$, and $S$ uses $(9/5 - \text{gap}(A?)^2)(9/5 - \text{gap}(B?)^2)$, along with rescaling and an offset, to construct numbers that relate the two pure questions, $q(A)$ and $q(B)$. The lines cross over each other in many places, showing that the transformation from $\text{gap}(A?)$ and $\text{gap}(B?)$ co-ordinates to $T$ and $S$ co-ordinates folds the unit square as well as stretches it out into a long structure. The $Y$ variable multiplies $S$ by $\text{gap}(A?)\text{gap}(B?)$, stretching the narrow structure out into the curved triangular structure shown in the middle-right panel of Figure 7.3.}
    \label{fig:figure-7.4}
\end{figure}

As Figure \ref{fig:figure-7.3} shows, when the co-ordinates change from $\text{gap}(A?)$ and $\text{gap}(B?)$ to $S$ and $T$, the result is a long, narrow diagonal shape. The fine structure of this shape is shown in \ref{fig:figure-7.4}. There, we see that the lines corresponding to constant values of $P(B)$ intersect many times within the extent of the shape. 

The next transformation involves introducing $Y$, which multiplies $S$ by $\text{gap}(A?) \cdot \text{gap}(B?)$, stretching out the narrow shape into a triangle with a curved base, when either $T$ or $S$ is plotted against $Y$. The quantity $U$ turns out to be purely imaginary, while $Y$ is real, so $Y + U$ is a complex number which is shown next. The cube root of this number is a rotated version of the final shape. This is then rotated in the complex plane by 120 degrees, through multiplication by $w_2$, to yield $V$.

\subsection{The Connecting Question}
\label{subsec:connecting-question}

When $A \sim B$, the discrepancy of $P(AB)$ from $P(A)P(B)$ is one-third of the discrepancy of $\text{Re}(V)$ from $\text{gap}(A?)\text{gap}(B?)$.

$\text{Re}(V)$ appears in this equation as a signed probability gap, playing the role of the gap of an askable question. We can investigate the properties that such a ``question" would need to have, and see whether it is, in fact, a question as we have defined it.

Let's call it the connecting question, and denote it by $C(A, B)?$, or just $C?$ when convenient. Then:
\[
\text{Re}(V) = \text{gap}(C?)
\]

\noindent so we have:
\[
x - a \cdot b = \frac{\text{gap}(C?) - \text{gap}(A?) \cdot \text{gap}(B?)}{3},
\]

\[
x - a \cdot b = \frac{\text{gap}(A? * B?) - \text{gap}(A?) \cdot \text{gap}(B?)}{4}.
\]

From the above two equations, we can see that $\text{gap}(C?)$ is always in between $\text{gap}(A?) \cdot \text{gap}(B?)$ and $\text{gap}(A? * B?)$. For $\text{gap}(C?)$ to be equal to 1, both $\text{gap}(A?) \text{gap}(B?)$ and $\text{gap}(A? * B?)$ must be equal to 1.

So the connecting question can only be known to have an answer of Yes if:
\begin{itemize}
    \item $\text{gap}(A? * B?) = 1$, which implies that the answers to $A?$ and $B?$ must be known to be the same, and
    \item $\text{gap}(A?) \text{gap}(B?) = 1$, which implies that either $\text{gap}(A?) = \text{gap}(B?) = 1$ or $\text{gap}(A?) =\text{gap}(B?) = -1$. This implies that the answers must not only be known to be the same, but must also actually be known.
\end{itemize}

Similarly, if $\text{gap}(C?) = -1$, then $\text{gap}(A? * B?)$ and $\text{gap}(A?) \text{gap}(B?)$ must be equal to $-1$. In other words, if the answer to the connecting question is known to be No, then it must be known that the answers to $A?$ and $B?$ are opposites, and those answers must be known.

In order to ``answer" the connecting question, it is therefore necessary to specify one of the four alternatives, $AB$, $A\neg B$, $\neg AB$, or $\neg A\neg B$. It is a ``question with four answers", which means that it is not the askable part of a proposition.

When we break a proposition into an askable question and an answer, we always get a Yes/No question, and a Yes or No answer. The connecting question is a different type of object. We can't use the $*$ operation with it in the way that we do with askable questions, because $*$ relies on questions having two answers.

\subsection{$\sim$ Uses Complex Numbers to Mix Properties of Pure and Askable Questions}
\label{subsec:tilde-complex-numbers}

We now have an approximate idea of what the mathematical form of the tilde relation is doing. It is using cubic operations to transform the square of possible values of $(\text{gap}(A?), \text{gap}(B?))$ into a triangular shape in the complex plane whose real part can be interpreted as a gap of a hypothetical question that connects $A?$ and $B?$ in a complex way.

$\sim$ is satisfied when the gap of this hypothetical question differs from $\text{gap}(A?)\text{gap}(B?)$ by three times the difference between $P(AB)$ and $P(A)P(B)$. When this occurs, the pure questions, $q(A)$ and $q(B)$, can be treated as independent subjects when calculating geometric mean probabilities, but if one question is settled and the other is not, then raising the settled question raises both questions at the same time.

There is, however, much that we still don't understand. We don't know what algebra the hypothetical connecting question belongs to, and we don't even know if the real part of $V$ corresponds to the gap of an object that actually exists. The convoluted formula for $\text{Re}(V)$ makes it very difficult to interpret as a property of a logical object.

We don't know why $3 - \text{gap}(A?)^2$ and $9/5 - \text{gap}(A?)^2$ play such important roles in the calculation of $V$. We can say that these quantities are both properties of the pure question, $a$, because $\text{gap}(A?)^2 = \text{gap}(q(A))^2$. From this, we can say that:
\[
T = \frac{1}{8} \cdot (3 - \text{gap}(A?)^2) \cdot (3 - \text{gap}(B?)^2) - \frac{3}{2}
\]

\noindent and:
\[
S = -\frac{5}{32}  \cdot \left( \left( \frac{9}{5} - \text{gap}(A?)^2 \right) \cdot \left( \frac{9}{5} - \text{gap}(B?)^2 \right) - \left( \frac{9}{5} \right)^2 + 9 \right)
\]

\noindent are quantities that relate the two pure questions, $q(A)$ and $q(B)$, while:
\[
Y = \text{gap}(A?) \cdot \text{gap}(B?) \cdot S
\]

\noindent is a quantity that relates the signed questions, $A?$ and $B?$, because it changes sign when $A$ is replaced by $\neg A$ or $B$ is replaced by $\neg B$. 

On the other hand, $U$ is a complex quantity that relates the pure questions, $q(A)$ and $q(B)$, because $Y$ appears in it squared:
\[
U = \sqrt{T^3 + Y^2}.
\]

That is, both $Y^2$ and $T^3$ are quantities that relate pure questions, so the square root of their sum also relates pure questions, remaining unchanged when either proposition is negated. $T$ is always negative, and $T^3$ is always greater than or equal to $Y^2$ in magnitude, with the result that $T^3 + Y^2$ is always negative or zero. As Figure \ref{fig:figure-7.3} shows, it's zero when $Y$ is 1 or $-1$ and $T$ is $-1$.

Intuitively, $T^3 + Y^2$ is always negative because $T$ is kept close to $-1$ by the subtraction of $3/2$ in its formula. $T$'s maximum value is $-3/8$, and its minimum value is $-1$. $Y$, on the other hand, approaches zero when $\text{gap}(A?)$ or $\text{gap}(B?)$ approach zero. The offsets and the scaling factors in the formulae for $T$, $S$, and $Y$, ensure that $|T^3| \geq |Y^2|$.

$U$ is consequently a purely imaginary quantity relating the two pure questions, while $Y$ relates the signed questions. Interestingly, the equation $T^3 + Y^2 = -|U|^2$, for a constant value of $U \neq 0$, specifies an elliptic curve whose points have $T$ and $Y$ co-ordinates\footnote{The standard form of an elliptic curve is $y^2 = x^3 + ax + b$. This corresponds to $y = Y$, $x = -T$, $a = 0$, and $b = -|U|^2$, in the formula for the tilde relation.}. Elliptic curves are symmetric around an axis -- the $Y$-axis in this case. They also have a group structure: Two distinct points on the curve can be added by drawing a straight line connecting them. The point where the continuation of this line intersects the curve for a third time, reflected across the axis of symmetry of the curve, is the result of adding the two points. 

This suggests that a form of reasoning, reflected in the group structure of the elliptic curve, could be involved. The magnitude of $U$ indexes a family of elliptic curves, and the magnitude of $U$ specifies $U$ entirely, since $U = i \sqrt{|T^3 + Y^2|} = i |U|$. 

The sum, $Y + U$, which appears in the expression for $V$, combines a real part that relates the signed questions with an imaginary part that relates the unsigned questions. This is an interesting quantity because of how it behaves under negation. If we write $U = i R$, then:
\begin{align}
A \to \neg A : Y(\neg A, B) + i R(\neg A, B) = -Y(A, B) + i R(A, B).
\end{align}

So $Y + U$ doesn't change sign when $A$ is replaced by $\neg A$, but it does change sign if that substitution is accompanied by complex conjugation:
\begin{align}
A \to \neg A, i \to -i : Y(\neg A, B) + i R(\neg A, B) = -Y(A, B) - i R(A, B).
\end{align}

This shows that the quantitative properties of pure questions can be considered to be the imaginary parts of quantitative properties of askable questions. We just need to amend the definition of a quantitative property of an askable question:
\begin{itemize}
    \item $s(A)$ is a property of the askable question, $A?$, if $s(A) = -s(\neg A)^*$
\end{itemize}

With this definition, the symmetric and the antisymmetric properties of $A$ can be represented as complex-valued antisymmetric properties of the askable question, $A?$. The real part of such a complex-valued property is a real-valued property of $A?$, and the imaginary part is a property of the pure question, $q(A)$.

Let's consider whether the complex number, $V=2 w_2 \sqrt[3]{Y + U}$, is a valid property of $A?$ according to this definition:

\begin{align}
A \to \neg A, i \to -i : \sqrt[3]{Y + U} \to \sqrt[3]{-Y - U} = -w \sqrt[3]{Y + U}.
\end{align}

\noindent where $w$ is either 1, $w_2$, or $w_1 = w_2^*$, which are the cube roots of 1. So:

\begin{align}
A \to \neg A, i \to -i : w_2 \sqrt[3]{Y + U} \to& \ w_1 \sqrt[3]{-Y - U}\notag \\
& = -w_1 w \sqrt[3]{Y + U} \notag \\
& = -w_2 \sqrt[3]{Y + U}.
\end{align}

\noindent where we have chosen $w = w_1$, and used $w_1^2 = w_2$ to reach the final result.

This means that:
\[
V = 2 w_2 \sqrt[3]{Y + U}
\]

\noindent is a quantity that changes sign when negation of a proposition is combined with complex conjugation.

This is a complex-valued quantity that encodes a relation between the askable questions, $A?$ and $B?$, in its real part, and also a relation between the pure questions, $q(A)$ and $q(B)$, in its imaginary part.

This shows that what is discarded when the real part of $V$ is taken, in the solution:
\[
x - a \cdot b = \frac{\text{Re}(V) - \text{gap}(A?) \cdot \text{gap}(B?)}{3}
\]

\noindent is the part of $V$ that relates the pure questions, leaving a real quantity that relates $A?$ and $B?$.

\section{Complex-Valued Properties of Questions}
\label{sec:complex-valued-properties}

The tilde relation has taught us that we can represent properties of pure questions as the imaginary parts of properties of askable questions. This also works the other way around: If $s(A) = s(\neg A)$, and $t(A) = -t(\neg A)$, then the quantity $s(A) + i t(A)$ is symmetric when negation is combined with complex conjugation:
\[
A \to \neg A, i \to -i : s(A) + i t(A) \to s(A) - i t(\neg A) = s(A) + i t(A).
\]

So, if we define the quantitative properties of the pure question of $A$ to be those for which $s(A) = s(\neg A)^*$, then properties of the askable question can be represented as imaginary components of properties of the pure question.

If we choose to represent a complex-valued property of a question using an antisymmetric real part and a symmetric imaginary part, which is what $\sim$ does, then a complex number, $z(A) = x(A) + i y(A)$, is a property of the askable question of $A$ if:
\[
z(\neg A) = x(\neg A) + i y(\neg A) = -x(A) + i y(A) = -z(A)^*.
\]

Complex numbers can therefore be used to specify both a property of an askable question and a property of the corresponding pure question, in a single number. This suggests that the algebra of complex numbers is relevant to relations between questions where both the pure question and the askable question are involved in the relation. 

For example, if $z(A?)$ is a complex-valued property of $A?$, encoding information about $q(A)$ as well as $A?$, then $z(A?) = -z(\neg A?)^*$. If there is a function, $f$, which maps complex numbers to complex numbers, and $f(-x + i y)^* = -f(x + i y)$, then $f(z)$ is also a property of $A?$.

The condition $f(-x + i y)^* = -f(x + i y)$ requires the imaginary part of $f$ to be even in $x$ and the real part to be odd in $x$. It's a strong constraint, and most complex functions don't satisfy it. Expressed in terms of $z = x + i y$, the constraint is $f(-z^*)^* = -f(z)$, or, equivalently, $f(-z^*) = -f(z)^*$.

The functions that do satisfy $f(-x + i y)^* = -f(x + i y)$ are operations that can be performed on complex-valued quantitative properties of questions, and which yield other complex-valued properties, in a way that handles the symmetric and antisymmetric quantities involved correctly.

A randomly chosen complex function, such as $f(z) = z^2$, will not satisfy the condition. The function $z^2$ doesn't preserve the symmetric and antisymmetric parts of a complex-valued property of a question:
\[
z^2 = (x + i y)^2 = x^2 + 2i x y - y^2.
\]

Under the operation $x \to -x$, $i \to -i$, $z^2$ remains unchanged. Neither its real nor imaginary parts change when negation is combined with conjugation. This means that it can't be a complex-valued property of an askable question, because it satisfies $f(-z^*)^* = f(z)$ rather than $f(-z^*)^* = -f(z)$.

Only functions that treat the real part of a quantity antisymmetrically, and treat the imaginary part of it symmetrically, relate complex-valued properties of questions in a coherent way. Complex addition does this: if the real parts of $z_1(A)$ and $z_2(A)$ are both antisymmetric under interchange of $A$ with $\neg A$, while their imaginary parts are symmetric, then $z_1 + z_2$ also shares these properties.

Multiplication by a non-zero real number also preserves the structure. This means that the complex-valued properties of a question form a real vector space. However, there are additional, non-linear operations that can also be performed. An example of a non-trivial function that satisfies this constraint is $f(z) = z^3$:
\[
(x + i y)^3 = x^3 - 3x y^2 + 3i x^2 y - i y^3.
\]

The real part of this, $x(x^2 - 3y^2)$, is antisymmetric in $x$, while the imaginary component, $i x^2 y - 3i y^3 = i y (3x^2 - y^2)$ is symmetric in $x$. In the above expression for $f(z) = z^3 = (x + i y)^3$, replacing $x$ with $-x$ and $i$ with $-i$ yields $-z^3$, so $f(-z^*)^* = -f(z)$.

This means that taking the cube of a complex-valued property of a question yields another valid complex-valued property of the same question.

Similarly, multiplying three or any odd number of complex-valued properties yields another valid property, but multiplying two or any even number doesn't.

\subsection{Cube Roots of Complex-Valued Properties of Questions}
\label{subsec:cube-roots-properties}

The formula for the tilde relation contains an example of an operation that satisfies $f(-z^*)^* = -f(z)$, namely:
\[
f(z) = w_2 z^{1/3}.
\]

This operation takes the principal cube root of $z$, $z^{1/3}$, which is the root with the smallest angle with the $x$-axis, and multiplies it by $w_2 = \frac{-1 - i \sqrt{3}}{2}$, which is a cube root of 1.

$f(z)$ is a valid cube root of $z$, because $f(z)^3 = w_2^3 z = z$, but it is not the principal root, $z^{1/3}=e^{\log(z)/3}$.

As demonstrated in the code provided, $w_2 z^{1/3}$ only satisfies $f(-z^*)^* = -f(z)$ in the upper half of the complex plane, while $w_1 z^{1/3}$ satisfies it in the lower half. The principal root, $z^{1/3}$, doesn't satisfy it in either half-plane.

The result of $w_2 z^{1/3}$ always has a negative imaginary component, because, when $z$ is expressed as $r e^{i \theta}$, $z^{1/3}$ is $r^{1/3} e^{i \theta / 3}$, so the angle of $z^{1/3}$ is one-third of the angle of $z$, which brings $z^{1/3}$ into the range of angles between $-60$ degrees and 60 degrees. Rotating this through multiplication by $w_2$ results in $w_2 z^{1/3}$ having an angle between $-180$ degrees and $-60$ degrees, yielding a negative imaginary component. Similarly, $w_1 z^{1/3}$ always has a positive imaginary component. 

So there are two cube root operations for complex-valued properties of questions:
\begin{itemize}
    \item $w_2 z^{1/3}$ is valid in the upper half-plane, and yields a cube root of $z$ which is in the lower half-plane.
    \item $w_1 z^{1/3}$ is valid in the lower half-plane, and yields a cube root of $z$ which is in the upper half-plane.
\end{itemize}

\begin{figure}[h!]
    \centering
    \includegraphics[width=0.45\textwidth]{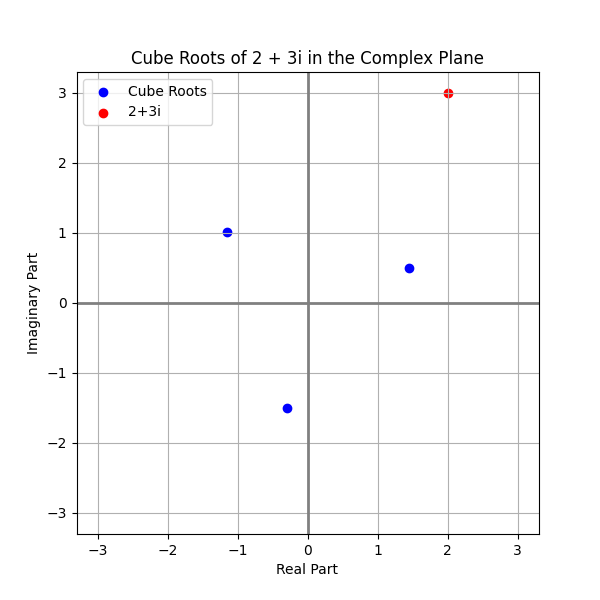}
    \caption{\footnotesize There are three complex numbers which are valid cube roots of a given complex number. If the original number is not real, then two of the cube roots lie in one half-plane (the principal root and one other), while the third root lies in its own half-plane. The root in its own half-plane is the single cube root of a complex-valued property of a question that yields another valid complex-valued property of the same question.}
    \label{fig:figure-8.1}
\end{figure}

These can be considered to be a single operation that selects the root that is in its own half-plane, as shown in Figure \ref{fig:figure-8.1}. So, within the algebra of complex transformations of complex-valued properties of questions, there is one operation that takes the cube of a complex-valued property, and one operation that takes the cube root of a complex-valued property.

This is a feature that complex-valued properties of questions share with real numbers: There is only one real number which is the cube root of a given real number. If there is a complex number, $z$, and we need to know which cube root of $z$ to take, then knowing that the result should be a real number, or a property of a question, is enough to specify a unique root.

\subsection{Complex Functions that Relate Properties of Questions}
\label{subsec:complex-functions-properties}

We've seen above that functions, $f(z)$, whose real parts are antisymmetric and whose imaginary parts are symmetric, are valid relations between complex-valued properties of questions. The properties can be added and multiplied by real numbers to yield other valid properties of the same question, forming a vector space. There are valid cube and cube root operations.

The following functions also satisfy $f(-z^*) = -f(z)^*$:
\begin{itemize}
    \item $e^{-i \pi / 2n} z^{1/n}$ in the upper half-plane, and $e^{i \pi / 2n} z^{1/n}$ in the lower half-plane,
    \item $z^n z^{*m}$, where $n + m$ is odd,
    \item $i z^n z^{*m}$, where $n + m$ is even, and,
    \item Sums and power series of the terms above, with real coefficients.
\end{itemize}

This includes functions like $\sin(z)$ and $\sinh(z)$, which are odd functions expressible as power series in $z$, as well as $i \cos(z)$ and $i \cosh(z)$, which are even functions multiplied by $i$. It also includes $i e^{i z} = i \cos(z) - \sin(z)$. This generates a rich algebra of functions relating valid complex-valued properties of questions.

\section{Whole Questions}
\label{subsec:whole-questions}

If we multiply a valid complex-valued property of an askable question by $i$, we get a valid complex-valued property of the pure question, and vice versa. So even though pure questions don't contain any preference for either of the two possible outcomes, a preference can be encoded into the sign of the pure question's properties. 

This makes it possible to effectively treat the pure question and the askable question as two manifestations of an underlying ``whole question", whose properties we can define as:
\begin{itemize}
    \item A function, $g(A)$, of the proposition, $A$, is a property of the whole question of $A$ if $g(A) = e^{i \phi} f(A)$, where $f(\neg A) = -f(A)^*$.
\end{itemize}

If $\phi = 0$, then $g$ is a property of the askable question. If $\phi = \pi/2$, then it's a property of the pure question. Identifying a valid property of the whole question identifies a unique valid property of the askable question, up to an overall sign, since $-f$ and $f$ are both valid properties achievable by shifting the phase of $g$.

The concept of a whole question expands the class of functions that qualify as properties of questions. Any function, $g$, that satisfies $g(\neg A) = e^{i \theta} g(A)^*$, for any value of $\theta$, qualifies as a property of the whole question. For any valid property of the askable question, there's a circle of valid properties of the whole question.

The properties of the parts of a proposition are properties of the proposition too. For example, a property of the pure question, $a$, is a property, $f(A)$, of the proposition, $A$, that satisfies $f(\neg A) = f(A)^*$. A property of the askable question satisfies $f(\neg A) = -f(A)^*$.

Similarly, the whole question is a part of the proposition. A property, $g(A)$, of the proposition, $A$, is a property of the whole question of $A$ if $g(\neg A) = e^{i \theta} g(A)^*$. The whole question is not as large as the proposition, $A$, because not every function of $A$ satisfies $g(\neg A) = e^{i \theta} g(A)^*$. However, it contains the askable and pure questions as parts, since their properties are also its properties.

The simplest set that can represent the whole question is the set, $(A, ¬A)=\{\{A\}, \{A, \neg A\}\}$, that represents the askable question. The proposition, $A$, can be recovered from $(A, \neg A)$, and so any function of $A$ is a function of $(A, \neg A)$. 

Pure questions are uniquely embodied by the set $\{A, \neg A\}$. Any function of that set must be symmetric under interchange of $A$ with $\neg A$, because the set itself doesn't change. However, the set $A?=\{\{A\}, \{A, \neg A\}\}$ allows $f(A?)$ and $f(\neg A?)$ to be different, and doesn't enforce $f(\neg A) = -f(A)$ or any other condition. 

When we specify askable questions in terms of sets, we can use $(A, \neg A)$ to represent $A?$, but we also need to specify that its properties are those for which $f((A, \neg A))=-f((\neg A, A))$. We need to put restrictions on the functions of the set that qualify as properties, as well as specifying the set. In the case of pure questions, we can just specify the set.

The whole question is represented by the same set, $\{\{A\}, \{A, \neg A\}\}$, as the askable question, but with a different restriction on the functions that qualify as its properties, namely $g(\neg A) = e^{i \theta} g(A)^*$.

These two distinct objects are represented by the same set in the formalism we're using here, because we are using logical propositions as the urelements of the sets. We're constructing other objects out of propositions and sets. This means that any set from which the proposition, $A$, can be extracted by a function, has all of the properties of $A$, if we suppose that every function of the set is a property.

Because of this, we can't construct different sets for the askable question and the whole question. We can only restrict the functions on those sets that qualify as properties of the two different objects.

We can use the notation, $\underline{A?}$, to represent the whole question of $A$. Whole questions inherit the group operation and group structure of askable questions:
\begin{align}
\underline{A?} * \underline{B?} = \underline{(AB\text{ or }\neg A \neg B)?}.
\end{align}

\subsection{Properties of Whole Questions}
\label{sec:properties-whole-questions}

A function, $g(A)$, that satisfies $g(\neg A)=e^{i \theta}g(A)^*$, for any value of $\theta$, is a property of the whole question of A. These are functions of the form $g(A)=e^{i\phi}f(A)$, where $f(\neg A)=-f(A)^*$.

We can see immediately that the product of two properties of a whole question is another valid property:
\begin{align}
g_1(\neg A) g_2(\neg A) &=  e^{i \theta_1}g_1(A)^* e^{i \theta_2}g_2(A)^* \notag \\
&= e^{i (\theta_1+\theta_2)} g_1(A)^* g_2(A)^*.
\end{align}

However, the sum of two properties is not always a valid property: In general, $e^{i\phi_1}f_1(A) + e^{i\phi_2}f_2(A)$ can't be factorized into a phase and a function that satisfies $f(\neg A)=-f(A)^*$. The phases, $\phi_1$ and $\phi_2$, need to be equal, or differ by $\pi$, to allow us to write the sum as $e^{i \phi}(f_1(A)+f_2(A))$ or $e^{i \phi}(f_1(A)-f_2(A))$.

Unlike pure and askable questions, the properties of whole questions don't form a vector space. In order to add two valid properties of a whole question, we need to first ensure that their phases are either aligned or anti-aligned. 

If there are two valid properties of the whole question of $A$, $g_1(A)=e^{i\phi_1}f_1(A)$ and $g_2(A)=e^{i\phi_2}f_2(A)$, then there are two sums of the form $g_1(A)+e^{i \phi}g_2(A)$ which are also valid properties, namely:

\begin{itemize}
    \item $g_1(A)+e^{i(\phi_1-\phi_2)}g_2(A) = e^{i\phi_1}f_1(A) + e^{i(\phi_1-\phi_2+\phi_2)}f_2(A) = e^{i\phi_1}(f_1(A)+f_2(A))$, and
    \item $g_1(A)+e^{i(\phi_1-\phi_2+\pi)}g_2(A) = e^{i\phi_1}f_1(A) + e^{i(\phi_1+\pi)}f_2(A) = e^{i\phi_1}(f_1(A)-f_2(A))$.
\end{itemize}

This means that, in order to write a general property of a whole question as a sum of other properties, it is necessary to use complex coefficients: $g_1(A)+g_2(A)$ isn't always a valid property, but there are two values of $\phi$ for which $g_1(A)+e^{i \phi}g_2(A)$ is a valid property.

\subsection{Values of Properties of Whole Questions}
\label{sec:values-properties-whole-questions}

A complex-valued property of a pure question is a function that satisfies $f(A)=f(\neg A)^*$. The function only needs to have two values, $f(A)$ and $f(\neg A)$. However, in probability theory, propositions have probabilities, and so those values can be functions of $P(A)$.

For example, the geometric mean probability of $q(A)$ is a property that can be expressed as $\text{gmp}(q(A))=\sqrt{P(A)P(\neg A)}=\sqrt{P(A)(1-P(A))}$. For this property to become a number (a value of a property) instead of a function (a property), we need to specify a value for $P(A)$.

If we do specify a value for $P(A)$, then we get a number, which is real in the case of geometric mean probabilities, but will be a complex number for a complex-valued property, such as $\text{gap}(A?)+i\text{gmp}(q(A))$. In fact, we get two numbers, $f(A)$ and $f(\neg A)$.

So, after probabilities have been specified, a property, $f$, will resolve to a two-dimensional complex vector, $(f(A), f(\neg A))$, specifying the values of this property for $A$ and for $\neg A$.

In the case of pure questions, $f(A)=f(\neg A)^*$, so if we specify $f(A)$, then it fixes the value of $f(\neg A)$. The possible pairs, $(f(A), f(\neg A))$, are therefore in one-to-one correspondence with the complex numbers, which are all of the possible values of $f(A)$.

Similarly, for askable questions, the pairs of complex numbers, $(f(A), f(\neg A))$, are in one-to-one correspondence with $\mathbb{C}$, because $f(\neg A)=-f(A)^*$.

Two-dimensional complex vectors are known as Weyl spinors, or, more commonly, just spinors. The possible  pairs of values of a complex-valued property of an askable question form a subspace of the space of spinors isomorphic to $\mathbb{C}$. The same is true for pure questions.

The possible pairs of values of properties of whole questions form a subset, but not a subspace of the space of Weyl spinors. The pairs of values of a property of a whole question form a spinor of the form 
\(  \begin{pmatrix} a+ib \\ e^{i\phi}(-a+ib) \end{pmatrix} \). This means that any spinor, \( \begin{pmatrix} z_1 \\ z_2 \end{pmatrix} \), for which $|z_1|=|z_2|$, specifies a possible pair of values of a property of a whole question.

These spinors are closed under pointwise multiplication:
\begin{align}
|z_1|=|z_2|, |w_1|=|w_2| \Rightarrow |z_1 w_1| = |z_2 w_2|
\end{align}
\noindent but not under pointwise addition:
\begin{align}
|z_1|=|z_2|, |w_1|=|w_2| \nRightarrow |z_1 + w_1| = |z_2 + w_2|.
\end{align}

They therefore do not form a subspace of $\mathbb{C}^2$, but they form a manifold on which multiplication of points is defined. This reflects how properties of whole questions can themselves be multiplied but not added.

\section{Summary of the Properties of the $\sim$ Relation}
\label{sec:summary-tilde-properties}

To recap, the properties that we could identify without knowing anything about questions, given in Section \ref{sec:unexpected-discovery}, were:
\begin{itemize}
    \item It is the only non-trivial way in which there can be an information balance between two propositions, in which, for the four possible conjunctions, $AB$, $A\neg B$, $\neg AB$, and $\neg A\neg B$, the information redundancies for the two more-likely-than-independence conjunctions cancel out the excess implied information from the two less-likely-than-independence conjunctions.
    \item It is a cubic interpolation of a set of simple rules for updating the probability of one proposition when the other is given.
    \item It describes a relationship in which $A$ and $B$ are mostly independent, except that if one of them has probability 0 and is given, or has probability 1 and its negation is given, the probability of the other proposition switches place with the probability of its negation.
    \item It has a complicated mathematical formula, involving complex numbers and cubic operations, but it can be used for symbolic reasoning without the formula in the special cases when the probability of one of the propositions is equal to 1, 0, or $1/2$.
\end{itemize}

Now that we have studied questions, we can say a lot more about the tilde relation:
\begin{itemize}
    \item It is the unique relation between questions that permits them to be treated as independent subjects, when calculating geometric mean probabilities and information values of subjects, but which does not permit them to be treated as independent questions, when calculating probability gaps and quantities of doubt.
\end{itemize}

It's worth noting that there is no comparable relation for propositions: Propositions don't have geometric mean probabilities; they only have probabilities. If propositions can be treated as independent from the point of view of information, then they are fully independent:
\[
i(AB) = i(A) + i(B) \quad \Rightarrow \quad P(AB) = P(A)P(B)
\]

The introduction of geometric mean probabilities, and their logarithms, information values, permits a new form of reasoning using addition and subtraction. $\sim$ is fundamentally tied in to this new form of reasoning, and has no analogue in classical logic:
\[
i(ab) = i(a) + i(b) \quad \Rightarrow \quad P(AB) = P(A)P(B) \text{ or } A \sim B
\]

\begin{itemize}
    \item It binds the two pure questions, $a$ and $b$, into a ``natural subject", so that raising a settled question raises both questions:
    \[
    A \sim B \text{ and } 0 < P(B) < 1 \text{ and } P(A) \in \{0, 1\} \quad \Rightarrow \quad P(X|a) = P(X|ab)
    \]
\end{itemize}

This equation, and the equation above, show that $\sim$ plays the role of an elementary logical/informational relationship from the point of view of subjects. It appears as an elementary symbol inside our system of reasoning as soon as we try to draw the consequences of $i(ab) = i(a) + i(b)$. 

\begin{itemize}
    \item Expressed using signed probability gaps, $\text{gap}(A?) = 2a - 1$ and $\text{gap}(B?) = 2b - 1$, where $a$ and $b$ are the probabilities of $A$ and $B$, the complex expression for $\sim$ given by Wolfram Alpha simplifies to:
\end{itemize}

\begin{align*}
T &= \frac{1}{8} \cdot (3 - \text{gap}(A?)^2) \cdot (3 - \text{gap}(B?)^2) - \frac{3}{2} \\
S &= -\frac{5}{32} \cdot \left( \left( \frac{9}{5} - \text{gap}(A?)^2 \right) \cdot \left( \frac{9}{5} - \text{gap}(B?)^2 \right) - \left( \frac{9}{5} \right)^2 + 9 \right) \\
Y &= \text{gap}(A?) \cdot \text{gap}(B?) \cdot S \\
U &= \sqrt{T^3 + Y^2} \\
V &= 2 \cdot w_2 \cdot \sqrt[3]{Y + U} \\
x - a \cdot b &= \frac{\text{Re}(V) - \text{gap}(A?) \cdot \text{gap}(B?)}{3}
\end{align*}

\begin{itemize}
    \item It combines a property of a pure question and a property of an askable question into a single complex number, $Y + U$, and then takes a cube root of this complex number, to obtain a final complex number, $V$, whose real part plays the role of a probability gap of a hypothetical ``connecting question", which is only answered if both $A?$ and $B?$ are answered.
    \item It reveals the rule that complex functions must satisfy if they are valid relations between complex-valued properties of questions, namely $f(-z^*) = -f(z)^*$. This is a strong constraint on complex functions, and the simplest non-linear function that satisfies it is $f(z) = z^3$. There is a cube root operation which also satisfies the constraint, which is $f(z) = w_2 z^{1/3}$ in the upper half-plane, and $f(z) = w_1 z^{1/3}$ in the lower half-plane.
\end{itemize}

Overall, then, the attempt to understand $\sim$ by studying questions has been largely successful. Although the mathematical expression for $\sim$ remains mysterious -- we do not know the significance of the numbers 3 and $9/5$, used in the calculation of $S$ and $T$ -- we have nonetheless understood the role of $\sim$ as it relates to the algebra of questions. 

We can now use $\sim$ along with simple rules to calculate geometric mean probabilities, and, in special cases, to calculate the result of raising a question or giving a proposition. Using the simplified but still complex mathematical formula, we can also calculate $P(AB)$ from $P(A)$ and $P(B)$.

\section{The Structure of Knowledge}
\label{sec:structure-knowledge}

In classical probability theory, without questions, the information that we know is represented by the set of propositions that have been given. When we write $P(C|AB)$, the set $\{A, B\}$ specifies the known information used to determine the probability of $C$. The proposition $A$ and $B$ is a single proposition expressing everything that is known.

Since $A$ has been given, it is impossible for $\neg A$ to be given, because then the single proposition would become $A$ and $B$ and $\neg A$, which is a contradiction. More generally, if $P(X) = 0$ for any proposition, $X$, then $P(Y|X)$ does not need to be defined. The formula $P(Y|X) = P(YX)/P(X)$ yields $0/0$, and we can say that $X$ will never be given, since its probability is zero. It can be defined, but it doesn't need to be.

If two propositions, $X$ and $Y$, have probability 1, then they are independent according to the criterion $P(XY) = P(X)P(Y)$. This means that, in classical probability, we are not forced to represent dependencies between known information. Everything which is known can be represented as independent of everything else that's known.

Classical probability is required to represent the structure of unknown information: If $P(X)$ and $P(Y)$ are strictly between 0 and 1, then $X$ implies $Y$ if and only if $P(XY) = P(X)$. It is not required, however, to represent the structure of known information, because everything that is known can be represented by a simple set of propositions, all with probability 1, and with no relations between them.

Introducing questions into probability theory makes it necessary to define $P(Y|X)$ even if $P(X) = 0$. The action of raising the pure question, $x$, is $P(Y|x) = (P(Y|X) + P(Y|\neg X))/2$, which is only defined if both $P(Y|X)$ and $P(Y|\neg X)$ are defined. We need to specify both $P(Y|X)$ and $P(Y|\neg X)$, even if $P(X)$ or $P(\neg X)$ is zero.

What classical probability lacks, that appears when questions are introduced, is a strict requirement to specify every conditional probability, including conditioning on propositions with probability zero. Something can be known to be false, but we must understand what the consequences would be if it turned out to be true, or became true.

The symbols that appear on the right-hand side of the conditional bar, $|$, in probability theory augmented with questions, do not simply specify a set of given propositions. They specify a sequence of actions on the probability distribution, carried out in order from left to right.

In the special case when the sequence of actions consists only of propositions, and none of them are negations of other propositions in the sequence, probability theory augmented with questions coincides with classical probability theory: $P(C|AB)$ is the probability of $C$ given the propositions $A$ and $B$. It doesn't matter whether $AB$ is regarded as a sequence of actions or a conjunction of propositions.

In the new system that appears when questions are introduced, $\neg A$ can be given even if $A$ has already been given: $P(X|A\neg A) = P(X|\neg A)$, because, if $A$ is given, and then $\neg A$ is given, the action of giving $\neg A$ undoes the action of giving $A$. That is, setting $P(A)$ to 0 undoes the action of setting it to 1.

Augmenting probability theory with questions therefore requires us to specify the dependencies between propositions that are known to be true. If $A$ and $B$ both have probability 1, then $P(A|\neg B)$ must be specified. 

If they both have probability 1, then the $P(AB) = P(A)P(B)$ criterion for independence is satisfied, $P(AB) = 1 = P(A)P(B)$, even if $P(A|\neg B)$ is not equal to 1, which implies a dependency. A different criterion of independence is therefore necessary to deal with propositions with probability 1. 

We can say that $B$ is independent of $A$ if $P(B|A) = P(B|\neg A)$, and we can say that $A$ is independent of $B$ if $P(A|B) = P(A|\neg B)$, and we can say that they are independent of each other if both of these are true. 

When none of the probabilities involved are 1 or 0, $P(B|A) = P(B|\neg A)$ implies and is implied by $P(A|B) = P(A|\neg B)$, because they are both equivalent to $P(AB) = P(A)P(B)$. So, in classical probability, $A$ is independent of $B$ if $B$ is independent of $A$. Propositions are either independent or not.

When probabilities are equal to 1, though, there can be one-directional dependencies: $B$ can depend on $A$, while $A$ does not depend on $B$: $\neg A$ can imply $\neg B$, $P(B|\neg A) = 0$, while $A$ is independent of $B$, $P(A|\neg B) = P(A|B) = 1$.

We are dealing with probabilities of propositions, not sets, and so measure theory, and the usual formulation of probability theory in terms of Kolmogorov's axioms \cite{Kolmogorov1933}, are not necessary. We are, however, entitled to imagine measurable sets that correspond to a proposition being true or false, and whose measure corresponds to a probability. 

If we do that, then the requirement of well-defined actions of questions on the probability distribution can be thought of as the requirement that probabilities should be defined even when conditioned on sets with zero measure.

Curiously, any two propositions with probability 1 satisfy the $\sim$ relation: $P(AB) = 1 \Rightarrow A \sim B$. This is because, as shown in Figure~\ref{fig:figure2.4} in Section \ref{sec:unexpected-discovery}, the tilde relation is satisfied when $P(A) = 0$, $P(B) = 1$, and $P(B|A)$ takes any value between 0 and 1. 

When we refer to $\sim$, we don't mean the complex formula provided by Wolfram Alpha, which specifies $P(AB) = 0$ at $P(A) = 0$, yielding $P(AB)/P(A) = 0/0$. We mean its natural extension, through continuity, which specifies a value for $P(B|A)$ at $P(A) = 0$, namely $P(\neg B)$, except when $P(B)$ is 0 or 1, in which case $P(B|A)$ can take any value.

By replacing $A$ with $\neg A$, we can see that, when $P(A) = P(B) = 1$, $P(B|\neg A)$ can also take any value, and still satisfy $\sim$. This means that everything that is known to be true is related to everything else that is known to be true by $\sim$. 

We can summarize the relationship of $\sim$ to the structure of knowledge as:
\begin{itemize}
    \item With unknown information, for which probabilities are neither 1 nor 0, the tilde relation is a balance of information that permits reasoning using geometric mean probabilities.
    \item When relating known information to unknown information, $\sim$ specifies how to update the unknown information when the known information is questioned.
    \item With known information, $\sim$ is universal.
\end{itemize}

\section{The Geometry of Questions - The Two-State Quantum System}
\label{sec:geometry-questions}

In previous sections, we developed the mathematics of questions which were either fully independent, $\text{gap}(A? * B?) = \text{gap}(A?) \text{gap}(B?)$, or were independent as subjects, $A \sim B$.

Now, we turn our attention to explicitly non-independent questions. Specifically, we'll consider questions whose interrelationship can be described geometrically, as an angle, $\theta$, between the two questions.

Let's consider the following scenario:
\begin{itemize}
    \item There are two askable questions, $X?$ and $W?$, at an angle of $\theta$ with respect to each other.
    \item Initially, nothing is known: $\text{gap}(X?) = \text{gap}(W?) = 0$.
    \item Then the question $W?$ is answered, and the answer is Yes: $\text{gap}(W?) \to 1$.
    \item We ask: What is the value of $\text{gap}(X?)$, given that $\text{gap}(W?) = 1$ and nothing else is known?
\end{itemize}

\begin{figure}[t]
    \centering
    \includegraphics[width=0.5\textwidth]{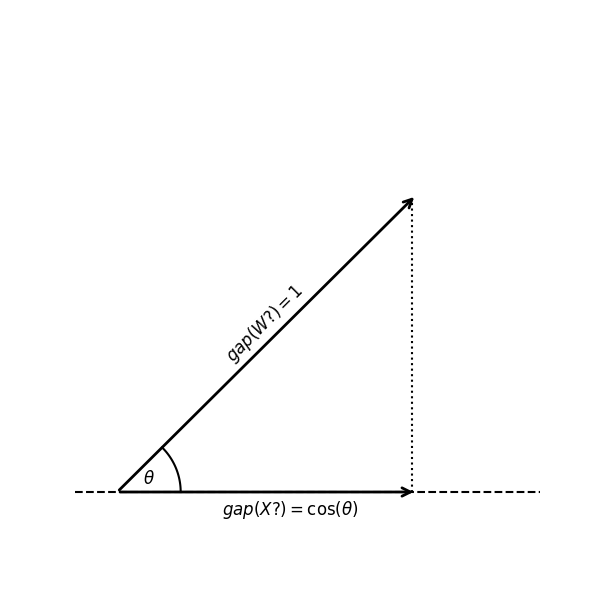}
    \caption{\footnotesize The simplest way that probability gaps can be related to each other geometrically is through cosine projection. When $W$ is known to be true, the probability gap of $W?$ is 1, and, when projected onto the $x$-axis, it yields a gap of $\cos(\theta)$ for $X?$.}
    \label{fig:figure-12.1}
\end{figure}

As Figure \ref{fig:figure-12.1} shows, the simplest and most natural value that $\text{gap}(X?)$ can have is $\cos(\theta)$. This is the component of $\text{gap}(W?)$ that lies along the $x$-direction.
\begin{align}
\text{gap}(X?) = \cos(\theta)
\end{align}

From this, we can derive $P(X)$:
\begin{align}
\text{gap}(X?) = 2 P(X) - 1 = \cos(\theta) \quad \Rightarrow \quad P(X) = \frac{1 + \cos(\theta)}{2}.
\end{align}

In quantum mechanics, this value of $P(X)$ is the probability that a spin-1/2 particle's spin will be found to be ``up" along the $x$-direction, if the initial state is ``up" along the $w$-direction \cite{DiracQM,Sakurai1994}.

This is quite a remarkable result. In standard quantum mechanics, the quantity, $P(X) = (1 + \cos(\theta))/2$, is derived using complex matrices. The fact that the same probability can be derived very simply from askable questions at an angle of $\theta$ with respect to each other seems to hint that quantum mechanics may be dealing with questions behind the scenes.

In the case of spin measurements on spin-1/2 particles:
\begin{itemize}
    \item $X$ is the proposition that, if the particle's spin is measured along the $x$-direction, the result will be ``up".
    \item $X?$ is the askable question that asks if the result will be ``up".
    \item The formula for $P(X)$ is not simple, intuitive or obvious.
    \item The formula for $\text{gap}(X?)$ is simple, intuitive and obvious.
\end{itemize}

We will see later that it is possible to derive the Hilbert space representation of the two-state quantum system, which describes spin-1/2 particles, from the geometry of askable questions. For now, we will focus on a purely real-valued representation of this quantum system.

\subsection{The Bloch Sphere}
\label{subsec:bloch-sphere}

The Bloch sphere is a unit 2-dimensional sphere whose surface corresponds to the set of possible pure states of a two-state quantum system \cite{Bloch1946,Fano1957}. Antipodal points on the sphere correspond to orthogonal quantum states. For example, the north pole of the sphere corresponds to the ``up" state along the $z$-axis, while the south pole corresponds to the ``down" state along the $z$-axis.

From our point of view, though, a point on the surface of the Bloch sphere represents a probability gap of 1, $\text{gap}(W?) = 1$, where $W?$ is an askable question corresponding to a direction in physical space.

The Bloch sphere provides a representation of quantum states in real, 3-dimensional, space. It makes it possible to calculate the probability of a measurement result using a simple projection of a vector onto the axis of measurement.

For example, if the initial state is ``up" along the $z$-axis, then the projection of this vector (the north pole of the Bloch sphere), onto the $x$-axis has a length of zero, indicating a 50\% probability that the result of a measurement of the particle's spin along the $x$-axis will be ``up".

The lengths of these vectors and of their projections onto different axes correspond to probability gaps, $2p - 1$, rather than probabilities. The entire probability distribution, $P$, which specifies the probabilities of measurement results along any axis, is encoded into a single real 3-dimensional vector representing the quantum state.

From this, we can see that the Bloch sphere representation of two-state quantum systems is not merely a way to represent states as vectors in real space, but is fundamentally a representation of how the askable questions involved are related to each other.

It is the simplest way that askable questions can be oriented in space and related to each other geometrically. The fact that the simplest quantum system, the two-state system, has the same structure as a sphere of askable questions suggests that further insights into quantum mechanics might be achieved by considering the properties of questions.

\section{Classical and Quantum Measurements}
\label{sec:classical-quantum-measurements}

In classical physics, we can make a measurement without disturbing the system. The only effect that the measurement has is that we gain information. There's a logical proposition, $X$, whose initial probability is greater than 0 and less than 1. After the measurement, $P(X) = 1$. This updates the probability distribution simply by giving $X$: $P(W) \to P(W|X)$.

Classical probability theory, in which propositions are the only objects that act on probability distributions, is therefore sufficient to describe how probability distributions change in classical physics.

In quantum mechanics, measurements don't simply yield information; they also change the state of the system. Information that we gained in previous measurements may no longer be accurate after a new measurement.

For example, if we measure the spin of a spin-1/2 particle along the $y$-axis, and get ``up" as the result, and subsequently measure its spin along the $x$-axis, then the second measurement destroys the information we previously had about the spin along the $y$-axis \cite{WheelerZurek}.

We can represent this loss of information as the raising of the pure question, $y = q(Y)$, where $Y$ is the proposition that asserts that a measurement along the $y$-axis will yield ``up":
\begin{description}
    \item[First measurement:] $P(W) \to P(W|Y)$
    \item[Second measurement:] $P(W|Y) \to P(W|YyX) = P(W|X)$
\end{description}

The effect of the second quantum measurement on the probability distribution is an element, $yX$, of the combined algebra of actions of pure questions and propositions on probability distributions.

When we express the effect of the measurement as $yX$, we are specifying which information is lost (by specifying $y$), as well as the information gained ($X$). In the case of spin-1/2 particles, all information about the previous state of the system is lost when a new measurement is performed. We can express this as:
\begin{align}
\label{eq:singleparticlespinmeasurement}
P(W) \to P\bigg(W\bigg| \prod_{\theta, \phi} r_{\theta, \phi} \cdot X\bigg)
\end{align}

\noindent where $R_{\theta, \phi}$ is the proposition that a measurement along the $(\theta, \phi)$ axis will yield ``up" as the result, $r_{\theta, \phi}$ is the corresponding pure question, and ${\footnotesize\prod\limits_{\theta, \phi}}$ indicates that this pure question should be raised for every value of $\theta$ and $\phi$\footnote{This is simply a choice of notation to represent the simultaneous raising of all questions. It isn't intended to imply that the action of an uncountable sequence of actions on the probability distribution is well-defined.}.

That is, we can describe the effect of a measurement that yields ``up" along the $x$-axis as a two-part process: First, every pure question, $r_{\theta, \phi}$, is raised, and then the result, $X$, is given. This allows us to express the effect of the measurement without making any reference to the previous state of the system.

It's worth noting that we must express the effects of quantum measurements as a loss of information followed by a subsequent gain of information. If the $w$-axis is at an angle of 45 degrees with respect to the $x$-axis, then:
\[
P(W|Xw) = \frac{1}{2}
\]

\noindent but:
\[
P(W|wX) > \frac{1}{2}
\]

\noindent since the $w$-direction is positively correlated with the $x$-direction. In the special case when the axes are orthogonal, the actions on the probability distribution commute, $Xy = yX$, but only the latter form, $yX$, generalizes to non-orthogonal axes.

In fact, the action of $Xw$ yields a probability distribution that isn't compatible with a valid quantum state of the system. Since the two directions are correlated, it isn't possible to have full certainty along the $x$-direction and complete uncertainty along the $w$-direction. Only actions that consist of very specific question-proposition combinations yield distributions compatible with valid quantum states.

To summarize, in both classical and quantum mechanics, performing a measurement corresponds to asking an askable question, $X?$. They differ because, in classical mechanics, the resulting effect is simply the action of a proposition, $X$ or $\neg X$, whereas, in quantum mechanics, the resulting effect, $yX$ or $y\neg X$, also raises questions.

Our analysis provides a technical but not a philosophical insight into the measurement problem in quantum mechanics, which asks how a superposition of possible measurement results resolves to a single result. We have shown that the subtraction of information that accompanies wavefunction collapse, for two-state systems, can be algebraically represented as raising a question. 

The deeper philosophical question of how the components of the state orthogonal to the measurement result, which, before the measurement, were physically relevant, and in fact detectable through interference experiments, can just disappear from existence and become physically irrelevant, is much more profound.

\subsection{Pure States, Mixed States, and Information}
\label{subsec:pure-mixed-states}

If we initially know nothing at all about the spin of a spin-1/2 particle, $P(X) = P(Y) = P(Z) = 1/2$, then the particle is in a mixed state, and, when we perform a measurement, we gain information but lose none, since we had no information to begin with: $P(W) \to P(W|X)$. Specifically, we gain 1 bit of information.

On the other hand, if we know as much as we can possibly know (1 bit), then the initial state of the system is a pure state, in which there is an axis, say the $y$-axis, along which we have certainty about the result of a measurement, $P(Y) = 1$. A subsequent measurement along the $x$-axis which yields ``up", $P(W|Y) \to P(W|YyX) = P(W|X)$, removes 1 bit of information and adds 1 bit.

A measurement on the maximally mixed state, in which we know nothing initially, acts like a classical measurement. Information is gained but not lost. With a pure state, by contrast, the amount of information lost is equal to amount of information gained.

From the point of view of information, then, a pure state is one in which we have as much information as we can possibly have about the system. We cannot gain more information without losing the same amount.

The two-state quantum system (the qubit) has an information capacity of 1 bit. We can know less than that, but we can't know more. It's the simplest quantum system, with the smallest information capacity.

Mixed states are those states in which the amount of information in the probability distribution is less than the information capacity of the system, while, for pure states, the probability distribution has as much information as it can possibly have.

\section{Deriving The Hilbert Space Representation from Questions}
\label{sec:hilbert-space-from-questions}

\subsection{Geometric Operations on the Sphere of Askable Questions}
\label{subsec:geometric-operations-sphere}

In the case of the Bloch sphere, its expression in terms of questions was very straightforward, requiring no more than a trivial relabeling of points on the surface of the sphere as askable questions.

The Hilbert space is a more complex structure, and how questions relate to it is not trivial or obvious. Before we examine the structure of the Hilbert space, let's see what natural operations act on the sphere of questions. 

For this purpose, we assume that we know nothing about quantum mechanics, but that we start with a collection of questions arranged on the surface of a sphere, whose probability gaps are related geometrically through cosine projections.

\begin{figure}[h!]
    \centering
    \includegraphics[width=0.6\textwidth]{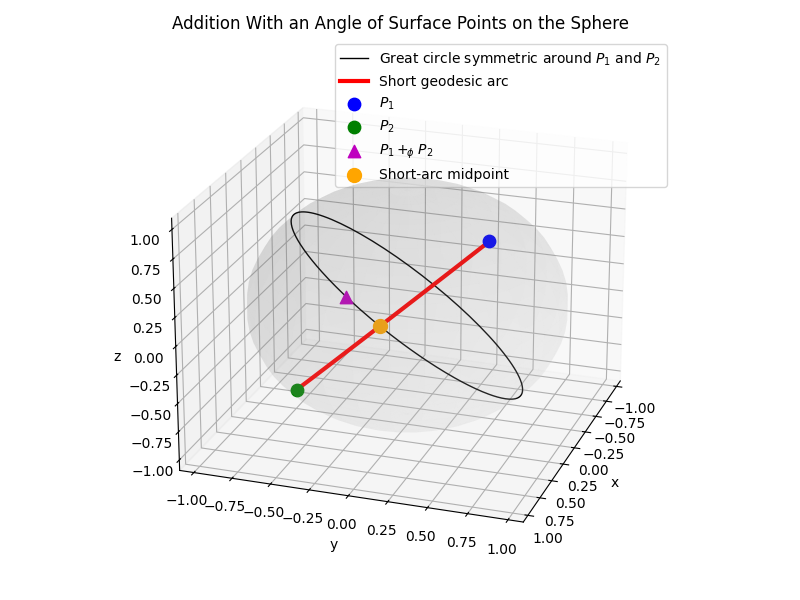}
    \caption{\footnotesize The points that are symmetrically positioned on the sphere with respect to $P_1$ and $P_2$ form a great circle, which intersects the shortest arc connecting $P_1$ and $P_2$ at its midpoint. Any point, $P_3$, on this circle can be obtained by starting at the shortest-arc midpoint, and then rotating around the great circle by an angle, $\phi$. This makes it possible to define an operation of ``addition of points with an angle", $P_3 = P_1 +_\phi P_2$.}
    \label{fig:figure-14.1}
\end{figure}

Given two points on the sphere, there is a great circle of points which are equidistant from the two points. The perpendicular bisector of the chord connecting the two points is a plane that passes through the center of the sphere. The plane's intersection with the sphere is a great circle, as shown in Figure \ref{fig:figure-14.1}.

The figure shows that any point on this great circle can be specified by specifying the angle around the great circle between that point and the intersection of the great circle with the shortest arc connecting the two points. This allows us to define an operation of ``addition with an angle":
\[
P_3 = P_1 +_\phi P_2.
\]

For every angle, $\phi$, there is a binary operation, $+_\phi$, on points of the sphere, with $+_0$ corresponding to the shortest-arc-midpoint operation.

This is a purely geometrical operation on points on the surface of the sphere, which can be considered to be a binary operation on the askable questions corresponding to those points:
\[
P_3? = P_1? +_\phi P_2?.
\]

There is, however, no clear algebraic structure that naturally arises from ``addition with an angle". There is no identity element nor are there inverses, so the $+_\phi$ operation doesn't naturally generate a group, for any value of $\phi$, such as $\phi = 0$. The arc-midpoint operation, $+_0$, yields a unique point, $P_3$, from $P_1$ and $P_2$, but doesn't generate any useful algebraic structures.

\subsection{Addition in the Hilbert Space}
\label{subsec:addition-hilbert-space}

In the previous subsection, we looked at geometric operations on points of a sphere, and found a family of addition operations, $+_\phi$. The Hilbert space has a single addition operation:
\[
\ket{\psi_1}, \ket{\psi_2} \to \ket{\psi_1} + \ket{\psi_2}.
\]

\begin{figure}[h!]
    \centering
    \includegraphics[width=0.6\textwidth]{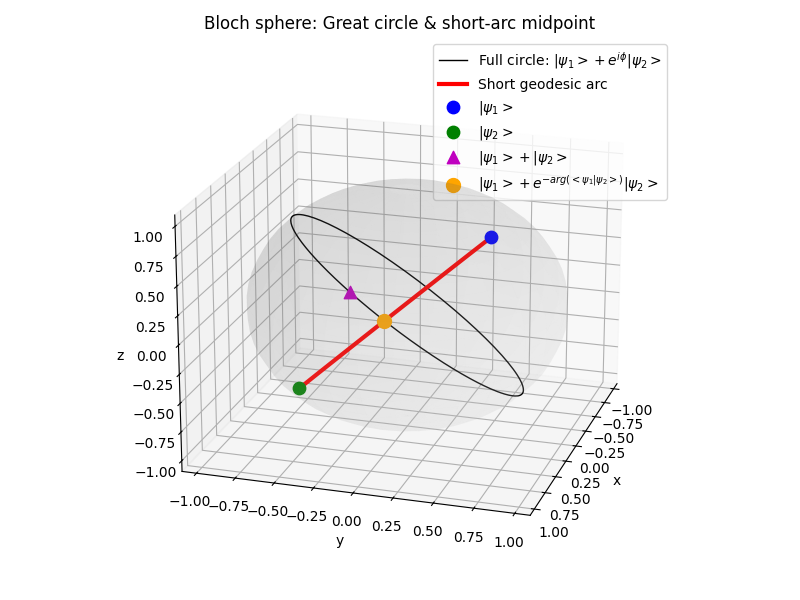}
    \caption{\footnotesize The same plot as Figure 14.1, relabelled using Hilbert space state vectors. Simple Hilbert space addition, $\ket{\psi_1} + \ket{\psi_2}$, corresponds to the ``addition with an angle" operation when the angle is specified by $\phi = \arg(\langle \psi_1 | \psi_2 \rangle)$.}
    \label{fig:figure-14.2}
\end{figure}

Figure \ref{fig:figure-14.2} shows that the point corresponding to $\ket{\psi_1} + \ket{\psi_2}$ on the surface of the sphere is on the great circle of points symmetric around $\ket{\psi_1}$ and $\ket{\psi_2}$. This means that it is expressible in the form $P_3 = P_1 +_\phi P_2$, where $P_1$ and $P_2$ are the points corresponding to $\ket{\psi_1}$ and $\ket{\psi_2}$.

As demonstrated by the code provided, the angle, $\phi$, for which $+_\phi$ coincides with addition in the Hilbert space is $\phi = \arg(\langle \psi_1 | \psi_2 \rangle)$. The quantity, $\arg(\langle \psi_1 | \psi_2 \rangle)$, is the relative phase difference between $\ket{\psi_1}$ and $\ket{\psi_2}$. Equivalently, it is the argument (phase) of the quantum amplitude, $\langle \psi_1 | \psi_2 \rangle$.

The midpoint of the shortest arc between the points corresponding to $\ket{\psi_1}$ and $\ket{\psi_2}$ is given by:
\begin{align}
|\psi_{\text{mid}}\rangle = \ket{\psi_1} + \exp(-i \arg(\langle \psi_1 | \psi_2 \rangle)) \ket{\psi_2}.
\end{align}

This is because $\ket{\psi_1}$ and $\ket{\psi_2}$ already have a phase difference, given by $\arg(\langle \psi_1 | \psi_2 \rangle)$, and multiplying $\ket{\psi_2}$ by $\exp(-i \arg(\langle \psi_1 | \psi_2 \rangle))$ sets the phase difference to 0, aligning the $+$ operation in the Hilbert space with the $+_0$ operation on points of the sphere.

So we can reconstruct the action of the $+$ operation in the Hilbert space on points of the sphere by specifying, in addition to the geometric operation $+_\phi$, the specific value $\phi = \arg(\langle \psi_1 | \psi_2 \rangle)$:
\begin{itemize}
    \item Simple addition in the Hilbert space, $\ket{\psi_1} + \ket{\psi_2}$, corresponds geometrically to rotating the midpoint of the shortest arc connecting the two points corresponding to $\ket{\psi_1}$ and $\ket{\psi_2}$, around the great circle orthogonal to that arc, by an angle which is equal to the phase of the quantum amplitude, $\langle \psi_1 | \psi_2 \rangle$.
\end{itemize}

The quantum amplitude, $\langle \psi_1 | \psi_2 \rangle$, encodes two real quantities in a single complex number:
\begin{itemize}
    \item Its magnitude encodes the probability that a particle prepared in the state $\ket{\psi_1}$ will be found to be in the state $\ket{\psi_2}$ when measured.
    \item Its phase encodes the angular discrepancy between addition in the Hilbert space and the shortest-arc-midpoint operation.
\end{itemize}

\subsection{From a Sphere of Questions to Hilbert Space}
\label{subsec:sphere-to-hilbert}

What we have found so far is that the action of Hilbert space addition on points on the surface of the sphere can be represented as a geometrical operation, corresponding to a binary operation on questions. We have not reconstructed Hilbert space addition, which acts on the linear vector space of unnormalized state vectors, rather than acting on the points of a sphere.

That is, we can say what point $\ket{\psi_1} + \ket{\psi_2}$ corresponds to, using what we've learned so far, but we can't identify a vector, $|\psi_3\rangle = \ket{\psi_1} + \ket{\psi_2}$, which is the result of adding two state vectors. If $|\psi_3\rangle$ corresponds to a point on the sphere, then $\alpha |\psi_3\rangle$ is a different vector that corresponds to the same point, where $\alpha$ is any non-zero complex number.

To get the Hilbert space representation, we need to:
\begin{itemize}
    \item Change from using points, $P_1$, on the sphere as the basic objects that operations act on, to using (point, complex number) pairs, which we can write in the form, $\alpha |P_1\rangle$, and
    \item Identify all pairs for which the complex number, $\alpha$, is zero, as the same object.
\end{itemize}

Having done that, we find that the purely geometric operation, $+_{\arg(\langle \psi_1 | \psi_2 \rangle)}$, extends to an algebraic addition operation, $+$, on the objects defined above, that naturally generates a vector space representation. The $+$ operation specifies the magnitude of $\ket{\psi_1} + \ket{\psi_2}$ in the vector space, as well as its phase, in addition to specifying the resulting point on the sphere.

That is, the Hilbert space representation results from the attempt to represent the geometrical relationships between the points on the surface of the sphere in a linear algebraic system. We know from uniqueness theorems that this is the only representation, up to isomorphism.

We can say, then, that starting from the sphere of questions, we reach the Hilbert space representation by identifying the unique two-dimensional complex vector space that represents the geometry of the questions.

Since we used a geometrical argument, it's reasonable to ask: To what extent have we shown a relationship between the Hilbert space and a sphere of askable questions, rather than just a sphere?

The answer is that askable questions have probability gaps, and points on the sphere don't. The inner products of the vector space, $\langle \psi_1 | \psi_2 \rangle$, specify the same probabilities as the geometric relationships between the questions, through cosine projection.

So, for two-state systems, the Hilbert space formalism can be seen as a method of calculating probability gaps of askable questions using linear algebra, rather than as merely a representation of the geometry of the sphere.

\section{The Born Rule}
\label{sec:born-rule}

A pure state in a spin-1/2 system has a known spin along a particular direction, $(\theta, \phi)$, in physical space, and is represented in Hilbert space as:
\begin{align}
\cos\left(\frac{\theta}{2}\right) \ket{0} + \sin\left(\frac{\theta}{2}\right) \exp(i \phi) \ket{1}
\end{align}

\noindent where $\ket{0}$ and $\ket{1}$ correspond to ``up" and ``down" along the $z$-axis, respectively.

The coefficient, $\cos(\theta/2)$, is a quantum amplitude, and the Born rule says that the square of its magnitude is the probability that the result of measuring along the $z$-axis will be ``up": 
\begin{align}
P(Z) = \cos\left(\frac{\theta}{2}\right)^2.
\end{align}

From the point of view of questions, there's a probability gap for each direction. Along the $(\theta, \phi)$ direction, the probability gap is 1. Projected onto the $z$-axis, it's $\cos(\theta)$:
\begin{align}
\text{gap}(Z?) = \cos(\theta)
\end{align}

\noindent which implies that:
\begin{align}
\cos(\theta) = 2 P(Z) - 1.
\end{align}

In trigonometry, the half-angle formula for cosine can be expressed as:
\begin{align}
\cos(\theta) = 2 \cos\left(\frac{\theta}{2}\right)^2 - 1.
\end{align}

Comparing the two equations above yields $P(Z) = \cos(\theta/2)^2$. 

What this means is that we now understand why the probability is the square of the magnitude of a quantum amplitude. The questions are at an angle of $\theta$, so the probability must be the square of $\cos(\theta/2)$.

We don't need to say, ``The Born rule tells us how to compute the probability from the quantum amplitude." We already know how to compute the probability, because we are computing probability gaps. We have derived the Born rule: 
\begin{itemize}
    \item The probability must be the square of the magnitude of the quantum amplitude, because of its trigonometric relationship to the angle between the questions.
\end{itemize}

\subsection{Quantum Amplitudes in Spin-1/2 Systems}
\label{subsec:quantum-amplitudes-spin}

The coefficients, $\cos(\theta/2)$ and $\sin(\theta/2)\exp(i \phi)$, in the expression for the pure state whose spin is up along the $(\theta, \phi)$ direction, are complex numbers that relate questions. The states $\ket{0}$ and $\ket{1}$ correspond to askable questions, and the state:
\[
\cos\left(\frac{\theta}{2}\right) \ket{0} + \sin\left(\frac{\theta}{2}\right) \exp(i \phi) \ket{1}
\]

\noindent corresponds to another askable question, expressed as a linear sum of the questions corresponding to $\ket{0}$ and $\ket{1}$.

We can consider whether these complex numbers are valid complex-valued properties of the questions they relate. If a quantum amplitude, $\psi$, is a property of a question, $A?$, then $\psi(\neg A?)$ must be equal to $-\psi(A?)^*$.

In the two-state system, replacing $A?$ with $\neg A?$ corresponds to swapping the orthogonal states, $\ket{0}$ and $\ket{1}$, because if the particle's spin is not found to be ``up" along a specific direction, it will be found to be down. The askable question of whether the result is not ``up" is the same as the question of whether it is ``down".

If the number, $\cos(\theta/2)$, is a complex-valued property of the question, $Z?$, of whether the spin is ``up" along the $z$-axis, since it is a coefficient of the corresponding state, $\ket{0}$, then $-\cos(\theta/2)^* = -\cos(\theta/2)$ would need to be the corresponding complex-valued property of $\neg Z?$.

We can see that this is not the case: $\cos(\theta/2)$ determines the probability of $Z$: $P(Z) = \cos(\theta/2)^2$. If the corresponding value for $\neg Z$ is $-\cos(\theta/2)$, then $P(\neg Z)$ would also equal $\cos(\theta/2)^2$, which is not true in general.

That is, quantum amplitudes are not law-abiding complex-valued properties of the questions that they relate, because they don't change in the right way when those questions are negated.

We can see if we can identify another question, to which the quantum amplitude belongs as a valid complex-valued property. We start from the fact that negating the question has the effect of flipping the sign of the quantum amplitude, and taking the complex conjugate:
\begin{align}
\psi(\neg A?) = -\psi(A?)^*.
\end{align}

In quantum mechanics, complex conjugation corresponds to time-reversal. The question therefore distinguishes between directions in time. It asks, partially, which orientation in time (forwards or backwards), is the relevant or correct orientation.

In addition to complex conjugation, negating the question flips the sign of the quantum amplitude. In spin-1/2 systems, a sign flip can be accomplished by rotating the system through a full 360 degree turn.

This allows us to specify an exact operation on the physical system that corresponds to negating the question to which the quantum amplitude belongs as a property: a 360 degree rotation plus time reversal.

So, in spin-1/2 systems, the question to which the quantum amplitude belongs appears to be: Is this the correct orientation in time and space, or should a 360 degree rotation plus a time reversal be applied?

\subsection{Wavefunctions}
\label{subsec:wavefunctions}

Spin-1/2 particles have an operation that corresponds to flipping the sign of the wavefunction, namely rotation through 360 degrees. Not every quantum system has a physical operation that flips the sign. We can ask what question the quantum amplitude belongs to more generally, when no such physical operation exists.

For example, if $x$ is a point in space, then $\psi(x)$ could be the value of a particle's wavefunction at $x$. Let's suppose the particle is spinless, and has only positional degrees of freedom.

If $\psi(x)$ is a property of a question, $A?$: $\psi(x) = \psi(x, A?)$, then: 
\begin{align}
\psi(x, \neg A?) = -\psi(x, A?)^*.
\end{align}

We can see that $A?$ cannot be the question, ``Is the particle at this position?", because $|\psi(x, A?)|^2$ is the probability that the answer to that question is Yes, and, if $|\psi(x, \neg A?)|^2$ is the corresponding probability that the answer is No, then the probability would have to be $1/2$, since $|\psi(x, A?)|^2 = |\psi(x, \neg A?)|^2$.

We can try to identify $A?$ by considering what the sign flip might mean. Complex conjugation once again corresponds to time reversal, but the flip in the sign has no obvious physical representation. A global change in the sign of the wavefunction can have no physically detectable consequences, because it has no effect on probabilities.

It's precisely because a 360 degree rotation has no physically detectable consequences that it was able to embody the sign flip in the case of spin-1/2 systems. When we have a wavefunction instead, the sign flip isn't embodied in an action on the system.

However, even though a global sign flip of the wavefunction has no effect on the physical state of the system, local sign flips do: $\ket{\uparrow} + \ket{\downarrow}$ is not the same state as $\ket{\uparrow} - \ket{\downarrow}$. 

The difference between $+\ket{\downarrow}$ and $-\ket{\downarrow}$ is that, when added to other states, such as $\ket{\uparrow}$, they produce different resulting states. The signs in the terms $+\ket{\downarrow}$ and $-\ket{\downarrow}$ don't specify different states; they specify how the single state $\ket{\downarrow}$ is to be used when combined with other states.

That is, the sign of a quantum amplitude is a logically meaningful representation of how that state should be used inside a larger system when it is combined with other states. 

Flipping the sign of a wavefunction flips this logical relationship between the state of the system and the environment outside it, which we can think of as ``the laboratory".

So we can say that negating the question, $A?$, to which the quantum amplitude belongs, has the effect of:
\begin{itemize}
    \item Reversing time, and
    \item Performing an involution in the relationship between the system and the laboratory.
\end{itemize}

This strongly suggests that switching $A?$ and $\neg A?$ swaps the preparation and detection of quantum states.

Reversing time reverses the order in which preparation and detection occur. Switching between $A?$ and $\neg A?$ not only does this, but also reverses an aspect of the relationship between the system and the laboratory.

This suggests an explanation of why quantum amplitudes play such an important role in physics: They are valid complex-valued properties of the question that distinguishes between preparation and detection of states. They respect the symmetry between preparation and detection. 

They make it possible to encode information about preparable and detectable phenomena in a form in which switching point of view between preparation and detection is accomplished by the transformation, $z \to -z^*$, which swaps complex-valued properties of a question with the properties of its negation. This does seem like a desirable property for a numerical representation of information about the physical world.

\subsection{The Global $U(1)$ Symmetry In Quantum Mechanics}
\label{subsec:u1-symmetry}

In the Hilbert space formulation of quantum mechanics, there's a global $U(1)$ symmetry, because multiplying the state vector of the entire system by an arbitrary phase factor doesn't change the physical state represented.

Each whole question has its own copy of $U(1)$, in the form of the possible phase shifts of its properties, $g(A) = \exp(i \phi) f(A)$, from a valid property of an askable question. 

Above, we identified the askable question to which the quantum amplitude could belong as a valid complex-valued property. Multiplying the quantum amplitude by $i$ would make it into a valid complex-valued property of the corresponding pure question. Multiplying it by $\exp(i \phi)$ would make it into a valid property of the whole question, and would not change the physical state represented.

This seems to indicate that, at least in two-state systems, the global $U(1)$ symmetry is a reflection of the $U(1)$ symmetry of whole questions. That is, quantum amplitudes may be better understood as properties of the whole question that distinguishes between preparation and detection.

\section{Quantum Non-Locality}
\label{sec:quantum-non-locality}

Spin measurements on entangled spin-1/2 particles can exhibit non-local effects, in which the choice of measurement made by one observer can apparently affect the results obtained by a distant observer. These effects are not limited by the speed of light, making it difficult to understand how they can be compatible with special relativity \cite{Bell1964,chsh1969,Aspect1982,Mermin1981}.

We'll look at a simplified explanation of quantum non-locality, and then we'll use questions to represent the actions on the probability distributions that are involved with two-state systems. This will give us an insight into what happens during measurements on entangled states that exhibit non-local correlations.

The correlation between $A$ and $B$, $c(A,B) = P(AB) + P(\neg A \neg B) - P(\neg A B) - P(A \neg B)$, is a valid real-valued property of the askable questions, $A?$ and $B?$, because $c(A,B) = -c(\neg A,B) = -c(A,\neg B) = c(\neg A,\neg B)$. In fact, $c(A,B) = \text{gap}(A? * B?)$, so when we deal with correlations in physics, we are in fact dealing with askable questions, their numerical properties, and their group operations, behind the scenes.

\subsection{A Simplified Version of Bell's Inequalities}
\label{subsec:bell-inequalities}

A simple Bell inequality can be derived from very few assumptions. We can consider the case where there are two experimenters, who can each perform a spin measurement, along an axis of their choosing, on one of two entangled particles, which are jointly in the singlet state\footnote{The normalized state vector would be $\frac{1}{\sqrt{2}}(\ket{\uparrow}\ket{\downarrow}-\ket{\downarrow}\ket{\uparrow})$. For clarity, we use unnormalized vectors to identify states when the normalization isn't relevant.}, $\ket{\uparrow}\ket{\downarrow}-\ket{\downarrow}\ket{\uparrow}$. We only need to assume that:

\begin{enumerate}
\item Experimenter 1's result does not depend on Experimenter 2's choice of axis,
\item Experimenter 2's result does not depend on Experimenter 1's choice of axis, and
\item If the two experimenters measure along the same axis, then they will get opposite results.
\end{enumerate}

From these assumptions, we can draw up Table \ref{tab:experiment}, which counts the number of particle pairs for which the experimenters would have gotten specific results. $N_1$ is the number of particle pairs for which Experimenter 1 would have gotten ``up" as a result, along the $x$, $y$ or $w$ axes, and for which Experimenter 2 would have gotten ``down".

\begin{table}[h]
    \centering
    \begin{tabular}{l *{3}{>{\centering\arraybackslash}p{0.5cm}} p{0.1cm} *{3}{>{\centering\arraybackslash}p{0.5cm}}}
        \toprule
        & \multicolumn{3}{c}{Experimenter 1} & & \multicolumn{3}{c}{Experimenter 2} \\
        \cmidrule(lr){2-4} \cmidrule(lr){6-8}
        & x & y & w & & x & y & w \\
        \midrule
        $N_1$ & + & + & + & & - & - & - \\
        $N_2$ & + & + & - & & - & - & + \\
        $N_3$ & + & - & + & & - & + & - \\
        $N_4$ & + & - & - & & - & + & + \\
        $N_5$ & - & + & + & & + & - & - \\
        $N_6$ & - & + & - & & + & - & + \\
        $N_7$ & - & - & + & & + & + & - \\
        $N_8$ & - & - & - & & + & + & + \\
        \bottomrule
    \end{tabular}
    \caption{\footnotesize The numbers of particle pairs that would have given specific measurement results, if those measurements had been made. + indicates a measurement result of ``up", and - indicates ``down." If the two experimenters measure along the same axis, they get opposite results. This table presumes that each experimenter's result depends on their own choice of axis, but not on the other experimenter's choice.}
    \label{tab:experiment}
\end{table}

Having constructed Table \ref{tab:experiment}, we can write the inequality:
\begin{align}
\label{eq:rawbell}
(N_1 + N_3) \ + \ (N_4 + N_8) \ \geq \ N_3 + N_4.
\end{align}

If we use the notation $N(x_1+,y_2+)$ to denote the number of instances for which Experimenter 1 would have gotten ``up" along the $x$-axis, and Experimenter 2 would have gotten ``up'' along the $y$-axis, then:
\begin{align}
N(x_1+,y_2+) = N_3+N_4.
\end{align}

Similarly,
\begin{align}
N(x_1+,w_2-) = N_1+N_3, \\
N(w_1-,y_2+) = N_4+N_8. 
\end{align}

We can use these to rewrite Equation \ref{eq:rawbell} as:
\begin{align}
\label{eq:bellinequality}
N(x_1+,w_2-) + N(w_1-,y_2+) \geq N(x_1+,y_2+).
\end{align}

Intuitively, this equation expresses the fact that, for every particle pair for which Experimenter 1 would get ``up" along the $x$-axis, and for which Experimenter 2 would get ``up" along the $y$-axis, either Experimenter 1 or Experimenter 2 would get ``down" along the $w$-axis.

That is, for every particle pair included in $N(x_1+,y_2+)$, either $w_2-$ is true, and the particle pair is included in $N(x_1+,w_2-)$, or $w_1-$ is true, and the pair is included in $N(w_1-,y_2+)$.

Quantum mechanics predicts a violation of Equation \ref{eq:bellinequality} when the $w$-axis is at an angle of 225 degrees with respect to the $x$-axis. The predicted probabilities of the corresponding measurement results are:

\begin{align}
\label{eq:bellinequalityviolation}
P(x_1+,y_2+) = 0.5 \times 0.5 = 0.25, \\
P(x_1+,w_2-) = 0.5 \cos(\frac{5}{8}\pi)^2 = 0.073, \\
P(w_1-,y_2+) = 0.5 \cos(\frac{3}{8}\pi)^2 = 0.073.
\end{align}

The conflict that leads to the violation can be thought of in the following way. We have chosen a $w$-axis that points away from both the $x$- and $y$-axes. The experimenters are likely to get the same results if one measures along the $w$-axis and the other measures along $x$ or $y$. This makes the combinations $(x_1+,w_2-)$ and $(w_1-,y_2+)$ very improbable. They don't occur frequently enough for all of the occasions for which $(x_1+,y_2+)$ is true to be a subset of those for which $(x_1+,w_2-)$ or $(w_1-,y_2+)$ is true.

\subsection{Consequences of the Violations}

There are very few assumptions needed to derive the inequality above. Quantum mechanics predicts that it is violated, and the violation has been experimentally observed. All of the loopholes in the experimental design have been closed \cite{Bennet2012,Hensen2015}, and the number of assumptions needed to derive the inequality has been reduced to three. Therefore, one of the three assumptions is wrong.

Let's consider their negations:

1. Experimenter 1 would have gotten different results if Experimenter 2 had made different choices.

2. Experimenter 2 would have gotten different results if Experimenter 1 had made different choices.

3. There were particle pairs for which, if the two experimenters had measured along the same axis, they would not have gotten opposite results.

One of these is true. 

There is a minority view in physics that the results of measurements that weren't performed should be regarded as undefined. In this view, none of the reasoning about results that an experimenter would have gotten, but didn't get, is valid. The number, $N(x_1+,y_2+)$, can't be broken up into $N_3+N_4$, because $N_3$ and $N_4$ are undefined.

This makes it possible to avoid the conclusion that one experimenter's choices were affecting the distant experimenter's results, but regarding unmeasured results as undefined has other consequences: If Experimenter 2 is informed that the result of the first measurement was ``up'' along the $x$-axis, then they can say with certainty, ``If I measure along $x$, the result will be `down'". However, if they measure along $y$ instead, they will not be able to truthfully say, ``If I had measured along $x$, the result would have been `down'".

The action of measuring along the $y$-axis would make the hypothetical result of measuring along the $x$-axis undefined, and unsuitable for use in logical reasoning, despite it being fully specified.

The mainstream point of view in physics is that it has been experimentally confirmed that measurements along the same axis always give opposite results, and therefore either assumption 1 is false, or assumption 2 is false, or both.

\subsection{Questions and Entangled States}
\label{subsec:questions-entangled-states}

The entangled state:
\[
\ket{\uparrow}\ket{\uparrow} + \ket{\downarrow}\ket{\downarrow}
\]

\noindent has the property that, if the first particle's spin is measured along the $x$-axis, and the result is ``up", then the state of the other particle's spin becomes ``up" along the same axis. If the result of the first measurement is ``down", the second particle's spin also becomes ``down".

Let's use $X_1$ to represent the proposition that, if the first particle's spin is measured along the $x$-axis, the result will be ``up", and let's use $X_2$ to represent the corresponding proposition for the second particle.

Then, in the initial state, the proposition $(X_1 X_2 \text{ or } \neg X_1 \neg X_2)$ is given. The askable question, $(X_1? * X_2?)$, has the answer Yes. Let's denote this question by $S_x?$ to indicate the fact that it asks whether $X_1?$ and $X_2?$ have the same answer.

Then the state, $\psi = \ket{\uparrow}\ket{\uparrow} + \ket{\downarrow}\ket{\downarrow}$, corresponds to a probability distribution, $P_\psi$, in which $S_x$ is given:
\begin{align}
P_\psi(W) = P_\psi(W|S_x).
\end{align}

The same is true for $S_y$, $S_z$, and for $S_{\theta, \phi}$, where $S_{\theta, \phi}$ is the proposition that measurements of the two spins along the $(\theta, \phi)$ direction will yield the same results. So:
\begin{align}
P_\psi(W) = P_\psi\bigg(W \ \bigg| \prod_{\theta, \phi} S_{\theta, \phi}\bigg).
\end{align}

The equation above expresses everything that the entangled state, $\ket{\uparrow}\ket{\uparrow} + \ket{\downarrow}\ket{\downarrow}$, tells us about the probabilities of the results of spin-1/2 measurements on the two particles\footnote{The $+$ sign in the expression for the entangled state, $\ket{\uparrow}\ket{\uparrow} + \ket{\downarrow}\ket{\downarrow}$, also encodes information about the total spin of the system, but this requires a different type of experiment, beyond spin-1/2 measurements, to detect.}.

If we measure the spin of the first particle along the $x$-axis, and get ``up" as the result, the state changes to $\ket{\uparrow_x}\ket{\uparrow_x}$. The effect of the measurement on the probability distribution can be expressed as:
\begin{align}
P\bigg(W \bigg| \prod_{\theta, \phi} S_{\theta, \phi}\bigg)
&\to P\bigg(W \bigg| \prod_{\theta, \phi} S_{\theta, \phi} \cdot \prod_{\theta, \phi} s_{\theta, \phi} \cdot S_x \cdot X_1\bigg) \notag \\
&= P(W | S_x X_1) = P(W | X_1 X_2).
\end{align}

That is, after the measurement, it is certain that, if the spin of either particle is measured along the $x$-axis, the result will be ``up" -- both $X_1$ and $X_2$ are given. However, it is no longer known whether measuring the two spins along any other axis will yield the same results. Previously, $S_y$ was given, but the measurement raised the pure question, $s_y = y_1 * y_2$, along with the pure questions corresponding to every other direction, consequently removing the entanglement between the spins.

The action on the probability distribution is $\prod_{\theta, \phi} s_{\theta, \phi} \ S_x \ X_1$, which can be compared to the action of a measurement on a single particle, $\prod_{\theta, \phi} r_{\theta, \phi} \ R_x$. In both cases, all of the questions of a certain type ($s_{\theta, \phi}$ or $r_{\theta, \phi}$) are raised, and one question is answered, namely the question relevant to the measurement that was made.

In the entangled case, the proposition, $S_x$, was already known before the measurement. The action of the measurement, in the form that we've written it, raises $s_x$ along with the other questions and then gives $S_x$ afterwards. This doesn't correspond to the receipt of new information; it just acts on the probability distribution to change it from $P(W| \prod_{\theta, \phi} S_{\theta, \phi})$ to $P(W|S_x)$, after which $X_1$ is given.

It's necessary to write it in this way because the propositions, $S_{\theta, \phi}$, don't commute with the questions, $s_{\theta, \phi}$, and so raising every question except $s_x$ would give $S_x \ \prod_{\theta, \phi} s_{(\theta, \phi)\neq x}$, instead of $S_x$. The fact that non-independent questions and propositions don't commute forces us to express the actions of measurements on entangled states in a specific format that matches how actions of measurements on single particles are expressed.

\subsection{Non-Local Questions}
\label{subsec:non-local-questions}

The question, $X_1?$, of whether the result of measuring the first particle's spin along the $x$-axis will be ``up", is a local question, because it asks about the local result of a local measurement. $X_2?$ is also a local question, but it refers to a different region of space, namely the neighborhood of the second particle.

On the other hand, the question, $X_1? * X_2?$, which asks whether $X_1?$ and $X_2?$ have the same answer, is a non-local askable question if the particles are in different locations, because then it refers to two distinct regions of space.

An entangled state can be prepared if the two particles are in the same place. The entangled state, $\ket{\uparrow}\ket{\downarrow} - \ket{\downarrow}\ket{\uparrow}$, can be prepared by selecting the particle pairs that have no overall angular momentum. This has the effect of giving all the propositions, $\neg S_{\theta, \phi}$, at the same time for every value of $\theta$ and $\phi$. 

$\ket{\uparrow}\ket{\downarrow} - \ket{\downarrow}\ket{\uparrow}$ is the singlet state for the particle pair. We used a different state, $\ket{\uparrow}\ket{\uparrow} + \ket{\downarrow}\ket{\downarrow}$, in the previous subsection, so that we could use the propositions, $S_{\theta, \phi}$, instead of $\neg S_{\theta, \phi}$ to describe the state. $\ket{\uparrow}\ket{\uparrow} + \ket{\downarrow}\ket{\downarrow}$ is one of three triplet states, which, together with the singlet state, $\ket{\uparrow}\ket{\downarrow} - \ket{\downarrow}\ket{\uparrow}$, form the four Bell states of the two-spin system.

The proposition, $S_x = X_1 X_2 \text{ or } \neg X_1 \neg X_2$, is a local proposition for as long as the two particles are in the same place. However, if they move apart, and their spins are measured in distant locations, then the propositions, $S_{\theta, \phi}$, become non-local propositions.

That is, propositions can only be given when they're local, because we ourselves are localized in space and can only receive information at our location, but they can evolve over time to become non-local.

When the measurement of the first particle's spin takes place, and the result is ``up", a local proposition, $X_1$, is given, from which $X_2$ is inferred by combining $X_1$ with the local knowledge, $S_x$, but the non-local questions, $s_{\theta, \phi}$, are raised.

The raising of these questions is a non-local action on the probability distribution. It doesn't just take information away; it subtracts it in a way that can't be expressed as a combination of local actions.

This means that non-local questions can be raised during experiments, while propositions can only be given locally. The non-local effects observed in measurements on entangled states arise from the non-local questions that are raised when a local measurement is made. 

The subtraction of information that accompanies the raising of these non-local questions is not simply a loss of information that we know about the system. It's a change in the state of the system itself, which remains in a pure state after the measurement, indicating that we still have full information.

It explains how $X$ and $Y$ can be uncorrelated when measured, while both are tightly correlated with $W$ when measured. A measurement along the $x$-axis actively decorrelates $Y$ from $X$, setting $\text{gap}(Y?)$ to zero. A measurement along the $w$-axis induces a correlation between $X$ and $Y$, which can never be observed, because only one of them can be measured.

\subsection{The Speed of Light}
\label{subsec:speed-of-light}

These non-local effects are not limited by the speed of light. They are actions on probability distributions that can't be expressed as a combination of local actions. 

When the inequalities are violated, it is indeed the case that the results obtained by one experimenter would have been different if the distant experimenter had made different choices about what axis to measure along, presuming we can meaningfully talk about the results that they would have gotten. This is difficult to reconcile conceptually with special relativity. If it's consistently possible to send signals faster than light, then it's possible to send signals backwards in time.

With entangled states, it's possible for one experimenter to affect the results that the other experimenter gets, across such a large distance that light can't travel between the two locations in the duration of the experiment, but it isn't possible to send a signal. The distant effect can be caused, but not controlled. 

From the point of view of questions, these non-local effects arise as a consequence of raising non-local questions, such as $s_x = x_1 * x_2$. This is a rank-1 pure question in the group of questions generated by the propositions corresponding to the askable questions oriented in space. 

This implies that raising $s_x$ can only take information away from the probability distribution that describes what we know about that system.

Raising a rank-2 question, such as $q(X_1 Y_2)$, would add information to the probability distribution: The probability of $X_1 Y_2$ would become $1/2$, making both $X_1$ and $Y_2$ more probable. If the first experimenter could raise the question, $q(X_1 Y_2)$, they could make $Y_2$ more probable, sending a signal faster than light to the distant experimenter\footnote{Actions on probability distributions, such as giving a proposition or raising a question, change what we know when they change the probability distribution. However, some changes in our knowledge can only occur when the system changes too. For example, if $A$ is the proposition that a switch is set to ``on", and $P(A)=0$, then we know that the switch is off, and setting $P(A)$ to 1 would require flipping the switch. The action of raising a rank-2 non-local question would require a similar change to the system, one which would allow faster-than-light communication.}.

The pure questions that can be raised through measurements on entangled spin-1/2 particles, though, are all rank-1 questions, ensuring that they only take information away. An experimenter with the power to raise a rank-1 non-local question will know less about the distant result if they use that power. They might change the result, but they can't increase their certainty that the distant experimenter will get any specific result.

\section{Discussion}
\label{sec:discussion}

In our investigation of the mathematics of questions, the tilde relation has acted as a Rosetta Stone. It drew our attention to pure questions and their algebras, and then to subjects and to askable questions. When we re-examined $\sim$ in the light of this new mathematics, it yielded another deep insight: That questions have complex-valued properties. It showed us how to construct these complex-valued properties, by combining properties of the pure and askable questions, and it showed us the constraint, $z(\neg A?) = -z(A?)^*$, that these properties must satisfy.

Although we have understood a lot about the tilde relation, there is more left to discover. It has revealed that $3 - \text{gap}(a)^2$ and $9/5 - \text{gap}(a)^2$ are important properties of the pure question, $a$. They combine with the corresponding properties of another pure question, $b$, to produce the numbers $S$ and $T$, which relate the two pure questions, and which play critical roles in the formula for the tilde relation.

If, in the future, we understand what the property $3 - \text{gap}(a)^2$ is, as a meaningful quantity, and what $9/5 - \text{gap}(a)^2$ is, then we will be able to trace our way through the formula for $\sim$, identifying meaningful quantities corresponding to $Y$ and $U$. If we can do that, then we will understand what complex-valued property the tilde relation constructs in the form $Y + U$, and we might be able to understand its cube root as a meaningful complex-valued property of the two questions.

If we can accomplish all that, then we will know exactly what the formula for $\sim$ states about relationships between meaningful quantities. The result is likely to be a profound insight into the relationship between information and logic. For now, it is beyond our reach.

We have, however, been able to understand what $\sim$ is from the point of view of the algebra of questions. We know that it allows questions to be independent as subjects, when calculating geometric mean probabilities, while not being independent as questions, when calculating probability gaps. We know that it binds questions together into a natural subject, so that raising a settled question raises the entire subject.

The irremovable presence of complex numbers in the formula for $\sim$ was reminiscent of quantum mechanics, where complex numbers also relate probabilities. There was some hope that an insight into quantum mechanics could be achieved by studying questions.

What we found was that, by considering questions that are related to each other geometrically, we can recover the essential formalism of the two-state quantum system, minus time-evolution. When probability theory is augmented with questions, the collapse of the wavefunction is representable as an element of the algebra of actions on probability distributions. It subtracts information from the probability distribution, by raising questions.

We also derived the Born rule, in the two-state case, namely that probabilities take the form of the squared magnitude of a quantum amplitude. We began by considering askable questions whose probability gaps were related to each other geometrically. We didn't need an additional rule to introduce probabilities.

For two-state systems, it's also possible to obtain the same result using purely symmetry-based arguments. We have just found another way to derive it. Whether we can extend the same concepts and recover the Born rule for higher-dimensional systems, as is done in Gleason's theorem \cite{Gleason1957}, remains an open question.

The fact that the two-state system can be derived this way has a philosophical consequence, which we can express as a theorem:

\vspace{2mm}

\textbf{Theorem:} In any lawful three-dimensional universe in which physicists find the simplest way to orient measurements in space, they will find the two-state quantum system.

\vspace{1mm}

\textbf{Proof:} The two-state quantum system is the simplest way to orient measurements in space.

\vspace{3mm}

This ``theorem" relies on the postulate that the sphere of askable questions corresponding to the Bloch sphere has a corresponding sphere of experimentally performable measurements in real space. In the case of the two-state system, in our universe, there is such a sphere of possible measurements, given by the directions along which spin can be measured.

Physicists in another three-dimensional universe might construct a sphere of possible experimentally performable measurements, corresponding to a sphere of askable questions. That sphere may or may not relate the probability gaps of those askable questions through the cosines of the angles between them, which is the simplest geometric relationship that they can have. If the gaps are related by the cosines of the angles between the questions, then the physicists will indeed find the two-state quantum system.

This gives us an insight into the question of whether quantum mechanics is a necessary feature of the universe. The answer is that it is logically possible for a universe to exist in which every piece of matter can always be subdivided into smaller pieces. Physicists might never be able to find the simplest possible phenomena, because the attempt to find the indivisible components of nature might never succeed.

However, any universe in which the physicists do succeed, and find the simplest way to orient measurements in space, will be a quantum universe. The physicists will find what we found when we first discovered how to algebraically represent quantum phenomena, the two-state system.

Quantum amplitudes, at least for the two-state system, are complex numbers that are relevant to askable questions, which correspond to possible measurements. However, as we have seen, quantum amplitudes are not valid complex-valued properties of the questions corresponding to the measurements. The way that they change under negation suggests that they are valid complex-valued properties of the question that distinguishes between preparation and detection.

From the point of view of questions, it makes intuitive sense that quantum amplitudes would be important in physics, because they can encode information about preparable and detectable phenomena in a way in which switching point of view between preparation and detection is accomplished by a transformation on complex numbers: $z \to -z^*$.

The mathematics of entangled states in two-state systems appears as soon as the $*$ operation is used to combine askable questions about spins of different particles. This suggests that, like the two-state system, quantum entanglement is a feature of every universe in which the simplest way to orient measurements in space can be practically realized.

Combining rank-1 questions using the $*$ operation yields another rank-1 question. Quantum entanglement, which, in two-state systems, can be represented using the $*$ operation on questions, produces non-local questions. These can be raised when a quantum measurement occurs, but they are rank-1, and therefore only take information away from the probability distribution. There's an effect that happens non-locally -- information is subtracted from the probability distribution in a non-local way -- but it can't be used to send a signal.

The introduction of questions doesn't change the uneasy coexistence of quantum non-locality with special relativity. Effects do take place across vast distances that light can't traverse during the duration of the experiment. In quantum mechanics, it was already known that no signal can be sent, because the effects are uncontrollable. Expressing entanglement in terms of questions just adds one insight: No signal can be sent, because the non-local action on the probability distribution is a pure subtraction of information.

There are many directions for further research. We have completely ignored the role of time-evolution, establishing only the mathematical equivalence of the geometry of questions and the two-state system. Expressing more complex quantum systems in terms of questions may yield new insights.

We have only studied the two-state system here, which is particularly well-suited to a representation using Yes/No questions. Extending the application of questions to, for example, the three-state system, in which a spin measurement can have three results, would force us to consider whether we should try to define questions with more than two answers, or stick with binary questions and try to describe the system using those.

The physical arrangement of the relevant experiment gives a possible insight: A single beam of spin-1 particles is split into three beams of particles, with spin -1, 0, and +1 along the measurement axis. To detect particles in the spin=+1 beam, a detector needs to be placed in the path of the beam, blocking it, and allowing the two other beams to continue, unless those beams are also blocked by detectors.

What this means is that, when a spin-1 particle's spin is measured along an axis, it's not accurate to describe what happens physically as the asking of an askable question with three possible answers. There's one binary askable question for each beam, and placing a detector in the path of the beam asks that question. A single detecting screen can be used to block all three beams, but that screen has different parts, which act as different detectors, performing different, but not independent, binary measurements.

This suggests that extending questions to higher-spin, and possibly even more complex quantum systems, might not require abandoning the use of binary questions. Whether it does or not will determine how powerful the application of questions to quantum mechanics truly is. The generally accepted definition of a logical proposition is that it's a statement that can be true or false. Introducing a third option would have the consequence that we are no longer dealing with logical propositions.

If Yes/No questions need to be abandoned to represent more complex quantum systems, then the fact that the geometry of questions coincides with the two-state system will appear to be nothing more than an interesting alignment of one quantum system with a system of reasoning. We would not be able to say that we have found a connection between reasoning and quantum mechanics in general.

Today, we haven't found that connection beyond the two-state system, but the physics of possible measurements indicates that, although a spin measurement in the three-state system has three possible results, in practice there are three binary questions involved. This suggests that it may be possible to extend the representation of quantum states in terms of questions to more complex systems without losing the connection to reasoning.

The introduction of questions expands the power of probability theory, by permitting the representation of the subtraction of information. It is likely that applications of the mathematics of questions can be found in many areas in which probability theory is used today.

Complex-valued properties of questions are a new development. They involve a new form of reasoning. They are relevant in quantum physics. It's conceivable that they could have applications outside physics.

We don't have a single example of anything that satisfies the complex formula for the tilde relation, in nature or in logic. We don't know whether $\sim$ plays any role at all in quantum mechanics, but we know that, like quantum mechanics, it makes use of complex-valued properties of questions.

Nonetheless, $\sim$ is usable within a form of symbolic reasoning, based on geometric mean probabilities, and using the algebra of subjects. We are not very familiar with this form of reasoning, and have just found its basic elements here. It's hard to predict the consequences of a new form of reasoning.

If, in the future, we find ourselves using the fact that geometric mean probabilities multiply for combinations of independent subjects in our calculations, then we will need to be conscious of the fact that independence is not the only thing that our calculations are describing. We will need to worry about $\sim$, even if we don't find an example or an application of it\footnote{There would be an immediate practical application of the tilde relation if it was common practice for programmers to check for independence using geometric mean probabilities. A hacker could use the tilde relation to fake independence. In practice, nobody checks for independence this way, and nobody should.}.

Overall, our investigation into questions has revealed a detailed and complex mathematical structure, which is still mostly unexplored. We have found the basic group structures and the constraints on numerical properties, and also the actions on probability distributions. We've found a connection to the two-state quantum system, but questions have their own structure, which is not identical to quantum mechanics. We haven't fully understood the two-question system yet. There is much left to discover.

\subsection{Previous Work}
\label{subsec:previous-work}

A deep insight into the relationship between quantum mechanics and reasoning was achieved by von Neumann and Birkhoff in their classic paper on quantum logic \cite{BirkhoffVonNeumann}. Subsequent research has revealed a rich mathematical structure, encompassing quantum information theory and quantum computation. The current work presents a novel view of how logic and quantum mechanics are related. Identifying how the group structures and numerical properties of questions relate to existing frameworks in quantum logic remains a task for the future.

Bayesian probability theory is the immediate precursor to the structure developed here. By clarifying that probabilities are natural properties of logical propositions rather than sets \cite{Jaynes2003}, and by introducing the action, $P(W) \to P(W|X)$, of propositions on probability distributions, it set the stage for the introduction of a second action, $P(W) \to P(W|x)$, of pure questions. It's not much of an exaggeration to say that everything we've found above is the result of introducing subtraction of information into Bayesian probability.

The attempt to understand quantum mechanics as a form of Bayesian inference is known as quantum Bayesianism \cite{caves2002}. Quantum analogues of Bayes' rule have been well-studied \cite{fuchs2013}, using structures such as positive-operator-valued measures to express the collapse of the wavefunction \cite{schack2001}. We've found a way to augment Bayesian inference with actions of questions, and reproduce the simplest quantum system. This could indicate a possible new path forward for quantum Bayesianism. Introducing questions is a smaller change to Bayesian probability theory than introducing POVMs.

Questions could potentially provide a bridge between the epistemic and realist views of quantum mechanics. A free agent who raises a pure question voluntarily forgets certain information. A rational agent would only do that when the information is no longer accurate, because the system has changed. So a rational agent who uses both propositions and questions to update their probability distribution can handle both incoming information and changes to the system within a single algebra.

The concept of a question has already been used to derive features of quantum mechanics within a toy model \cite{spekkens2007}. Here, we've derived the essential structure of the two-state system from the geometry of askable questions. In combination, these results seem to suggest that questions are not merely a useful tool for thinking about quantum systems, but may be part of the structure of quantum mechanics itself.

Boole first showed how logical propositions could be governed by the same algebraic rules as numerical quantities in his foundational work, \textit{The Laws of Thought} \cite{Boole1854}. We've found that questions, which are components of propositions, have algebras of their own. They also have natural numerical properties whose arithmetic reflects the algebraic properties of the questions. What we've found here validates and extends Boole's foundational vision, but the complexity of the tilde relation seems to indicate that there is a long way to go before we fully understand how numbers relate to reasoning.

Erotetic logic is the branch of logic devoted to the analysis of questions. It was already well-understood by erotetic logicians that every proposition provides an answer to a question \cite{Collingwood1939,Wisniewski1995}. Over the last century, the field has developed and identified many structures related to questions \cite{BelnapSteel1976,Harrah2002,Aqvist1965}. The results presented here connect questions to probability theory, establishing the relevance of group theory, probability gaps, geometric mean probabilities, and even the $\sim$ relation, to erotetic logic.

\section*{Acknowledgements}

The tilde relation and the concept of questions as symmetric components of propositions were first discovered in the neuroscience lab of the late Professor Charles F. Stevens, with his support. The research was funded in part by Howard Hughes Medical Institute.

\pagebreak


\section*{Appendix A: Mathematical Results}

\subsection*{A.1. Definitions and Representations}

\begin{table}[h!]
\scriptsize
\centering
\begin{tabular}{c|c|c}
\toprule
\textbf{Object} & \textbf{Set Representation} & \textbf{Notation} \\
\midrule
\textbf{Pure Question} $q(A)$
  & $\{A, \neg A\}$ & $q(A) = \{A, \neg A\}$ \\
\textbf{Askable Question} $A?$
  & $(A, \neg A)$ & $A? = (A, \neg A)$ \\
\textbf{Subject} (Rank-2 e.g.)
  & $\{\,AB,\;A\neg B,\;\neg A\,B,\;\neg A\,\neg B\}$ & $ab = q(A)\circ q(B)$ \\
\textbf{Answers}
  & $\mathrm{Yes}((x,y))=x,\;\mathrm{No}((x,y))=y$ &  $\mathrm{Yes}(A?)=A$\\
\bottomrule
\end{tabular}
\end{table}

\begin{table}[h!]
\tiny
\centering
\begin{tabular}{c|c|c|c|c}
\toprule
\textbf{Type of Question}
 & \textbf{Set Representation}
 & \textbf{Condition on Properties}
 & \textbf{Pairs of Values}
 & \textbf{Shape Formed by Value Pairs} \\
\midrule
\textbf{Pure}
 & $\{A,\neg A\}$
 & $f(\neg A)=f(A)^*$
 & $(z,\,z^*)$
 & $\mathbb{C}$ \\
\textbf{Askable}
 & $(A,\neg A)$
 & $f(\neg A)=-\,f(A)^*$
 & $(z,\,-z^*)$
 & $\mathbb{C}$ \\
\textbf{Whole}
 & $(A,\neg A)$
 & $g(\neg A)=e^{i\theta}\,g(A)^*$
 & $(z,\,-e^{i\phi}z^*)$
 & $\mathbb{R}_+ \times S^1 \times S^1$ \\
\bottomrule
\end{tabular}
\end{table}

\FloatBarrier
\subsection*{A.2. Algebraic Structures}

\begin{table}[h]
\scriptsize
\centering
\begin{tabular}{c|c|c|c|c}
\toprule
\textbf{Structure}
 & \textbf{Group}
 & \textbf{Operation}
 & \textbf{Identity}
 & \textbf{Size} \\
\midrule
\textbf{Pure Questions}
 & $Q(N)$
 & $q(A)*q(B)=q(A\,\mathrm{xor}\,B)$
 & $I=\{\text{True},\text{False}\}$
 & $2^{2^N -1}$ \\

\textbf{Askable Questions}
 & $K(N)\cong Q(N)\times\mathbb{Z}_2$
 & $A?*B?=\bigl(AB\lor\neg A\,\neg B\bigr)?$
 & $\text{True}?$
 & $2^{2^N}$ \\

\textbf{Subjects} (rank $m$)
 & $S_m(N)=Q_m(N)/Q_{m-1}(N)$
 & combine via $*$
 & $I$
 & $2^{\binom{N}{m}}$ \\

\textbf{Group of Subjects}
 & $S(N)$
 & $a \circ b$
 & $I$
 & $2^N$ \\
\bottomrule
\end{tabular}
\end{table}

\FloatBarrier
\subsection*{A.3. Core Equations}

\begin{itemize}
    \scriptsize
\item \textbf{Probability $\leftrightarrow$ Information:}
\[
i(A) \;=\; -\log \,P(A).
\]

\item \textbf{Information Between Two Propositions:}
\[
i(A,B)\;=\;\log \,\frac{P(A B)}{P(A)\,P(B)}.
\]

\item \textbf{Information Balance Condition:}
\[
i(A,B)+i(A,\neg B)+i(\neg A,B)+i(\neg A,\neg B) \;=\;0.
\]

\item \textbf{Quartic Form:}
\[
x\,(a - x)\,(b - x)\,(1 - a - b + x)\;=\;a^2\,b^2\,(1 - a)^2\,(1 - b)^2,
\]
where $x = P(A B)$, $a=P(A)$, $b=P(B)$.

\item \textbf{Gap (askable):}
\[
\mathrm{gap}(A?)=2\,P(A)-1.
\]
\item \textbf{Gap (pure):}
\[
\mathrm{gap}(q(A))=\bigl|P(A)-P(\neg A)\bigr|.
\]
\item \textbf{Geometric Mean Probability:}
\[
\mathrm{gmp}(a) \;=\;\sqrt{P(A)P(\neg A)}.
\]
\[
\mathrm{gmp}(ab) \;=\;\sqrt[\,4]{\,P(A B)\,P(A \neg B)\,P(\neg A\,B)\,P(\neg A\,\neg B)}.
\]

\end{itemize}

\FloatBarrier
\subsubsection*{A.4. Independence vs. Tilde Relation}

\begin{table}[h!]
\scriptsize
\centering
\begin{tabular}{c|c|c}
\toprule
 & \textbf{Independence} & \textbf{Tilde Relation} \\
\midrule
$P(A)=1$                & $P(B\mid A)=P(B)$ & $P(B\mid A)=P(B)$ \\
$P(A)=\tfrac12$         & $P(B\mid A)=P(B)$ & $P(B\mid A)=P(B)$ \\
$P(A)=0,\;0<P(B)<1$     & $P(B\mid A)=P(B)$ & $P(B\mid A)=P(\neg B)$ \\
$P(A)=0,\;P(B)\in\{0,1\}$ & $P(B\mid A)=P(B)$ & $P(B\mid A)$ \text{ unconstrained} \\
\bottomrule
\end{tabular}
\end{table}

\FloatBarrier
\subsection*{A.5. Numerical Properties}

\begin{table}[h!]
\scriptsize
\centering
\begin{tabular}{c|c}
\toprule
\textbf{Property} & \textbf{Formula / Definition} \\
\midrule
\textbf{Gap (askable)} & $\mathrm{gap}(A?)=2\,P(A)-1$. \\
\textbf{Gap (pure)}    & $\mathrm{gap}\bigl(q(A)\bigr)=\bigl|\,P(A)-P(\neg A)\bigr|$. \\
\textbf{Geometric Mean Probability} & $\mathrm{gmp}(ab)=\sqrt[\,4]{\,P(AB)\,P(A\neg B)\,P(\neg A\,B)\,P(\neg A\,\neg B)\,}$. \\
\textbf{Information} & $i(A)=-\log\,P(A)$. \\
\textbf{Overlap} & $i(A,B)=\log \bigl(\tfrac{P(AB)}{P(A)P(B)}\bigr)$. \\
\textbf{Evidence} & $e(A?)=\log \tfrac{P(\neg A)}{P(A)}$. \\
\textbf{Doubt (Real)} & $d(A?)=-\log\!\bigl|\mathrm{gap}(A?)\bigr|$. \\
\textbf{Doubt (Complex)} & $d(A?)=-\log\mathrm{gap}(A?)$ (imag.\ part $=\pi$ if gap$<0$). \\
\bottomrule
\end{tabular}
\end{table}

\FloatBarrier
\subsection*{A.6. Independence Conditions}

\begin{table}[h!]
\scriptsize
\centering
\begin{tabular}{c|c|c}
\toprule
\textbf{Object Type} & \textbf{Numerical Independence Condition} & \textbf{Implication} \\
\midrule
\textbf{Propositions} & $i(AB)=i(A)+i(B)$ & $P(AB)=P(A)P(B)$ \\
\textbf{Questions}    & $d(a*b)=d(a)+d(b)$ & $P(AB)=P(A)P(B)$ \\
\textbf{Subjects}     & $i(ab)=i(a)+i(b)$ & $P(AB)=P(A)P(B)\;\text{or}\;A\sim B$ \\
\bottomrule
\end{tabular}
\end{table}

\FloatBarrier
\subsection*{A.7. Complex-Valued Properties}
{\scriptsize
\textbf{Definition:}
A function, $z(A)$, is a valid complex-valued property of the question $A?$ if:
\[
z(\neg A)=-\,z(A)^*.
\]

\noindent
\textbf{Example:}
\[
z(A)=\mathrm{gap}(A?)+i\,\mathrm{gmp}(q(A)).
\]

\noindent
\textbf{Allowed Non-Linear Operations} (satisfying $f(-z^*)=-f(z)^*$):
\[
\begin{aligned}
& f(z)=z^n\,z^{*m},\quad n+m\;\text{odd} \\
& f(z)=i\,z^n\,z^{*m},\quad n+m\;\text{even} \\
& f(z)=e^{-i\pi/(2n)}\,z^{1/n}\ (\text{upper half-plane})\\
& f(z)=e^{i\pi/(2n)}\,z^{1/n}\ (\text{lower half-plane}) \\
& \text{(Plus sums of the above terms with real coefficients.)}
\end{aligned}
\]
}

\FloatBarrier
\subsection*{A.8. Actions on Probability Distributions}

\begin{table}[h!]
\tiny
\centering
\begin{tabular}{c|c|c}
\toprule
\textbf{Action} & \textbf{Formula} & \textbf{Effect} \\
\midrule
\textbf{Give Proposition $X$}
 & $P(W)\to P(W\mid X)$
 & Sets $P(X)=1$ (adds information). \\

\textbf{Raise Pure Question $q(A)$}
 & $P(W)\to \tfrac{1}{2}\bigl[P(W\mid A)+P(W\mid\neg A)\bigr]$
 & Sets $\mathrm{gap}(q(A))=0$ (removes info). \\

\textbf{Raise Subject $ab$}
 & $P(W)\to \tfrac{1}{4}\sum P(W\mid\cdot)\ \text{(four terms)}$
 & Uniform over four alternatives of $a\circ b$. \\

\textbf{Ask Askable Question $A?$ (classical)}
 & $P(W)\to
   \begin{cases}
   P(W\mid X), & \text{with prob }P(X),\\
   P(W\mid\neg X), & \text{with prob }P(\neg X),
   \end{cases}$
 & Measurement: gain info, no info loss. \\

\textbf{Ask (2-state quantum measure)}
 & $P(W)\to P\bigl(W\mid \prod_{\theta,\phi}r_{\theta,\phi}\,X\bigr)$
 & Gains new info $X$, removes old spin info. \\
\bottomrule
\end{tabular}
\end{table}


\pagebreak

\section*{Appendix B: Physical Results}
\FloatBarrier
\subsection*{B.1. Two-State Quantum System}

\begin{table}[h!]
\scriptsize
\centering
\begin{tabular}{c|c}
\toprule
\textbf{General State Vector}
 & $\displaystyle \ket{\psi}=\cos\bigl(\tfrac{\theta}{2}\bigr)\ket{0}+e^{i\phi}\,\sin\bigl(\tfrac{\theta}{2}\bigr)\ket{1}.$ \\

\textbf{Probability Gap}
 & $\mathrm{gap}(W?)=\cos(\theta).$ \\

\textbf{Measurement Probability (Born Rule, 2-state)}
 & $\displaystyle P(W)=\cos^2\bigl(\tfrac{\theta}{2}\bigr)=\tfrac{1+\cos(\theta)}{2}.$ \\

\textbf{Collapse Example}
 & $P(W\mid Y)\;\to\;P(W\mid Y\,y\,X)=P(W\mid X).$ \\

\textbf{Post-Measurement Update}
 & $\prod_{\theta,\phi}r_{\theta,\phi}\;X$ (removes old info, gives $X$). \\
\bottomrule
\end{tabular}
\end{table}

\FloatBarrier
\subsection*{B.2. Entangled States}

\begin{table}[h!]
\scriptsize
\centering
\begin{tabular}{c|c}
\toprule
\textbf{Triplet Example}
 & $\ket{\uparrow}\ket{\uparrow}+\ket{\downarrow}\ket{\downarrow}$ (spins match along every axis). \\
\textbf{Initial Probability}
 & $P_{\psi}(W)=P_{\psi}\Bigl(W\Bigl|\prod_{\theta,\phi}S_{\theta,\phi}\Bigr)$. \\
\textbf{Non-Local Question}
 & $X_1?*X_2?$ (Asks if distant spins match along $x$). \\
\textbf{Correlation Function}
 & $c(A,B)=P(AB)+P(\neg A\,\neg B)-P(\neg A\,B)-P(A\,\neg B)=\mathrm{gap}(A?*B?)$. \\
\textbf{Post-Measurement Update}
 & $\prod_{\theta,\phi}s_{\theta,\phi}\;S_x\;X_1$ (removes old entanglement, sets new local info). \\
\bottomrule
\end{tabular}
\end{table}

\pagebreak


\begin{thebibliography}{99}
\bibitem{Shannon1948}
C. E. Shannon,
A Mathematical Theory of Communication,
\textit{Bell System Technical Journal} \textbf{27}, 379–423 (1948), DOI: \href{https://doi.org/10.1002/j.1538-7305.1948.tb01338.x}{10.1002/j.1538-7305.1948.tb01338.x}

\bibitem{wolframalpha}
Wolfram Alpha LLC. Retrieved from \url{https://www.wolframalpha.com/input?i=x\%28a-x\%29\%28b-x\%29\%281-a-b\%2Bx\%29+\%3D+a\%5E2+b\%5E2+\%281-a\%29\%5E2+\%281-b\%29\%5E2}, accessed on March 10, 2025.

\bibitem{Good1950}
I.J. Good,
\textit{Probability and the Weighing of Evidence},
Charles Griffin \& Co., 1950.

\bibitem{Kolmogorov1933}
A. N. Kolmogorov,
\textit{Foundations of the Theory of Probability},
Chelsea Publishing, 1956 (Russian original 1933).

\bibitem{Cox1961}
R. T. Cox,
\textit{The Algebra of Probable Inference},
Johns Hopkins Press, 1961.

\bibitem{Jaynes2003}
E. T. Jaynes,
\textit{Probability Theory: The Logic of Science},
Cambridge University Press, 2003, DOI: \href{https://doi.org/10.1017/CBO9780511790423}{10.1017/CBO9780511790423}.

\bibitem{Boole1854}
G. Boole,
\textit{An Investigation of the Laws of Thought on Which Are Founded
the Mathematical Theories of Logic and Probabilities},
Macmillan, 1854.

\bibitem{HammingLogic}
R. W. Hamming,
\textit{Coding and Information Theory}, 2nd ed.,
Prentice Hall, 1986. 

\bibitem{DiracQM}
P. A. M. Dirac,
\textit{The Principles of Quantum Mechanics},
4th ed., Oxford University Press, 1958.

\bibitem{Sakurai1994}
J. J. Sakurai,
\textit{Modern Quantum Mechanics},
Addison–Wesley, 1994 (1st ed. 1985).

\bibitem{Bloch1946}
F. Bloch,
Nuclear Induction,
\textit{Physical Review} \textbf{70}, 460–474 (1946), DOI: \href{https://doi.org/10.1103/PhysRev.70.460}{10.1103/PhysRev.70.460}.

\bibitem{Fano1957}
U. Fano,
Description of States in Quantum Mechanics by Density Matrix and Operator Techniques,
\textit{Reviews of Modern Physics} \textbf{29}, 74–93 (1957), DOI: \href{https://doi.org/10.1103/RevModPhys.29.74}{10.1103/RevModPhys.29.74}.

\bibitem{Bell1964}
J. S. Bell,
On the Einstein Podolsky Rosen Paradox,
\textit{Physics} \textbf{1}, 195–200 (1964), DOI: \href{https://doi.org/10.1103/PhysicsPhysiqueFizika.1.195}{10.1103/PhysicsPhysiqueFizika.1.195}.

\bibitem{chsh1969}
J.~F. Clauser, M.~A. Horne, A. Shimony, and R.~A. Holt,
Proposed Experiment to Test Local Hidden-Variable Theories,
\emph{Physical Review Letters}, \textbf{23}, 15, 880--884, (1969), DOI: \href{https://doi.org/10.1103/PhysRevLett.23.880}{10.1103/PhysRevLett.23.880}.

\bibitem{Aspect1982}
A. Aspect, P. Grangier, and G. Roger,
Experimental Realization of Einstein–Podolsky–Rosen–Bohm Gedankenexperiment:
A New Violation of Bell's Inequalities,
\textit{Physical Review Letters} \textbf{49}, 91–94 (1982), DOI: \href{https://doi.org/10.1103/PhysRevLett.49.91}{10.1103/PhysRevLett.49.91}.

\bibitem{Bennet2012}
A. J. Bennet, D. A. Evans, D. J. Saunders, C. Branciard, E. G. Cavalcanti, H. M. Wiseman, and G. J. Pryde,
Arbitrarily Loss-Tolerant Einstein-Podolsky-Rosen Steering Allowing a Demonstration over 1 km of Optical Fiber with No Detection Loophole,
\textit{Physical Review X} \textbf{2}, 031003 (2012), DOI: \href{https://doi.org/10.1103/PhysRevX.2.031003}{10.1103/PhysRevX.2.031003}.

\bibitem{Hensen2015}
B. Hensen, H. Bernien, A. Dreau et al. 
Loophole-free Bell inequality violation using electron spins separated by 1.3 kilometres.
\textit{Nature} \textbf{526}, 682–686 (2015), DOI: \href{https://doi.org/10.1038/nature15759}{10.1038/nature15759}.

\bibitem{Mermin1981}
N. D. Mermin,
Bringing Home the Atomic World: Quantum Mysteries for Anybody,
\textit{American Journal of Physics} \textbf{49}, 940 (1981), DOI: \href{https://doi.org/10.1119/1.12594}{10.1119/1.12594}.

\bibitem{WheelerZurek}
J. A. Wheeler and W. H. Zurek (eds.),
\textit{Quantum Theory and Measurement},
Princeton University Press, 1983.

\bibitem{Gleason1957}
A. M. Gleason,
Measures on the Closed Subspaces of a Hilbert Space,
\textit{Journal of Mathematics and Mechanics} \textbf{6}, 885–893 (1957).

\bibitem{BirkhoffVonNeumann}
G. Birkhoff and J. von Neumann,
The Logic of Quantum Mechanics,
\textit{Annals of Mathematics} \textbf{37}, 823–843 (1936).

\bibitem{caves2002}
C.~M. Caves, C.~A. Fuchs, and R.~Schack, Quantum probabilities as Bayesian probabilities, \textit{Physical Review A}, \textbf{65}, 2, Jan. 2002. DOI: \href{https://doi.org/10.1103/PhysRevA.65.022305}{10.1103/PhysRevA.65.022305}.

\bibitem{fuchs2013}
C. A. Fuchs and R. Schack,
Quantum-Bayesian coherence,
\textit{Rev. Mod. Phys.} \textbf{85}, 4, 1693–1715 (2013), DOI: \href{https://doi.org/10.1103/RevModPhys.85.1693}{10.1103/RevModPhys.85.1693}.

\bibitem{schack2001}
R.~Schack, T.~A. Brun, and C.~M. Caves, Quantum Bayes rule, \textit{Physical Review A}, \textbf{64}, 1, Jun. 2001. DOI: \href{https://doi.org/10.1103/PhysRevA.64.014305}{10.1103/PhysRevA.64.014305}.

\bibitem{spekkens2007}
R. W. Spekkens,
Evidence for the epistemic view of quantum states: A toy theory,
\textit{Phys. Rev. A} \textbf{75}, 3, 032110 (2007), DOI: \href{https://doi.org/10.1103/PhysRevA.75.032110}{10.1103/PhysRevA.75.032110}.

\bibitem{Collingwood1939}
R. G. Collingwood,
\textit{An Essay on Metaphysics},
Clarendon Press, 1939.

\bibitem{Wisniewski1995}
A. Wi\'sniewski,
\textit{The Posing of Questions: Logical Foundations of Erotetic Inferences},
Springer, 1995.

\bibitem{BelnapSteel1976} N. Belnap and T. B. Steel, \textit{The Logic of Questions and Answers}. Yale University Press, 1976.

\bibitem{Harrah2002} D. Harrah, The Logic of Questions, in \textit{Handbook of Philosophical Logic} (D. Gabbay and F. Guenthner, eds.), 2nd ed., vol. 8, Kluwer, 2002, pp. 1--60.

\bibitem{Aqvist1965} I. \r{A}qvist, \textit{A New Approach to the Logical Theory of Interrogatives,} Uppsala, 1965.



\end{thebibliography}
\end{document}